\documentclass[12pt, a4paper, twoside, openright]{report}

\usepackage[a4paper]{geometry}
\usepackage{fullpage}
\usepackage{epsfig,float,subfig}
\usepackage[usenames]{color}
\usepackage[font=footnotesize]{caption}

\usepackage{lmodern}

\usepackage{amssymb}
\usepackage{amsmath}
\usepackage{MnSymbol}
\usepackage{array}
\usepackage{mathrsfs}

\usepackage{quotchap}

\usepackage[usenames]{color}
\usepackage[sort&compress,numbers]{natbib}
\usepackage{enumerate}
\usepackage[breaklinks=true,a4paper=true,pagebackref=false]{hyperref}
\usepackage{lineno}

\usepackage{atbegshi}
\AtBeginDocument{\AtBeginShipoutNext{\AtBeginShipoutDiscard}}

\newcommand{\ve}[1]{\mathbf{#1}}

\newcommand{\avr}[1]{\langle #1\rangle}
\definecolor{!R}{rgb}{1,0,0}
\definecolor{!G}{rgb}{0,1,0}
\definecolor{!B}{rgb}{0,0,1}
\newcommand{\rem}[1]{#1}
\newcommand{\remB}[1]{#1}

\newcommand{\intlim}[3]{\int\limits_{#2}^{#3}\!\!\mathrm{d}#1}
\newcommand{\intpos}[1]{\int\limits_{#1}^+\!\!}
\newcommand{\Int}[1]{\int\limits_{#1}\!\!}
\newcommand{\re}{\mathrm{Re}}
\newcommand{\im}{\mathrm{Im}}

\newcommand{\kelphi}[2]{\varphi_{#1}^{(#2)}}
\newcommand{\kelphid}[2]{(\varphi_{#1}^\dagger)^{(#2)}}
\newcommand{\MX}[4]{\ensuremath{\left(\begin{array}{cc} #1 & #2 \\ #3 & #4 \end{array}\right)}}

\newcommand{\LNNL}{}
\newcommand{\LNFIG}{}

\newcommand{\FINAL}[1]{#1}

\usepackage[T1]{fontenc}
\usepackage{url}
\usepackage[svgnames]{xcolor}
\ifpdf
\usepackage{pdfcolmk}
\fi
\ifxetex
\usepackage{fontspec}
\fi
\usepackage{graphicx}

\usepackage{tikz}
\usepackage{pifont}
\providecommand{\HUGE}{\Huge}
\newlength{\drop}

\definecolor{Dark}{gray}{0.2}
\definecolor{MedDark}{gray}{0.4}
\definecolor{Medium}{gray}{0.6}
\definecolor{Light}{gray}{0.8}
\makeatletter
\DeclareRobustCommand\ltseries
{\not@math@alphabet\ltseries\relax
\fontseries\ltdefault\selectfont}
\newcommand{\ltdefault}{l}
\DeclareTextFontCommand{\textlt}{\ltseries}
\DeclareRobustCommand\hbseries
{\not@math@alphabet\hbseries\relax
\fontseries\hbdefault\selectfont}
\newcommand{\hbdefault}{hb}
\DeclareTextFontCommand{\texthb}{\hbseries}
\makeatother
\newcommand*{\titleGM}{\begingroup
\vspace*{2.5\baselineskip}
\vfill
\hbox{%
\hspace*{0.1\textwidth}%
\rule{1pt}{1.0\textheight}
\hspace*{0.05\textwidth}%
\parbox[b]{0.9\textwidth}{
\vbox{%
\vspace{\drop}
{\noindent\HUGE\bfseries Statistical and \\[0.45\baselineskip]
\noindent\HUGE\bfseries Thermodynamical \\[0.3\baselineskip]
\noindent\HUGE\bfseries Studies of the \\[0.5\baselineskip]
\noindent\HUGE\bfseries Strongly Interacting Matter\\ [0.5\baselineskip]
}\\[1\baselineskip]
{\Large PhD thesis}\\[8\baselineskip]
{\Large {MIKL\'OS HORV\'ATH}}\\[2\baselineskip]
{\large \textit{Advisor:} Prof. Tam\'as S\'andor Bir\'o \\ \textit{Consultant:} Prof. Antal Jakov\'ac}\par
\vspace{0.15\textheight}
{\noindent \itshape \href{http://wigner.mta.hu}{Wigner Research Centre for Physics} \& \\
\href{http://www.bme.hu}{Budapest University of Technology and Economics} \\ \itshape 2016}\\[6\baselineskip]
}
}
}
\vfill
\null
\endgroup}

\begin{document}
\pagestyle{empty}
\setlength{\parskip}{0cm}

%
%

\titleGM
\clearpage

%
%

\vspace*{0.25\textheight}
\hspace*{0.6\textwidth} {\large\textit{To my parents \& my brother}}\\
\newpage
\chapter*{Abstract}
\thispagestyle{empty}
In this thesis we discuss three separate analysis of various phenomenological aspects of heavy-ion collisions (HIC). The first one is a possible generalization of the kinetic theory framework for dense systems. We investigate its long-time behaviour and the properties of the equilibrium. The second discussion is about the phenomenological analysis of the azimuthal asymmetry of the particle yields in a HIC, where we link the initial stage geometrical asymmetry to the particle yields and examine the possible organizing mechanisms that could be responsible for such a relation. The third, and also the most thorough part of the thesis is about the relation of the spectral density of quasi-particle states and the macroscopic fluidity measure ($\eta/s$) and other transport properties of the system. We extensively study the liquid--gas crossover with the help of model spectral functions. The main conclusion is that the relative intensification of the continuum of the scattering states compared to the quasi-particle peak makes the matter more fluent. \par
These different effective models have an important feature in common: the phenomenology in all three cases is rooted in the medium-modification of the microscopic degrees of freedom.
\addtocontents{toc}{\protect\thispagestyle{empty}}
\tableofcontents
\newpage
\pagestyle{plain}
\setcounter{page}{1}
\section*{Abbrevations}
\begin{center}
\begin{tabular}{cl}
2PI & 2-particle irreducible \\
AdS/CFT & anti-de Sitter / conformal field theory \\
BE & Boltzmann equation \\
CEP & critical endpoint \\
CGC & color-glass condensate \\
CME & chiral magnetic effect \\
DoF & degree of freedom \\
DoS & density of states \\
EM & electromagnetic \\
EoM & equation of motion \\
EoS & equation of state \\
(E)QP & (extended) quasi-particle \\
FRG & functional renormalization group \\
HIC & heavy-ion collision\\
IR & infrared \\
KMS & Kubo-Martin-Schwinger \\
LHC & Large Hadron Collider \\
MBE & modified BE \\
MMBE & multi-component MBE \\
QFT & quantum field theory \\
QGP & quark-gluon plasma\\
QCD & quantum chromodynamics\\
RHIC & Relativistic Heavi-Ion Collider \\
\end{tabular}
\end{center}

\section*{Special functions}
\begin{center}
\begin{tabular}{cl}
$\theta(x)$ & Heaviside step-function, 1 for $x>0$, 0 elsewhere \\
$\theta_\epsilon(x)$ & approximation for the Heaviside step-function \\
$\Theta(x)$ & window-function, equals 1 if $x$ is true, 0 if it is false\\
$\delta(x)$ & Dirac-delta distribution\\
$\delta_\epsilon(x)$ & approximating delta-function \\
$J_0$, $J_1$, \dots & Bessel functions of the first kind \\
$K_0$, $K_1$, \dots & modified Bessel functions
\end{tabular}
\end{center}

\newpage
\section*{Notations \& conventions}
\begin{center}
\begin{tabular}{cl}
$p$ & In case of vector variables, we distinguish Lorentz four-vectors (italic)\\
$\ve{p}$ & and Euclidean three-vectors (bold) \\
$p^\mu$ & We use Greek indices for the components of four-vectors. \\
$p^i$ & Latin indices usually indicate three-vector components \\
& or the three-vector part of a four-vector. \\
$T^{\mu\nu}$, $T^{ij}$ & The same goes for tensors. \\
$\eta^{\mu\nu}$ & We use the signature $(+,\,-,\,-,\,-)$ for the metric tensor. \\
$u_\mu u^\mu$, & Repeated indices means summation, \\
$w_iw_i$ & irrespectively the position of the indices.
\end{tabular}
\end{center}

\section*{Units}
Unless it is specifically stated, the identification ${k_B =\hbar =c =1}$ system of units is used. This allows us to express all the units in energy dimension, for example in electronvolts:
\[\begin{array}{ccccc}
[\text{energy, mass}] = eV, & & [\text{distance}] =eV^{-1}, & & [\text{velocity}] =eV^0,
\end{array}\]
and velocity becomes dimensionless. We note that to change length units into energy, one should keep in mind ${1 GeV^{-1}\approx 0.197 fm}$, which will be used in Sec.~\ref{ellflow}.

\newpage
\chapter[Effective models in the physics of heavy-ion collisions]{Effective models in the\\ physics of heavy-ion collisions\label{introduction}}

The focus of high-energy physics in the recent few decades were undoubtedly about the properties of the strongly interacting matter. Conclusive evidences show, that on very high energy density, the so-called quark-gluon-plasma (QGP) is formed\footnote{Traces were first found in the RHIC experiment at BNL, USA \cite{QGPevidence}.}, a state of the matter ruled completely by the strong interaction. The details of the formation, material properties and dynamics of the QGP are, however, to be revealed. \par
The standard theoretical framework of the strong interaction is quantum chromodynamics (QCD). A peculiar feature of QCD, that the fundamental degrees of freedom i.e.\ the quarks and gluons are confined, therefore the dynamics of the observables is emergent. From the theoretical point of view, it is a great challenge to develop the methodology which is capable of dealing with the non-perturbative nature of QCD. The quantitatively most accurate tool, lattice QCD (lQCD), is not yet able to describe out-of-equilibrium situations. Therefore theorists construct effective models concerning the different behaviour of QCD on different energy scales.\par
From the experimental side, the main focus of attention is on the high-energy particle accelerator facilities. The purpose of particle accelerator experiments, putting it simply, is to make extreme conditions and to see what happens with the matter under such circumstances. Such investigations take place on extreme high energy density in the RHIC and the LHC experiments, but this can be also achieved on large nuclear density, as it is planned in the FAIR experiment. By every new experimental findings of these research projects, a further region of the phase diagram of the strongly interacting matter (SIM) is explored. We already have an approximate picture about the phases of the SIM, many details, however, still remain unknown. Is there a critical end point (CEP) of the deconfinement phase boundary, as non-perturbative investigation of QCD suggests? If so, on which temperature and chemical potential? Is the deconfinement transition a 1${}^\text{st}$ order one on this phase boundary? Programs, like the beam energy scan in RHIC are aimed to answer such questions.\par
In this chapter, we summarize some of the key concepts which nowadays drive the investigation of heavy-ion collisions (HIC). We also try to cover the most important prevailing problems and challenges from the theoretical point of view.

\subsection*{The typical history of a heavy-ion collision event}
\begin{figure}[!t]
\centering
\includegraphics[width=0.75\linewidth]{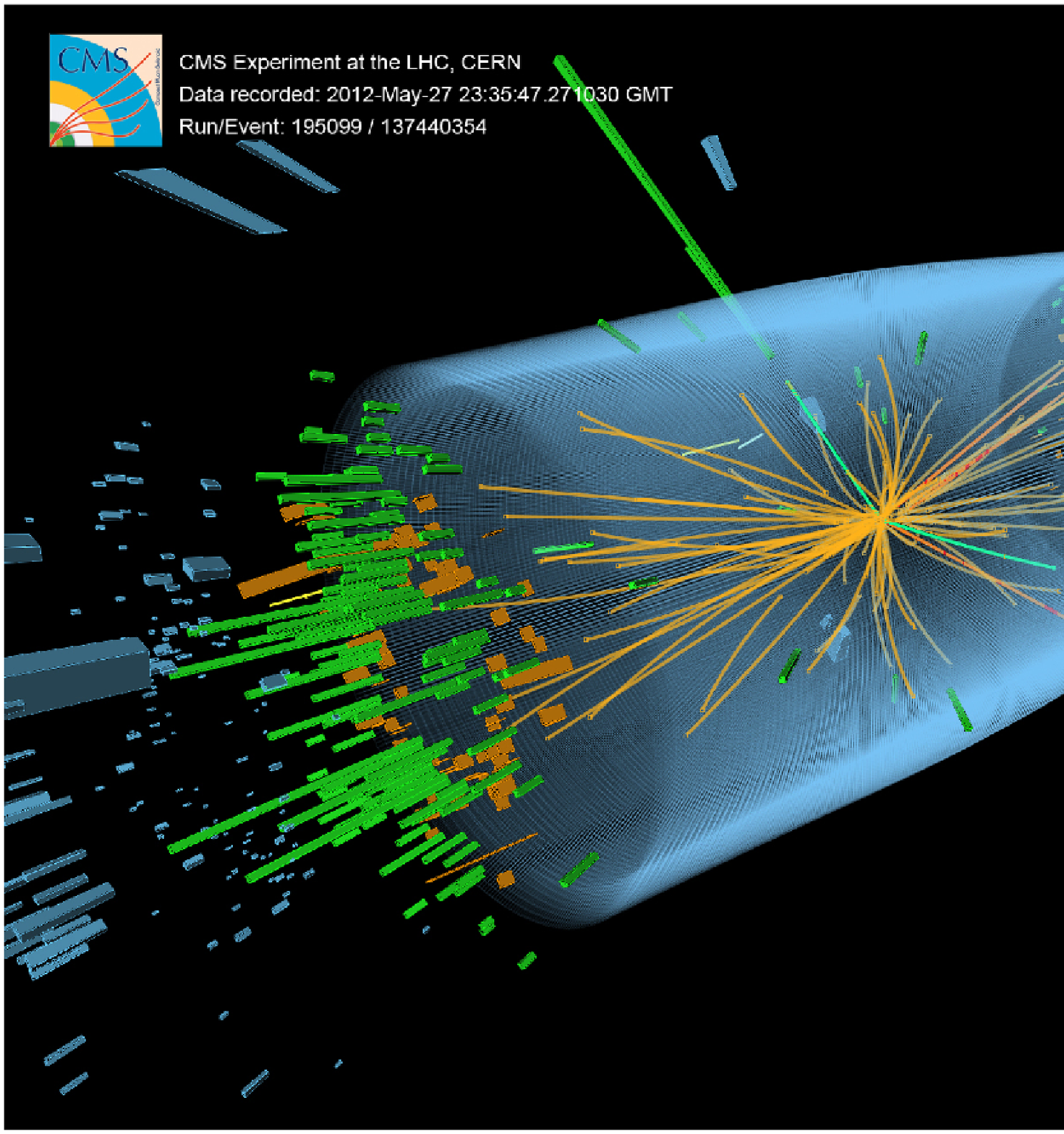}
\LNFIG
\caption{Particle traces recorded by the \href{http://cms.web.cern.ch/}{CMS experinment} at LHC}
\label{fig:particletraces}
\end{figure}
When two protons or heavy ions collide with nearly equal sized, but opposite velocities and kinetic energies comparable to their rest mass, it is likely according to QCD, that one finds newly formed hadrons in the final state. This is indeed the case, the experienced outcome of a collision has very rich structure both in the observed hadronic species and the distribution of those, see Fig.~\ref{fig:particletraces} as a demonstration. So it is not surprising that the few fm/c long history after the collision and before the hadron formation is also structured. The time-line of a HIC, the often called ''little bang'', is depicted on Fig.~\ref{fig:HIChistory}. Now we enumerate the important stages which requires different theoretical framework to care about. This chain of models usually called the standard model of the HIC:
\begin{figure}[!t]
\centering
\includegraphics[width=0.75\linewidth]{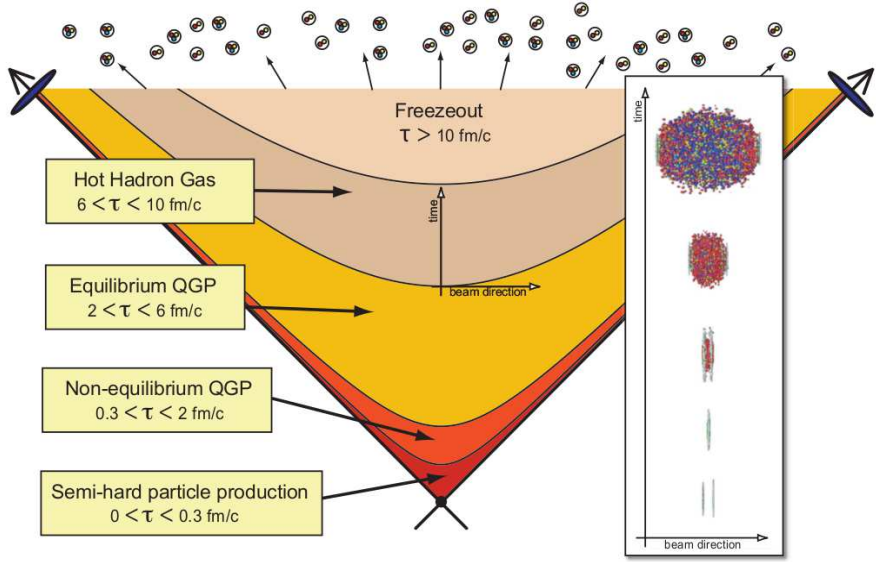}
\LNFIG
\caption{Various stages of the QGP created in a HIC. This cartoon depicts the space-time history of the plasma, indicating the different phases from the theoretical modelling point-of-view. The figure is taken from \cite{aHydro}.}
\label{fig:HIChistory}
\end{figure}
\begin{itemize}
\item[\it i.)] After the dissociation of the participating nuclei, the average energy density of a fairly big domain of the system -- compared to its overall size -- quickly exceeds $\Lambda_\text{QCD}^4$\footnote{The typical energy-scale of deconfinment.}. In this region, presumably, QGP is formed.
\end{itemize}
For a while, the QGP is in a highly non-equilibrium state in which quantum fluctuations possibly have an important role. This domain is poorly understood theoretically. Several attempts for its description exist, including semi-classical Yang-Mills lattice field theory\footnote{The so-called color glass condensate (CGC), for an introductory review see Ref.~\cite{introCGC} and the references therein.} and even kinetic theory\footnote{With processes like gluon fusion and quark decay, see Ref.~\cite{introBAMPS} for example.}. Perhaps the least well-explained feature at this stage is the surprisingly fast equilibration. This can be estimated only of course if the other stages of the time-evolution are somehow modelled.
\begin{itemize}
\item[\it ii.)] In the equilibrated plasma\footnote{Supposedly such a phase of the time evolution exists. Nevertheless, the hydrodynamic simulations suppose that the fluid is in local thermal equilibrium.}, the hydrodynamical modes seem to be ruling the next stage of the time-evolution. The QGP is expanding whilst cooling.
\end{itemize}
Using CGC-motivated coarse-grained initial conditions for the energy density, dissipative hydrodynamics describes the expanding stage very well with almost zero shear viscosity, see for example Ref.~\cite{introHydro} and also the references therein. This means that viscous hydrodynamics can be well embedded into the model-chain of the HIC. It reproduces the particle spectra and correlations if a simplified (Cooper--Frye-like) hadronization picture is applied. Moreover, it requires only one or two free parameters to fix -- the viscosities -- besides the initial densities.\par
However, it is not yet explained satisfactorily why the hydrodynamic degrees of freedom (DoF) catches the physics of this stage of the QGP that well. Given the fact, it basically means that one follows the time-evolution of the conserved charges and energy-momentum of the long-wavelength modes of the underlying QFT only. Another intriguing question is whether correlations not captured in the CGC can considerably grow in this expanding period \cite{colorantennas}.
\begin{itemize}
\item[\it iii.)] During the hydrodynamic evolution, the QGP expands and cools down. Moreover, as its energy density drops below $\Lambda_\text{QCD}^4$, the plasma goes through the confinement phase-transition, it hadronizes. This makes kinetic theory feasible to describe its evolution.
\end{itemize}
To match hadronic species in the kinetic theory to the final densities of the hydrodynamical simulation, usually the Cooper--Frye-formula is used \cite{introHydro}. This is based on the equality of conserved currents of hydrodynamics and those of kinetic theory on a space-like hypersurface \cite{introHIC}. There are also more detailed models of the hadronization, with long-established history (color ropes, etc.\!, see for example Ref.~\cite{colorRopes}). How important this details are concerning the overall picture, however, is still the subject of debates.
\begin{itemize}
\item[\it iv.)] As the last two noticeable events of the time-evolution, the hadronic gas-mixture freezes out. First chemically: inter-hadronic changes and decays stop, then kinetically: it becomes so dilute that no more scattering happens.
\end{itemize}
After the freeze-out, the hadronic products stream freely into the detectors, leaving there the traces of the final stage spatial and momentum distribution. The imprinted energy and momentum distributions of hadrons suggest, that the medium they left behind is in thermodynamic equilibrium. Transverse momentum spectra from proton-proton and heavy nucleus collisions, however, show a deviation on {high-$p_T$} from the exponential fall off in the Boltzmann distribution. This issue, which is not fully understood up to now, can be caused by several phenomena. Multiplicity and/or temperature fluctuations of a thermal bath with finite heat capacity could be responsible. Another possible cause may be the modified kinetic evolution due to the medium, which we discuss briefly in Chap.~\ref{MMBE}. We note for the sake of completeness, that the power-law high-momentum tail can be the direct effect of perturbative QCD evolution \cite{hadronizatonPartonRecomb}.\par
As one can see, following the whole history of a HIC, from the collision to it signals the detectors, is very complicated and needs the use of a whole variety of theoretical tools. The main difficulty here is to link the evolution of the various stages together. Since dividing the HIC into those stages is arbitrary to a certain degree and introduces further parameters to fix -- often by hand --, this may be the most serious source of uncertainty of the modelling approach.

\subsection*{About the conjectured phase diagram of the strongly interacting matter}
\begin{figure}[!t]
\centering
\subfloat[]{
\includegraphics[width=0.47\linewidth]{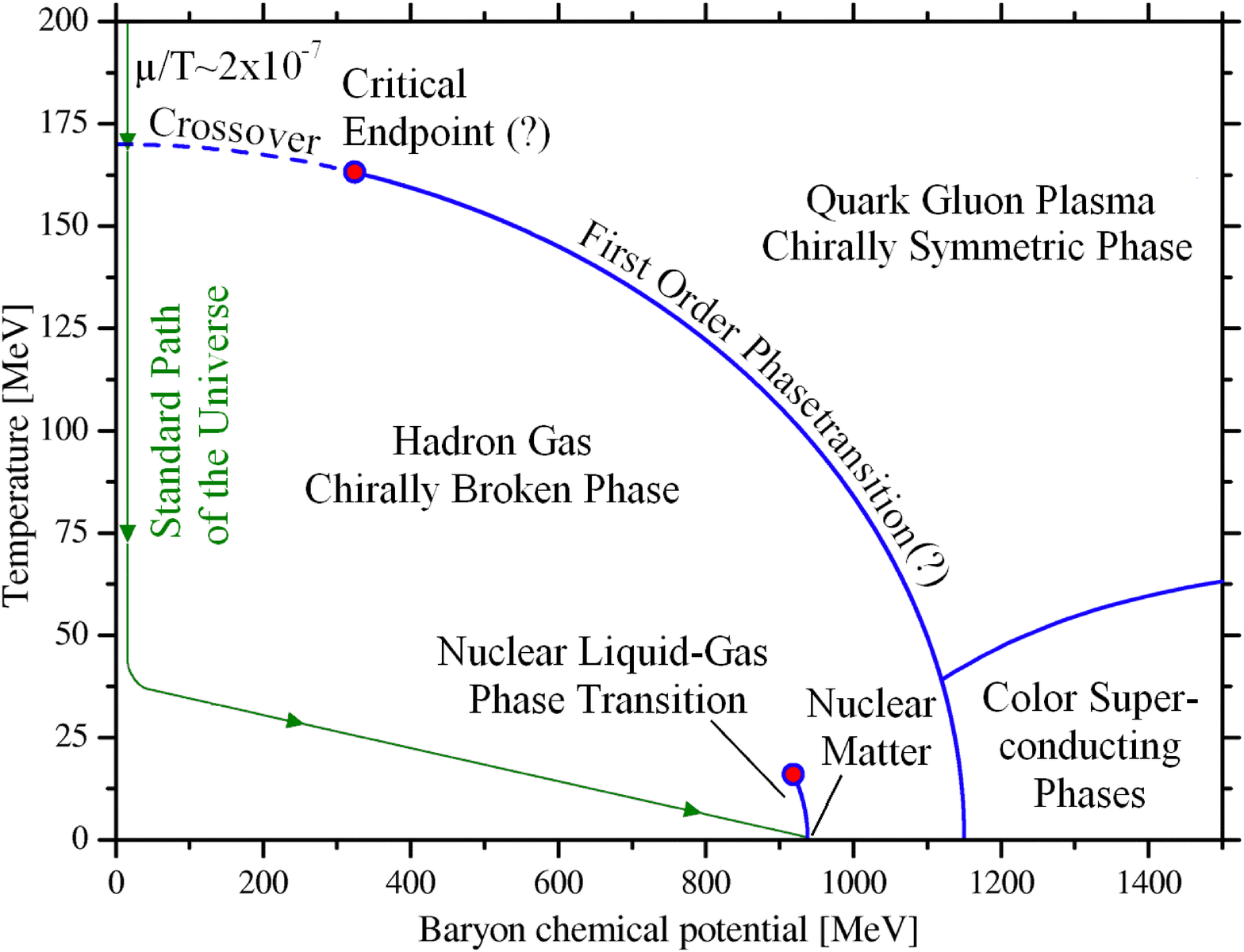}
}
\subfloat[]{
\includegraphics[width=0.53\linewidth]{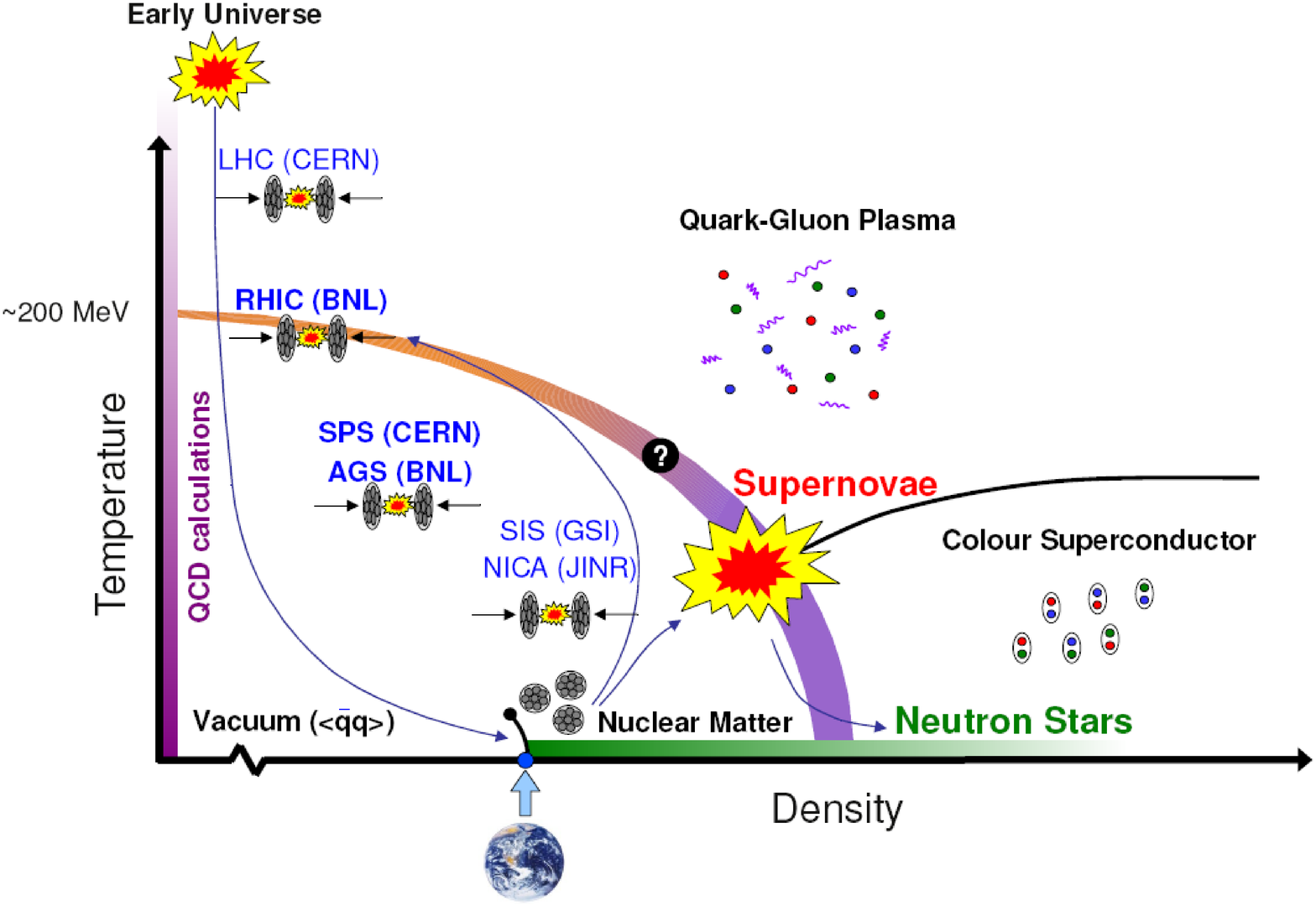}
}
\LNFIG
\caption{Sketches of the possible QCD phase diagram. Figure (a) is taken from \cite{QCDphases1}, (b) is from \cite{QCDphases2}.}
\label{fig:QCDphasediagram}
\end{figure}

As we outlined in the previous section, the so-called standard model of the HIC is not a first-principle description, although, is motivated by and uses the experience physicists gained by exploring QCD. We have also seen, that the non-perturbative analysis of QCD on a wide range of energy scales would be necessary in the way towards a full first-principle explanation. For example, it is not yet known, why the SIM behaves as a nearly perfect liquid for a considerable period of the time-evolution. In that sense, the investigation of data -- taken from collision experiments -- by hydrodynamical simulations serves as a tool to get a better understanding about the QCD phase diagram\footnote{The beam-energy scan program of RHIC is aimed to do that.}. This also complements the lattice field theory simulation of QCD, which is unfortunately unavailable for large bariochemical potentials $\mu_B$ at the present time\footnote{Since the hydrodynamical simulations often use the lattice equation of state (EoS) recorded at ${\mu_B=0}$, this raises questions about the reliability of these simulations near to the CEP.}. The most prominent feature of the phase diagram is that the QGP--hadron gas boundary ends with a critical endpoint (CEP) on finite $\mu_B$, see Fig.~\ref{fig:QCDphasediagram}. The existence of the CEP is widely accepted, since many effective low-energy model studies confirmed it \cite{cepKP1, cepKP2, cepKP3}, although no first-principle QCD calculations yet. \rem{Below this critical value of $\mu_B$, there is a crossover, as it was also shown by the ${\mu_B=0}$ lattice QCD simulations almost a decade ago. In this region, neither the Polyakov-loop\footnote{The Polyakov-loop serves as the order parameter of the deconfinement phase transition. However, it is a real order parameter only in the pure Yang-Mills theory, where it is connected to the spontaneous breaking of the center symmetry.} susceptibility diverges, nor the chiral one does\footnote{The so-called chiral condensate $\avr{\overline{\psi}\psi}$ of the fermion fields. The temperature of chiral symmetry restoration (when the chiral condensate becomes zero) happens to be very close to the critical temperature of the deconfinement transition. Until now, there is no theoretical explanation for this phenomenon.} \cite{latticeQCDresults}. \par
It is both an interesting and a highly non-trivial question, that which measurable quantities can map the phase-transition of the SIM to the measured yields and distribution of the produced particles. We emphasize few important faces of this experimental challenge here:
\begin{itemize}
\item[\textit{i.)}] The modified properties of hadrons in medium can indicate the restoration of the chiral symmetry. The in-medium spectral function of low-mass vector mesons like $\rho$, $\omega$ or $\phi$ can be measured directly via their decays into dilepton pairs. The investigation of particles containing charm quarks is also promising. The charmonium ($\avr{\overline{c}c}$-condensate) is sensitive to the screening effects present in the QGP. The suppression of charmonium is reflected in the relative decrease of the $J/\psi$ yield\footnote{This can be measured via its decay into electron-positron pairs.} and also in the change of the effective mass of the $D$-mesons\footnote{It is measurable through its hadronic decay products.} \cite{FAIRtechrep}. \par
\item[\textit{ii.)}] The increase in the production of strange particles can possibly signal the deconfinement transition. This can be reflected as the multiplicity enhancement of $\Xi$ and $\Omega$ hyperons as a function of the beam-energy (compared to pions, for example) \cite{FAIRtechrep}. \par
\item[\textit{iii.)}] The sudden, non-monotonic beam-energy dependence of event-by-event fluctuations can be suitable for the direct observation of a phase transition. These effects might be particularly pronounced near the CEP. Certain investigations of femtoscopic observables like the difference of Gaussian emission source radii revealed scaling behaviour similar to the finite-size scaling patterns near the critical point \cite{cepLacey}. \par
\end{itemize}
}
Phase diagrams of this kind are not rare in nature. For example, water and carbon dioxide have similar phase structure with a CEP. In Chapter \ref{etaOs} we present an effective model study aimed to describe the crossover region near to a possible CEP. \par

\section{Probing the quark-gluon plasma}
In a HIC, fluctuations and correlations of the very early times evolve to the macroscopic patterns we are able to detect. As these fluctuations propagate through the medium they serve as probes of the SIM. If we are able to identify processes which probe a specific aspect of the interaction of a ''test'' DoF (a quasi-particle) with the medium, we have the opportunity to compare the underlying theory to the experiment.\par
Supposing that QCD (possibly supplemented with QED) would be enough to describe a HIC on a first-principle basis, lattice field theory computations can be the framework of such a comparison for the SIM in thermal equilibrium only. The EoS\footnote{We mean the relation between the pressure and energy density by the equation of state.} of lQCD is reflected in the spectra of emitted particles. The emission of photons and leptons is connected to the electromagnetic response function of the QGP, also a possible subject of lattice computations. The color screening length (or the inverse Debye-mass), which is reflected in the lifetime of heavy quark flavours in the plasma, is also reliably accessible on the lattice.\par
The non-equilibrium properties like the diffusion and the linear energy loss of the fast partonic DoF (related to jet-quenching\footnote{Jets are narrow, cone-like showers of hadrons, initiated by energetic partons: They can loose energy and broaden while propagating through the QGP. Their contribution to the particle yield is thought to be mainly by intense gluon bremsstrahlung radiation}) are, however, out of the scope of lattice field theory. We mention here some of those important measurable quantities:
\begin{itemize}
\item The fluctuations of the final flow profile, i.e.\! the angular dependence of the yields and correlations, probes the expansion dynamics of the plasma. The result, however, is also affected by how we model the initial fluctuations (the most widespread is CGC) and possibly by final-state interactions.
\item \rem{The penetrating probes of photons and jets tries the medium properties of the plasma, but carry the effect of the interactions in the pre-plasma stage as well. Also, the medium properties presumably have a strong time-dependence even in the stage accessible by hydrodynamics, let alone in the regime between the strong and weak parton-medium coupling, for which we do not have any description at the time being.}
\item The analysis of the hadron yields, which serves as chemical probes, have an imprinted uncertainty caused by the model one uses for converting the energy-momentum and charge densities of the hydrodynamics into hadrons.
\end{itemize}
Thus, the main difficulty reveals itself when one tries to evaluate such observables probing the SIM for which a full treatment of the standard model of HIC is needed. Each of those quantities store information from many stages of the time-evolution. Therefore possible uncertainties arise also from our preconception in which stage of the collision we consider a given mechanism important enough to produce a measurable effect.\par
For a thorough overview about the probes of the QCD matter, see Ref.~\cite{QCDprobes} and the references therein.

\section{Effective modelling}
As we have seen so far, the description of a HIC is a complex task. Although the initial state of such events are relatively well-controlled, the final state is inevitably contaminated by collective effects. One should extract information about the QGP via particles captured by the detectors. These particles are formed in and influenced by various stages of the evolution of the plasma: the matter expands and cools down while it radiates photons. Than hadrons are formed, the decay products of hadrons reach the detectors alongside photons and other debris from the plasma. This imprint is mediated by effects of interactions out of the QGP-phase and also the collective motion of the system. Since HICs encompass such a rich phenomenology, the use of effective models is necessary. Although, a comprehensive theory based on QCD has not yet been established.\par
Using effective theories is often chosen to attack a given phenomenon in the field of high-energy particle physics anyway. A good effective model is able to tackle the key features of a physical phenomenon, yet reduces the number of degrees of freedom such that quantitative handling of the problem becomes possible. We emphasize that building an effective model is not the procrastination of a more determined investigation, on the contrary: it is a natural part of the modelling process. We use effective theories if the one thought to be fundamental is too difficult to tackle, but also when we do not yet have a clear conception what this underlying theory would have to be.\par
Let us give few examples, how one can usually gain an advantage from effective modelling, particularly in the context of HICs:
\begin{itemize}
\item An effective description deals only with those DoF which are essential in terms of the considered physics. It can happen that the elementary DoF are not the relevant ones or a ''reordering'' of them occurs below or above a given energy scale. QCD is a typical example, which has weakly interacting quarks and gluons at high energy density -- asymptotic freedom --, but the strongly coupled bound states, i.e.\! hadrons, at low energy density as relevant DoF -- confinement. \rem{Models of low-energy QCD are built on hadronic, i.e.\! color-neutral DoF: this way the strongly interacting theory of elementary objects becomes weakly interacting with composite ones from the viewpoint of the fundamental theory -- at least in the sense that it is tractable with perturbation theory.}
\item One is often satisfied if the effective theory in question is able to reproduce a certain phenomenon, even if it has no real predictive power. That is, by tuning a set of phenomenological parameters (i.e.\! parameters not necessarily linked to the fundamental theory) it describes experimental data. In other words our model is capable to reinterpret measurable quantities as a set of parameters with -- hopefully clear -- physical meaning. Our understanding of the phenomenon is then represented as how we motivated the choice of this particular set of variables and how those are possibly related to the fundamental theory -- if there exists any in the field of our investigation. In Chap.~\ref{ellflow} we show an example of this through the reproduction of the elliptic flow coefficient $v_2$ measured in HICs, using a phenomenological model.
\item An effective description might help to understand which features of the phenomenon are specific to the given situation and which ones are the most profound. Why is viscous hydrodynamics able to describe the expansion of the thermalized QGP? Is there a universal behaviour in the background which connects hydrodynamics also to other QFTs with a CEP in its phase diagram as well? In Chap.~\ref{etaOs} we raise the question whether a certain aspect of the quasi-particle (QP) spectrum of a field theory is reflected in its low-viscous liquid-like behaviour.
\end{itemize}
We close this short overview by mentioning few examples to effective models widely used nowadays in the description of HICs:
\begin{itemize}
\item The color glass condensate (CGC) model is an effective field theory (EFT) aimed to describe the pre-hydrodynamic non-equilibrium evolution of the QGP. It is motivated by the partonic picture of a relativistically fast nucleus. That is, in a relativistic HIC, from a rest frame of a participating nucleus, the quarks and gluons of the other one moving with relativistic speed can be treated as free particles for the time period of the collision. The reason of this is the relativistic time dilation of the time-scales of the fast moving nucleus. This eventually allows to view the constituents -- which are deeply off-shell in the rest frame of the same nucleus -- as on-shell particles. \rem{The CGC model utilizes that the valence quarks act as sources of the gluon field, which also can be approximated classically to leading order due to the high intensity of the field (or equivalently the high occupation number of the gluons). See Ref.~\cite{introCGC} for a detailed introduction.}
\item Viscous hydrodynamics is a widely accepted framework to follow the expansion of the equilibrated QGP. It catches the relevant dynamics of the strongly coupled plasma specified by only through its EoS and two viscosity coefficients. For a thorough review see Ref.~\cite{introHydro} and the references therein.
\item A reformulation of strongly coupled QFTs is perhaps possible in a wide range of theories as a weakly interacting (gravitational) theory due to the so-called holographic principles, as is the AdS/CFT duality \cite{maldacena1, maldacena2, qm2006mclerran}. The thermalization of a (CFT) plasma being out of equilibrium then can be paralleled with the dynamics of the black hole formation in the dual gravity theory.
\end{itemize}

\section{Outline of the thesis}
In this section we briefly summarize the topics we shall discuss in this thesis. In Chapters \ref{MMBE}, \ref{ellflow} and \ref{etaOs} one finds three independent studies, all of which are related to the phenomenology of HICs. The order of those chapters is chronological, starting with the most early topic the author has been involved with. Nonetheless, presumably Chapter \ref{etaOs} contains the most thorough investigation and is the most original part of this thesis, according to the author's somewhat biased opinion. That chapter is about the most recent results of the work in which the author has been engaged the longest coherent time period during his work so far. Since the different chapters are sought to be self-contained, it is up to the reader to tackle them in any order he or she feels pleasant.\par
Let us now emphasize the common aspects of these problems which allow us to join them in this thesis. The most obvious reason of the same motivation, namely the phenomenological description of HICs. But there is more than that. In all cases we try to address problems in which the underlying mechanism is not understood in terms of a fundamental theory. However, the in-medium effects have a crucial role in all three models we present. The dynamics tests itself -- so to say --, as the elastic scattering of particles is modified in Chap.~\ref{MMBE}, as the quasi-particle-like source pairs decelerates whilst propagating through the plasma created in the collision in Chap.~\ref{ellflow}, or as in Chap.~\ref{etaOs}, where the QP-properties of the DoF are fundamentally altered by colliding with the surrounding medium-particles. This motive unifies these investigations, and also puts a clear urge to understand the medium interactions as a requirement for an underlying theory of HICs.\par
So finally, here is what the interested reader can find:
\begin{itemize}
\item[--] In \textbf{Chapter \ref{MMBE}} we present a kinetic theory model where the elastic nature of the two-particle collisions is substituted by a different kinetic energy constraint. The result is the modified detailed balance distribution which a one-component gas achieves as an equilibrium state. On the contrary, for a two-component system, this modification leads to the lack of equilibrium solution. We present the analysis of the long-time evolution in that case, and also the possible recovery of the equilibration by an appropriate feedback of the kinetic energy constraint from the dynamics.\par
The results of this chapter are published in Ref.~\cite{MMBEpaper}.
\item[--] In \textbf{Chapter \ref{ellflow}} the focus is on the elliptic asymmetry factor $v_2$ of the particle yields of a HIC. A simple model is presented which can describe the measured $v_2$ with three phenomenological parameters. We discuss the physical motivations and the possible mechanisms behind our oversimplified description, and also the interpretation of the fitted parameters.\par
The results of this chapter are published in Ref.~\cite{v2paper}.
\item[--] In \textbf{Chapter \ref{etaOs}} we analyse the behaviour of the fluidity measure $\eta/s$ in light of certain properties of the two-particle spectrum of an effective field theory. We compute the macroscopic parameters like the shear viscosity $\eta$ and the entropy density $s$ in terms of the spectral density of states. We reproduce the phenomenological behaviour of $\eta/s$ in liquid--gas crossover, and observe the change of the robust property of the density of states during the transition. It tuns out, that the presence of a multi-particle continuum plays a crucial role in this manner: As the QP-peak melts into the continuum, the fluidity of the system is considerably increased. We also discuss the lower bound of $\eta/s$, which is non-universal and constrained by thermodynamics in our model, unlike the results of a theorized universal lower bound in the context of AdS/CFT duality. We point out further directions worthwhile investigating.\par
The results of this chapter are published in Ref.~\cite{etaOspaper}.
\end{itemize}
\clearpage
\chapter[Multi-component modified Boltzmann equations]{Multi-component modified \\Boltzmann equations}\label{MMBE}
\chaptermark{Multi-component modified BE}

The topic of this chapter is kinetic theory. Our goal here is to investigate a certain modification of the Boltzmann kinetic equation (BE). The main motive of this modification is to involve the effect of the medium, in which the particles move through and collide to each other. In order to do that, we are not going to use a mean-field description or put the particle ensemble into a potential. Instead, we modify the pair-interaction of colliding particles. We realize this by modifying the constraint on the kinetic energy in every binary collision. \rem{That is, the constraint relation for two particles with kinetic energy $E_1$ and $E_2$, respectively, ${E_1+E_2=\text{const.}}$ is replaced by ${E_1+E_2+aE_1E_2=\text{const.}}$ We call $a$ the modification parameter, and later we let it depend also on the species of the colliding particles or on their energy.} \par
Our main goal here is the analysis of the long-time behaviour of such a modified BE. Firstly, we motivate the actual form of the modified kinetic energy constraint by an argument beyond kinetic theory in Section~\ref{motivationMMBE}. Then we discuss the detailed balance solution of a one-component gas using the modified kinetic equation in Section~\ref{detailedBalanceMMBE}. Finally we turn to analyse multi-component systems. \par
The new scientific results summarized here are related to the multi-component modified Boltzmann equation (MMBE). In the case of two components, there are three different types of processes need to be balanced in equilibrium. For example, for particle species $A$ and $B$ there are collisions involving two $A$-type, two $B$-type or an $A$- and a $B$-type particle as well. One of the new results of this chapter is to show the non-existence of a detailed balance solution of the MMBE in case of non-equal modification parameters, say $a_{AA}$, $a_{BB}$ and $a_{AB}$ (see Sec.~\ref{toymodelMMBE}). However, a dynamical feedback of the average kinetic energy to the modification parameters can lead to equilibrium. In this case, an effectively one-component system emerges as the modification parameters of the different collision processes converge to each other. \par
The analytical and numerical investigation of the MMBE results a class of time-dependent, scaling solutions (Sec.~\ref{scalingSolMMBE}), which is the other new finding this chapter is aimed to present. These solutions can be parametrized by the average kinetic energy of the system. We also discuss the thermodynamic interpretation of such solutions. The results summarized here were firstly presented in Ref.~\cite{MMBEpaper}.\par

\section{Modifying the kinetic framework}\label{motivationMMBE}
Kinetic theory is an efficient and widely used approach to describe weakly interacting quasi-particle systems, like dilute gases. In these systems, the basic interaction events are instantaneous, binary collisions. The BE is the evolution equation of the phase-space density function $f^\alpha(\ve{p},\ve{r},t)$ of the particle species $\alpha$ with momentum $\ve{p}$ at the space-time point ${(t,\ve{r})}$. In a homogeneous approximation, the state of the system at a given time instant is characterized only by the particle momenta. The BE governs the time-evolution of $f^\alpha$ by summing up probabilities of events wherein a particle with a given momentum is scattered in or out of a small volume element of the phase-space in a unit time. Assuming that the particles forget their history between two consecutive collisions, the product of two density functions enters into the binary collision integral\footnote{This is the so-called \textit{hypothesis of molecular chaos} or \textit{Stosszahlansatz}.}:
\begin{equation}\label{BE}
\frac{\partial}{\partial t}f^\alpha_1=\sum_\beta\int_{234}\underbrace{\delta^{(4)}(\textnormal{constraints})w^{\alpha\beta}_{1234}}_{=:\mathcal{W}^{\alpha\beta}_{1234}}(f^\alpha_3f^\beta_4-f^\alpha_1f^\beta_2)=:\sum_\beta\mathcal{I}^{\alpha\beta}_1.
\end{equation}
\rem{The lower indices refer to phase-space coordinates, namely the momentum of the particle in a homogeneous system. The Greek letters are used to distinguish the different components of the gas. The density function $f^\alpha$ is normalized to unity at any time: ${\int_1f^\alpha_1=1}$. As usual in kinetic theory, the dynamics can be interpreted as a sequence of $~\{(E^\alpha_1,\ve{p}_1),(E^\beta_2,\ve{p}_2)\} \rightarrow \{(E^\alpha_3,\ve{p}_3),(E^\beta_4,\ve{p}_4)\}$ collisions. The number of particles is conserved for all species separately. Therefore changes of species like $\alpha\beta\leftrightarrow\beta\beta$, $\alpha\alpha\leftrightarrow\alpha\beta$ are not allowed in this model. \par
The probability rate that such an event happens is given by $$\mathcal{W}^{\alpha\beta}_{1234}=w_{1234}^{\alpha\beta}\delta^{(4)}(\mathrm{constraints}),$$ including now all of the constraints. $\mathcal{W}$ has the following symmetries in its indices simultaneously: $i)$ the interchangeability of incoming (1,2) and outgoing (3,4) collision partners: $1234 \leftrightarrow 2134$ with $\alpha\leftrightarrow\beta$ or $1234 \leftrightarrow 1243$ with $\alpha\leftrightarrow\beta$ and $ii)$ microscopic time-reversibility: $1234 \leftrightarrow 3412$. }
\par
There are many possible ways to elevate the strict restrictions of elastic collision and to investigate more complex dynamics, keeping the concept of binary collisions as elementary events. The need for modification emerges when the quasi-particles are not point-like, or long-range interactions are involved \cite{coulomb,spicka1,spicka2,spicka3,knoll2}. \par
Let us regard two possible modifications of the original form of the BE. One can $i)$ use other constraints instead of the ones of elastic collision, or $ii)$ abandon the product structure of the collision kernel. In the first case, the composition rule for conserved quantities are modified, while in the second case, two-particle correlations are taken into account in a non-trivial way. Several examples have been discussed in the literature \cite{biro3,kaniadakis,biro4,kaniadakis2,lima} on both cases. Here we exploit the modification of the energy composition rule in details, in particular its possible extension to multi-component systems. \par
In general, the assumption of instantaneous, pairwise collisions is getting worse as the interaction becomes stronger. In the kinetic theory, one aims to describe the system on a much longer time scale compared to the characteristic time period of a single collision. There are several examples for deriving the kinetic description from a microscopic theory, the most famous one is by Kadanoff and Baym \cite{kadanoff_baym}, but other examples can be found in the literature as well \cite{spicka3, greiner1, cassing1, knoll1}. These approaches distinguish between microscopic and mesoscopic time scales. However, it is not guaranteed, that all consistent approximation schemes lead to the BE. Let us mention a few examples, wherein the microscopic details modify the original picture:
\begin{itemize}
\item \textit{Non-instantaneous collision processes.} The authors of Ref.~\cite{spicka1} argue that the time and space arguments of the functions in the collision term of the BE differ because of non-local corrections. In case of long-range interactions, the particles can ''feel'' the presence of each other long before they get as close as it is comparable to their actual size. This increase of the effective particle size and/or the timescale of a collision demand higher terms in the gradient expansion of the pair-correlation to be taken into account.
\item \textit{Off-mass-shell scattering processes} can show increasing importance in dense systems (e.g. dense plasma of electrons, the Fermi-liquid in semiconductors or the kinetic description of high-energy nuclear collisions \cite{powerLawWilk, nonaddThermostat, thermoCompRules, hadronPowerLaw}). This effect is mainly due to that the quasi-particles get correlated with each other or with the environment. Therefore, the consecutive collisions do not link asymptotic states of the microscopic theory.
\end{itemize}
\begin{figure}[!t]
\centering
\subfloat[]{
\includegraphics[width=0.57\linewidth]{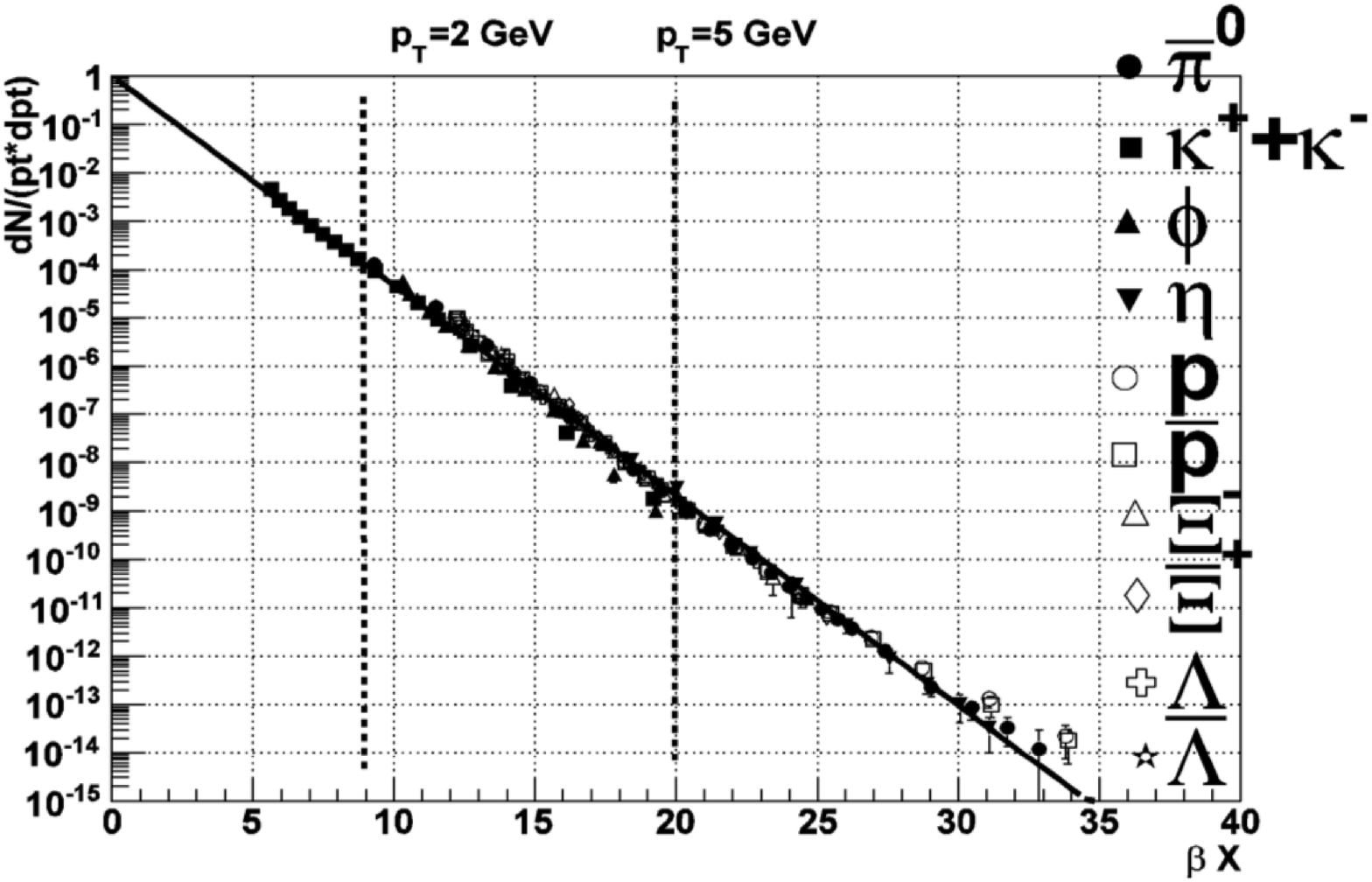}
\label{fig:powerSpectra1}
}
\subfloat[]{
\includegraphics[width=.38\linewidth]{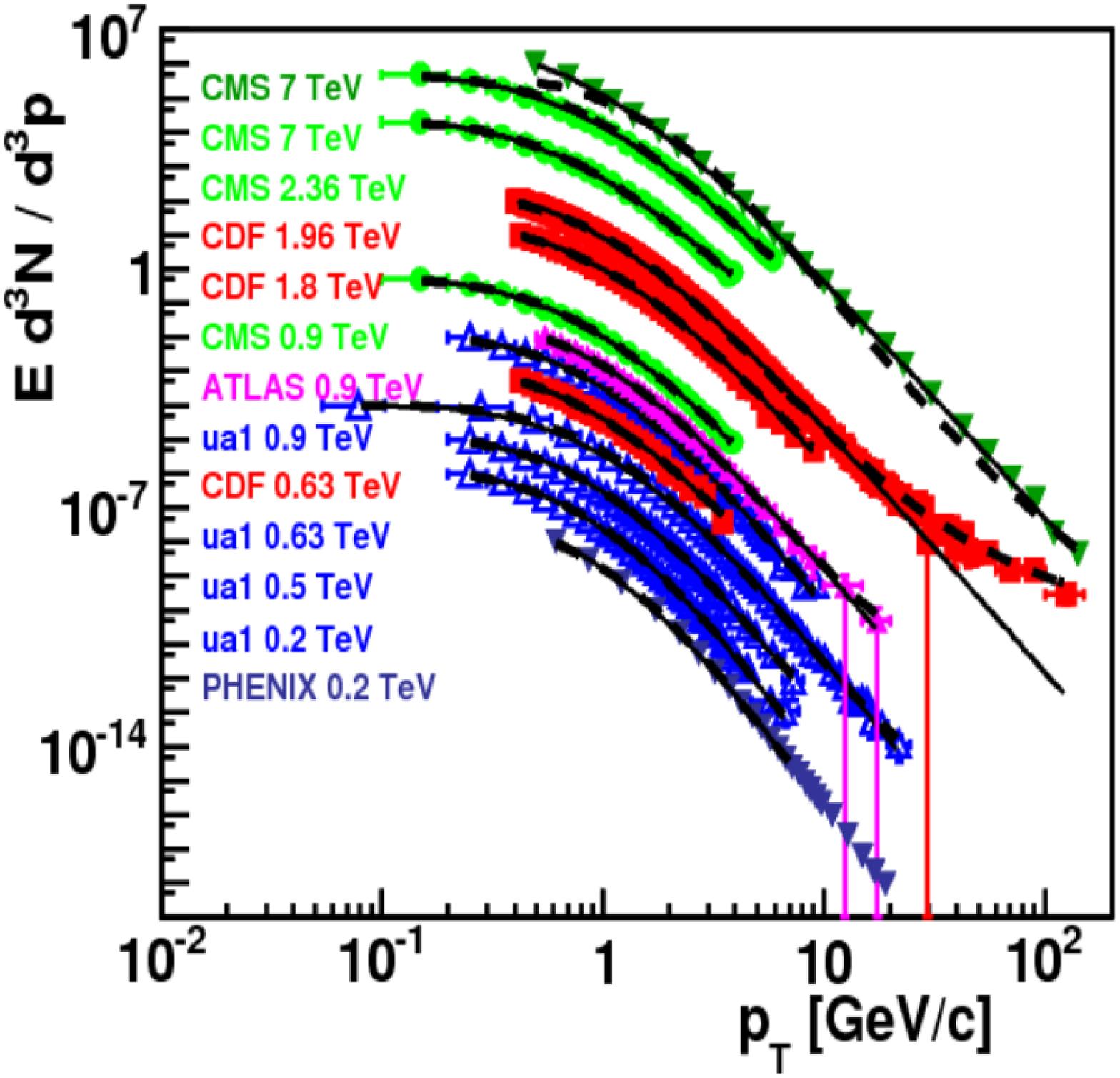}
\label{fig:powerSpectra2}
}
\LNFIG
\caption{Transverse momentum spectra in pp and heavy-ion collisions -- (a): Figure taken from \cite{thermoCompRules}. Tsallis fits to transverse momentum distributions of identified particles stemming from Au-Au collisions at ${\sqrt{s} = 200\text{AGeV}}$ collision energy. The expansion of the quark-gluon plasma is taken into account by the blast wave model. ${x=\frac{1}{a}\ln(1+a\gamma(m_T-vp_T))}$ and $\beta=\frac{1}{T}$, $\gamma$ is the Lorentz-factor $1/\sqrt{1-v^2}$ with the blast wave velocity $v$, and the transverse mass $m_T=\sqrt{m^2+p_T^2}$.
(b): Taken from \cite{hadronPowerLaw}: Experimental data plotted as a function of logarithmic transverse momenta. Points are: PHENIX data: blue full triangles, CERN UA1 data: blue open triangles, CDF data: red full squares, ATLAS data: magenta full stars, earlier CMS data: green full circles, upscaled CMS data: green full triangles. The $p_T$ bins and errors are also indicated on the data points with bars.}
\label{fig:powerSpectra}
\end{figure}
\rem{In this work, we are mainly motivated to give attention to the modified kinetic description because of the power-law like transverse momentum spectra observed in proton-proton and also in heavy-ion collisions (HIC). With the simple modified constraint on the kinetic energies ${E_1+E_2+aE_1E_2=\text{const.}}$ The equilibrium solution of the modified BE can be worked out easily by solving the detailed balance condition $$f(E_1)f(E_2)=\text{const.},\,\,\text{with}\,\, E_1+E_2+aE_1E_2=E.$$ Factorizing the kinetic energy constraint we get the product ${(1+aE_1)(1+aE_2)=1+aE}$, in which the factors depend only on one of the energies $E_1$ or $E_2$. Comparing this with the first equation we therefore conclude \LNNL
\begin{align}
f(E) & \sim \left(1+aE\right)^{-\frac{1}{aT}}\overset{a\rightarrow 0}{\longrightarrow} e^{-E/T},
\end{align}
which is the so-called Tsallis distribution with temperature $T$. The limit $a\rightarrow 0$ recovers the Boltzmann-Gibbs distribution. }
In Fig.~\ref{fig:powerSpectra} this function is fitted to experimental data collected in various high-energy experiments. \par
We devote the rest of this section to present a heuristic argument on how the modification of the energy composition can emerge. In the ordinary BE, all the particles are on-shell $p^0=E(\ve{p})$. Here, we suppose, that the particles do not propagate freely between two collisions.
The possible interaction with the medium, which is not included in the kinetic description, can alter the kinetic energy of the particles under consideration. \par
Consider one of the colliding particles off-shell with a spectral density of possible states in energy, which is -- near the mass-shell of the particle -- is parametrized by
\begin{equation}
\rho^\gamma(p^0-E_{\ve{p}}) = \frac{1}{\pi}\frac{\gamma}{(p^0-E_{\ve{p}})^2+\gamma^2} \xrightarrow{\gamma \rightarrow 0} \delta(p^0-E_{\ve{p}}).
\end{equation}
We suppose that $\gamma$ could take on several different values, stochastically collision-by-collision. Let us assume moreover, that the random variable $\gamma$ has small variance and its expectation value is large, compared to the typical energy scale of the collisions. Now, we perform an averaging over $\gamma$, similarly to the reasoning appeared in \cite{rafelski}. This results the collision integral $\mathcal{I}=\llangle \mathcal{I}^\gamma \rrangle:=\int_0^\infty\mathrm{d}\gamma g(\gamma)\mathcal{I}^\gamma$, with $g(\gamma)$ being the density function of the random variable $\gamma$. To analyse the effect of the averaging, we write the r.h.s. of the BE (\ref{BE}) into the following compact form: \LNNL
\begin{align}
\mathcal{I}^\gamma &=\int_{234}\!\rho^\gamma(E_1+E_2-E_3-E_4)\mathcal{K}_{1234},\,\,\text{where}\nonumber\label{avrCollIntMMBE}\\
\mathcal{K}_{1234} &=w_{1234}\delta^{(3)}({\sum_i}\ve{p}_1+\ve{p}_2-\ve{p}_3-\ve{p}_4)(f_3f_4-f_1f_2).
\end{align}
It is apparent that $\rho^\gamma$ takes over the role of the constraint on the kinetic energy. Assuming that the integration with respect to $\gamma$ and the phase-space variables can be interchanged, we arrive at\LNNL
\begin{align}
\mathcal{I} &= \Big\llangle\int_{234}\!\! \mathcal{K}_{1234}\rho^\gamma_2\Big\rrangle \approx \int_{234}\delta(E_1+E_2-E_3-E_4-\Delta)\mathcal{K}_{1234}, \label{averaging}
\end{align}
where $\Delta$ refers to the position of the peak of the smeared spectral density $\llangle\rho^\gamma\rrangle$ and ${\Omega=E_3+E_4-E_1}$ (for the detailed derivation see Appendix \ref{app1MMBE}). \par
We do not specify the mechanism encoded in the presence of the noise on $\gamma$. Some possible sources, however, can be mentioned:
\begin{enumerate}[\it i)]
\item In a medium, particles can have a thermal mass due to the heat bath \cite{biro5}.
\item An external force, stemming from a momentum-dependent, so called optical potential. Such momentum-space inhomogeneities can be observed in early-time non-Abelian plasmas due to the plasma instabilities \cite{strickland,ipp}.
\item When long-range interaction is present, the propagation of the quasi-particles could be disturbed by the long-wavelength modes of the system. As a consequence, the originally independent two-particle collisions start to ''communicate'' with each other. The effect of these, beyond-two-particle processes can be incorporated into a mean-field description \cite{biro6,arnold}. In cases when the leading order process is still the two-quasi-particle collision, one may deal with these events on the level of a kinetic description, but with modified kinetic energy addition rule.
\end{enumerate}
In the spirit of the argumentation given above, we consider a density of states showing two peaks and we get:
\begin{equation}\label{kinequ0}
\mathcal{I} \approx \int_{234}\!\! \left(\delta({\sum_i}' E_i-\Delta^A)+ \delta({\sum_i}' E_i-\Delta^B) \right)\mathcal{K}_{1234}.
\end{equation}
The two distinct constraints suggest a multi-component treatment, where the collisions between different particle species have different modifications for the respective kinetic energy sums.

\section{Detailed balance solution of the kinetic equations\\ with modified constraints}\label{detailedBalanceMMBE}
\rem{In this section we investigate the conditions of the existence of a detailed balance solution in a multi-component set-up. First we discuss how to formulate the detailed balance conditions in the case of a multi-component kinetic equation. Then we show for two components that detailed balance does not exist in the case when different constraint applies for the different collision types $AA$, $BB$ and $AB$ -- not even when two of those are the same, for example $AA$ and $BB$. }In the case of the modified kinetic energy constraint, the invariants of a binary collision are the total momentum ($\ve{p}_1+\ve{p}_2=\ve{p}_3+\ve{p}_4$, 3 constraints) and a quantity depending on the kinetic energies of the particles on the incoming or outgoing side of the process, respectively: $E_1^\alpha\oplus^{\alpha\beta}E_2^\beta=E_3^\alpha\oplus^{\alpha\beta}E_4^\beta~$ (1 constraint). The energy composition rule is represented here by the symbol $\oplus^{\alpha\beta}$. 
It can be approximated by the simple addition up to first order in its variables: $E\oplus^{\alpha\beta}E' \approx E+E'+\mathcal{O}^{\alpha\beta}(E^2,(E')^2,EE')$.\par
We introduce the function $L(E)$ to rewrite the energy composition rule in an additive form:
\begin{align}
& L^{\alpha\beta}(E_i^\alpha\oplus^{\alpha\beta}E_j^\beta)=L^{\alpha\beta}(E_i^\alpha)+L^{\alpha\beta}(E_j^\beta). \label{MMBEquasiE}
\end{align}
\rem{This maps the composition of the energies into the sum of single energy-dependent quantities, which we call \textit{quasi-energies} \cite{biro2}. Such a function can be constructed in the following way: Differentiating Eq.~\ref{MMBEquasiE} by one of its energy arguments, then taking the other zero leads us to $$(L^{\alpha\beta})'(x\oplus^{\alpha\beta}0)\left.\partial_y(x\oplus^{\alpha\beta}y)\right|_{y=0} =\left.L'(y)\right|_{y=0}.$$ Now we use that ${x\oplus^{\alpha\beta}0=x}$ and prescribe ${L'(y=0)=1}$. After integration we get
\begin{equation} \label{formlog}
L^{\alpha\beta}(x)=\int_0^x\frac{\mathrm{d}z}{\left.\partial_y (z\oplus^{\alpha\beta}y)\right|_{y=0}}.
\end{equation} }
\textit{Therefore one might view the binary collisions of the MMBE like the usual none-modified ones, but between particles with modified dispersion relation. However, in this case the dispersion relation depends not only on the particle itself, but also on the collision partner.} \par
\begin{figure}[!t]
\centering
\fbox{\includegraphics[width=.6\linewidth]{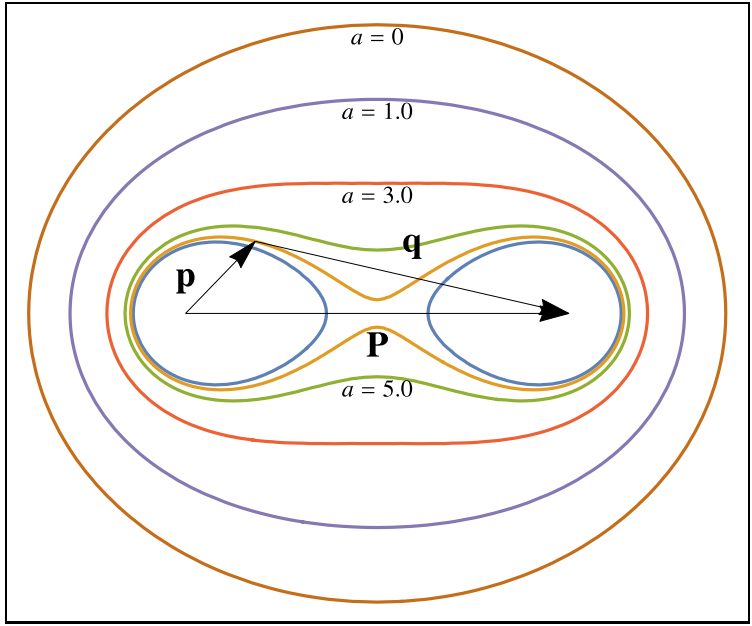}}
\LNFIG
\caption{Illustration for the two-dimensional sections of constraint surfaces $\mathcal{C}$ with various modification parameters $a$. The two colliding particles have momenta $\ve{p}$ and $\ve{q}$, respectively, such that ${\ve{p}+\ve{q}=\ve{P}}$. We used a massless dispersion relation ${E(\ve{p})=|\ve{p}|}$. The equienergetic curves are defined as $|\ve{p}|+|\ve{P}-\ve{p}|+a|\ve{p}||\ve{P}-\ve{p}|=K$, where we fixed $K=1.0$ and $|\ve{P}|=0.55$. The origin on this plot is assigned by $\ve{p}=0$. The unmodified case $a=0$ is also indicated.}
\label{fig:constrSurface}
\end{figure}
It is an irregular feature of the MMBE that detailed balance cannot be guaranteed in the general multi-component case. For a detailed balance solution, $\partial_t f^\alpha \equiv 0$ should be fulfilled for every $\alpha$ in such a way that all the kernels of the r.h.s. collision integrals in Eq.~(\ref{BE}) are zero, i.e. $f^\alpha_3 f^\beta_4 - f^\alpha_1f^\beta_2 =0$, for every pair of $\alpha$, $\beta$ and for every value of the phase-space variables 1, 2, 3 and 4 which are satisfying the constraints. We shall write the kinetic equation in a form more convenient for the analysis of the detailed balance state. Thinking of a given two-particle collision, we deal with a 6-dimensional phase-space (three dimension for each particles) which is however constrained by $i)$ the momentum conservation (three restrictions) and $ii)$ the suitable quasi-energy conservation depending on the type of the collision. We have two free parameters left, which means the phase-space is restricted to a surface in each collisions: ${~\mathcal{C}^{\alpha\beta}(\ve{P},K):=\left\{(\ve{p},\ve{p}') \,\,:\,\, \ve{p}+\ve{p}'=\ve{P},\,\, E(p)\oplus^{\alpha\beta}E(p')=K \right\}}$, see Fig.~\ref{fig:constrSurface} for illustration. After the four constraints have been integrated out, a 5-fold integral remains instead of the 9-fold one in Eq.~(\ref{BE}). The kinetic equation takes the form
\begin{equation} \label{kinequ2}
\frac{\partial}{\partial t} f^\alpha_1 = \sum\limits_\beta \int\!\!\mathrm{d}^3\ve{P}\iint\limits_{\mathcal{C}^{\alpha\beta}}\!\!\mathrm{d}^2\sigma^{\alpha\beta} w^{\alpha\beta}_{1234}\left.(f_3^\alpha f_4^\beta -f_1^\alpha f_2^\beta)\right|_{\left\{(\ve{p}_1,\ve{p}_2),\,(\ve{p}_3,\ve{p}_4)\right\} \in \mathcal{C}^{\alpha\beta}(\ve{P},K^{\alpha\beta})},
\end{equation}
where ${\ve{P}=\ve{p}_1+\ve{p}_2=\ve{p}_3+\ve{p}_4}$ and ${K^{\alpha\beta}=E_1\oplus^{\alpha\beta}E_2=E_3\oplus^{\alpha\beta}E_4}$. Detailed balance requires that the appropriate kernel vanishes on the constraint surface $\mathcal{C}^{\alpha\beta}$. In order to achieve this, the following equations have to hold for every $\alpha$ and $\beta$ pairs simultaneously, with momentum-independent constants $\mathcal{M}^{\alpha\beta}$:
\begin{eqnarray} \label{detbal}
\left. f^\alpha(E(p))f^\beta(E(p'))\right|_{(\ve{p},\ve{p}')\in\mathcal{C}^{\alpha\beta}(\ve{p}+\ve{p}',E\oplus^{\alpha\beta}E')} &=& \mathcal{M}^{\alpha\beta}.
\end{eqnarray}
Looking for isotropic equilibrium solution with the ansatz
\begin{equation}\label{detbal_ansatz}
f^\alpha(E) \sim e^{-L^{\alpha\alpha}(E)/T^\alpha},
\end{equation}
with constant $T^\alpha$. the conditions in Eq.~(\ref{detbal}) are automatically satisfied for $\alpha=\beta$. This is the familiar detailed balance solution of the single-component modified BE \cite{biro2}. Therefore, in case of a one-component system\LNNL
\begin{align}
f(p) &= \frac{1}{Z}e^{-\frac{L(E(p))}{T}}, \nonumber
\end{align}
where $T$ is the thermodynamic temperature of the system (the one which equalizes when two previously separated systems are put in thermal contact). \par
In our case, however, conditions with different $\alpha$ and $\beta$ must be satisfied, too. Let us investigate the case of two components, $\alpha\in\{A,B\}$. One has to deal with three kind of collisions then: $AA$, $BB$ and $AB=BA$, since all the constraints are commutative. The structure of the MMBE (\ref{kinequ2}) in this case reduces to:
\begin{equation}\begin{array}{cccc}
\frac{\partial}{\partial t} f^A_1 = \mathcal{I}^{AA}_1+\mathcal{I}^{AB}_1, & & &
\frac{\partial}{\partial t} f^B_1 = \mathcal{I}^{BA}_1+\mathcal{I}^{BB}_1.
\end{array}\end{equation}
In the state characterized by Eq.~(\ref{detbal_ansatz}), all the $\mathcal{I}^{\alpha\alpha} \equiv 0$ by construction of the density functions. The problem is, that supposing the detailed balance of a given kind of collision ($AA$, $BB$, $\mathcal{I}^{AA}\equiv 0$, $\mathcal{I}^{BB}\equiv 0$), the rest ($AB$) will not be fulfilled with the same type of density functions ($\mathcal{I}^{AB}\neq 0$). (There is, however, a trivial solution, namely when all the $L^{\alpha\beta}$ functions are the same.) \par
Let the collisions between the particles in the same type be in detailed balance ($\mathcal{I}^{AA}\equiv 0$, $\mathcal{I}^{BB}\equiv 0$). For the sake of simplicity, we consider the situation of $L^{AA} \equiv L^{BB}$. In this case $f^A_1 \sim f^B_1 \sim e^{-L^{AA}_1/T}$. Then the kernel of the mixed collisional term $\mathcal{I}^{AB}$ is proportional to
\begin{equation}
\mathrm{kernel\,\,of\,\,} \mathcal{I}^{AB} \sim e^{-\frac{1}{T}L^{AA}_3-\frac{1}{T}L^{AA}_4}-e^{-\frac{1}{T}L^{AA}_1-\frac{1}{T}L^{AA}_2}.
\end{equation}
In the detailed balance state it vanishes for all phase-space points lying on $\mathcal{C}^{AB}$. With no further assumptions for $L^{\alpha\beta}$, the achievement of such a state implies the relation $\mathcal{C}^{AA} \subseteq \mathcal{C}^{AB}$. Starting on the other hand from $\mathcal{I}^{AB} \equiv 0$, the above reasoning leads to $\mathcal{C}^{AB} \subseteq \mathcal{C}^{AA}$. In conclusion, if one does not have other conditions for the modification except the symmetry properties mentioned above, the only detailed balance solution is $\mathcal{C}^{AA}=\mathcal{C}^{AB}$, that is when all the quasi-energies are the same. This is fulfilled only for the one component matter.\par
This result does not imply, however, that the system can not saturate to a time-independent state: \LNNL
\begin{align}
f^{A,B}(p,t) &\xrightarrow{t\rightarrow\infty} f^{A,B}_\mathrm{Eq.}(p). \nonumber
\end{align}
According to the previous argument, when the modification is coupled to the dynamics in such a way that $~\mathcal{C}^{AB} \xrightarrow{t\rightarrow\infty} \mathcal{C}^{AA}$, saturation behaviour may arise.\par
We note here that some authors emphasized the non-universal nature of the modification of the energy addition, in the sense it should depend on dynamical details of the system \cite{wang}.

\section{Numerical results for the long-time behaviour}\label{longtimeMMBE}
\subsection{A two-component toy model}\label{toymodelMMBE}
In this section we investigate the time evolution of a two-component MMBE (\ref{BE}). For this purpose we use a simple toy model. From here on we consider only the isotropic case when all functions $f^\alpha$ depend on the phase-space position through the single-particle energy only. If the collision does not prefer any direction for some reason, or with other words $w_{1234}^{\alpha\beta}$ depends on $|\ve{p}_1-\ve{p}_2|$ and $|\ve{p}_3-\ve{p}_4|$ only, it is reasonable to expect an isotropic state after appropriately long evolution, irrespective to the initial state. If external fields are not present, isotropisation is usually much faster than equilibration. \par
Three elements of the model have to be specified: $i)$ the energy addition rule in the constraint, $ii)$ the properties of the particles building up the ensemble and $iii)$ the rate function $w^{\alpha\beta}_{1234}$ which incorporating the dynamical details of the collisions. Let us specify the last two first: we consider non-relativistic point particles with dispersion relation $E(p)=\frac{1}{2m}p^2$. We choose the rate function in a way that the system can reach every outgoing state which fulfil the kinematics (i.e. the modified constraint for the energy- and momentum-conservation) with equal probability.\par
Now we specify the energy addition rule. As it was emphasized in \cite{biro_comp_rule}, we use a rule which gives the simple addition for low energies. Therefore the simplest choice is
\begin{equation} \label{addrule}
E\oplus^{\alpha\beta} E'=E+E'+A^{\alpha\beta}EE'.
\end{equation}
The quantities $A^{\alpha\beta}$ may depend on the phase-space or other dynamical details. $A^{\alpha\beta}= 0$ means conventional addition. The rule (\ref{addrule}) is also easily tractable, namely its inverse function can be constructed analytically. Then the characteristic scales of the system are the total energy per particle $\avr{E} =\sum_\alpha\int\!\!\mathrm{d}^3\ve{p}E(p)f^\alpha(E(p))$ and $1/A^{\alpha\beta}$. \par
At this point the MMBE (\ref{kinequ2}) reads as:\LNNL
\begin{align}
\frac{\partial}{\partial t} f^\alpha(p) = \sum_\beta\int\!\!\mathrm{d}^3\ve{p}'\!\!\int\!\!\mathrm{d}^3\ve{q}& w^{\alpha\beta}(P,K^{\alpha\beta},x)\delta(E(p)\oplus^{\alpha\beta}E(p') -E(q)\oplus^{\alpha\beta}E(|\ve{P}-\ve{q}|))\times\nonumber \\
\times& \left\{f^\alpha(q)f^\beta(|\ve{P}-\ve{q}|)-f^\alpha(p)f^\beta(p')\right\}, \label{kinequ3}
\end{align}
where $\ve{P}=\ve{p}+\ve{p}'$ and $x$ is defined by $\ve{p}\cdot\ve{p}'=:pp'x$. We define the rate function $w^{\alpha\beta}(P,K^{\alpha\beta},x)$ by\LNNL
\begin{align}
\frac{1}{w^{\alpha\beta}(P,K^{\alpha\beta},x)} &= \int\!\!\mathrm{d}^3\ve{q}\delta(K^{\alpha\beta} -E(q)\oplus^{\alpha\beta}E(|\ve{P}-\ve{q}|))= \nonumber \\
&= \frac{2\pi}{P}\int\limits_0^\infty\!\!\mathrm{d}q \frac{q\,\Theta(|y^*|\leq 1)Q^{\alpha\beta}(q)}{g^{\alpha\beta}( E(q),E(Q^{\alpha\beta})) E'(Q^{\alpha\beta})}. \label{inverserate}
\end{align}
The quantity $Q^{\alpha\beta}$ is the solution of the equation $K^{\alpha\beta}=E(q)\oplus^{\alpha\beta}E(Q^{\alpha\beta})$, while $y^*$ is the value of $y=\frac{P^2+q^2-|\ve{P}-\ve{q}|^2}{2Pq}$, when ${|\ve{P}-\ve{q}|=Q^{\alpha\beta}}$. The function in the denominator is $g^{\alpha\beta}(u,v)=\partial_v u\oplus^{\alpha\beta}v$. The definition Eq.~(\ref{inverserate}) makes $1/w^{\alpha\beta}$ equal to the surface area of the constraint surface $\mathcal{C}^{\alpha\beta}$.\par
Because of the simple form of the rate function, only the kinematics restricts the collisions. The rate in Eq.~(\ref{inverserate}) defines a uniform distribution on $\mathcal{C}^{\alpha\beta}(P,K)$, which has a non-trivial density function in the energy variable. Let us denote this function by $\rho^{\alpha\beta}(\epsilon,P,K,x)$. Then Eq.~(\ref{inverserate}) can be interpreted as the normalization condition for $\rho^{\alpha\beta}$: $$ 1= \frac{1}{2\pi}\intlim{\epsilon}{0}{\infty}\rho^{\alpha\beta}(\epsilon,P,K,x):= \frac{1}{P}\int\limits_0^\infty\!\!\mathrm{d}q \frac{q\,\Theta(|y^*|\leq 1)Q^{\alpha\beta}(q)w^{\alpha\beta}(P,K,x)}{g^{\alpha\beta}( E(q),E(Q^{\alpha\beta})) E'(Q^{\alpha\beta})}.$$ Up to this point, we have not used the actual form of the energy composition rule. Using Eq.~(\ref{addrule}) with the notations $\epsilon=E(q)$ \FINAL{and $q=E^{-1}(\epsilon)$}, the density function reads as
\FINAL{
\begin{equation} \label{distronC}
\rho^{\alpha\beta}(\epsilon,P,K,x) =\frac{2\pi}{P}\frac{\Theta(|y^*(\epsilon,P,K,x)|\leq 1)w^{\alpha\beta}(P,K,x)E^{-1}(\epsilon)}{E'(E^{-1}(\epsilon))E'\left(E^{-1}\left(\frac{K-\epsilon}{1+a^{\alpha\beta}\epsilon}\right)\right)}\frac{K-\epsilon}{(1+a^{\alpha\beta}\epsilon)^2}.
\end{equation}
}
\rem{Now we briefly summarize the numerical method we used to solve (\ref{kinequ3}). Since our model is homogeneous in space, we are not going to deal with the propagation path of the particles. A cascade method, following the evolution of the system collision-by-collision, is also satisfactory. The usual method, i.e. considering the collision in the center-of-mass frame, is not convenient, because it is problematic to implement the modified energy composition rule. The problem manifests in the $\frac{1}{P}$ asymptotic of (\ref{inverserate}) -- ${P=0}$ corresponds to the center-of-mass frame. We rather use the lab frame, as it was described in Ref.~\cite{biro2}. The key element of this cascade method is the distribution $\rho^{\alpha\beta}$ defined on the constraint surface $\mathcal{C}^{\alpha\beta}$, parametrized by the energies of the outgoing particles.} The sampling of the kinetic energy $\epsilon$ on the constraint surface was implemented using the rejection method, see App.~\ref{rejMethMMBE} for details. We simply select the energy for one of the outgoing particles randomly according to $\rho^{\alpha\beta}$, the quasi-energy conservation provides the other one. \par

\subsection{Scaling solutions}\label{scalingSolMMBE}
\begin{figure}[!t]
\centering
\subfloat[]{
\includegraphics[width=.5\linewidth]{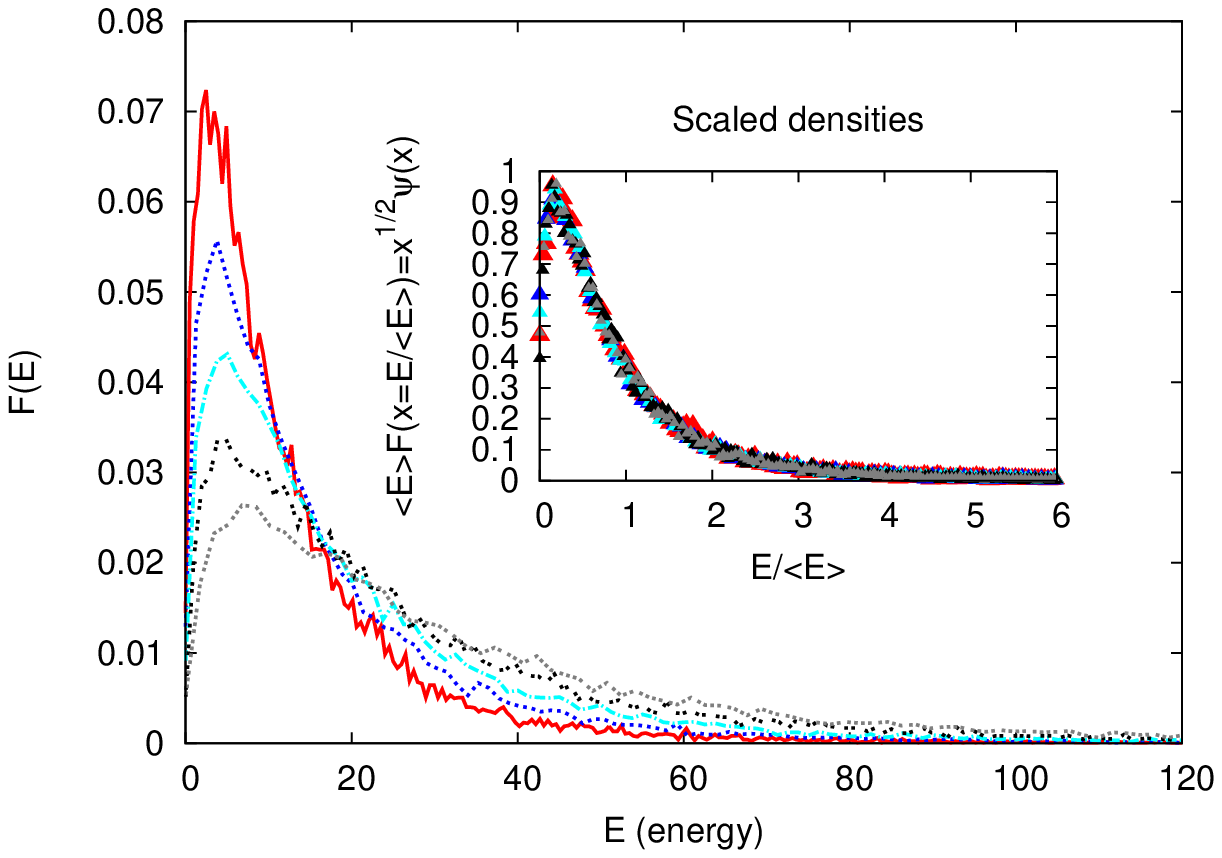}
\label{fig:scale_dens_func}
}
\subfloat[]{
\includegraphics[width=.5\linewidth]{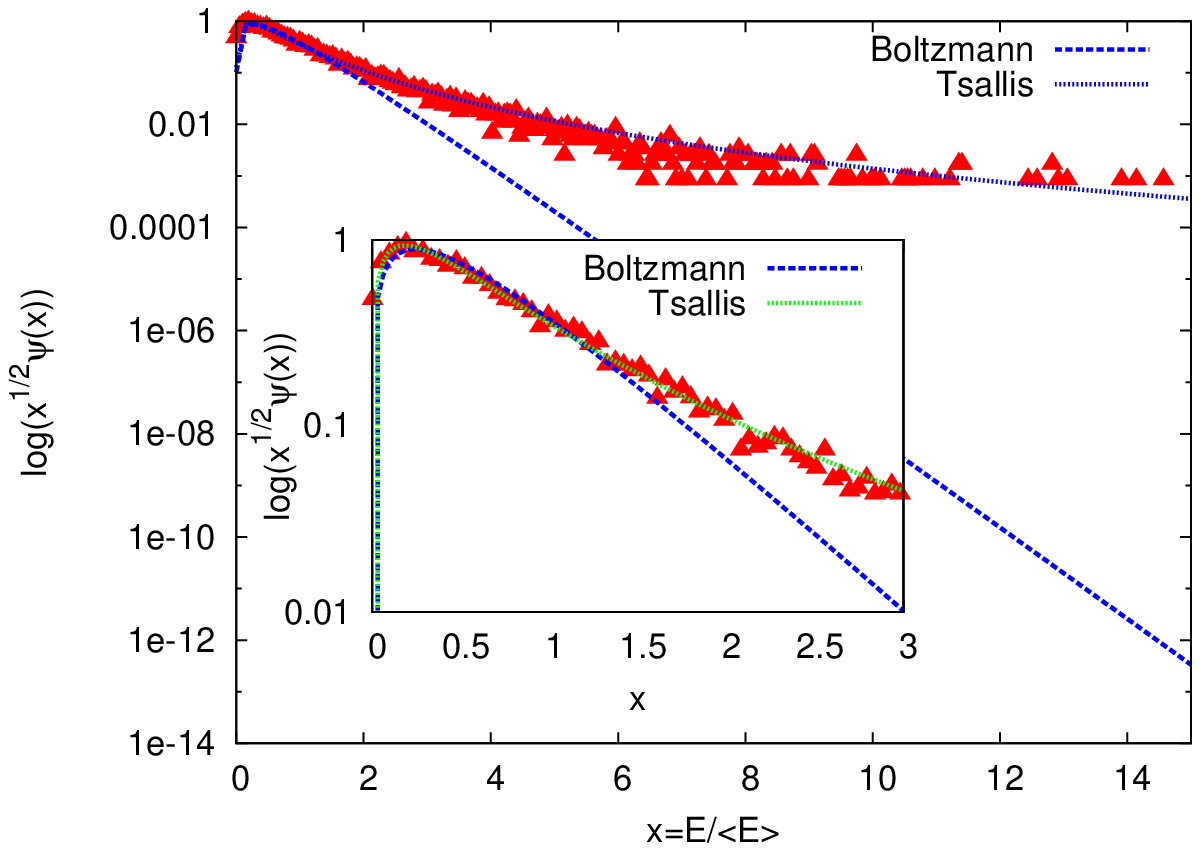}
\label{fig:dens_func_fit}
}
\LNFIG
\caption{(a): The density function $F(E,t)$ in different collision times, $a_0=0$, $a_1=1$. The scaling behaviour is apparent when one uses the relation (\ref{numscalingSOL}), as can be seen on the inner figure. (b): The asymptotic behaviour of $F(E,t)$ fitted by a Tsallis-like function ${\sim\sqrt{x}(1+\tilde{A}x)^{-\frac{\tilde{B}}{\tilde{A}}}}$. The Boltzmann-Gibbs behaviour (${\sim\sqrt{x}e^{-\tilde{B}x}}$) is also indicated.}
\end{figure}
In Section~\ref{detailedBalanceMMBE} we argued, that a detailed balance state does not exist for two components in the case when the modification parameters $a^{\alpha\beta}$ kept fixed. Nevertheless the question, what will happen long time after the initial state was prepared, should be answered. \rem{In this section we investigate the long-time evolution of Eq.~(\ref{kinequ3}) by the cascade method we have just introduced in Sec.~\ref{toymodelMMBE}. The strength of the modification parameters is linked to $1/\avr{E}$, where $\avr{E}$ is the average kinetic energy per particle. The main observation here is that after a short time of isotropization, the time-dependence of the distributions can be scaled out by $\avr{E}$, see Figs.~\ref{fig:scale_dens_func}~and~\ref{fig:dens_func_fit}.} Our cascade simulation provides the digitalized version of $f^\alpha(E(p))$ collision-by-collision. We use the moments of the density function for the qualitative analysis. We introduce the density $F^\alpha(E)$ depending on the kinetic energy and fulfilling the normalization condition\LNNL
$$ 
\int\!\!\mathrm{d}EF^\alpha(E) := \int\!\!\mathrm{d}^3\ve{p}f^\alpha(E(p))=1.
$$ 
The average kinetic energy per particle\LNNL
$$ 
\avr{E} :=\sum\limits_\alpha\int\!\!\mathrm{d}E EF^\alpha(E),
$$ 
and the entropy-like quantity\LNNL
$$ 
S_E := -\sum\limits_\alpha\int\!\!\mathrm{d}EF^\alpha(E)\ln F^\alpha(E)
$$ 
are the main subjects of the following investigation. \par
We analyse two cases as summarized in the Table~\ref{tab:tab1MMBE} (with $a_0$, $a_1$, $\Lambda_0 >0$ constants):
\begin{table}[!t]\centering
\begin{tabular}{c|c|c}
& $I.)$ ($a_0$, $a_1$) & $II.)$ ($a_0$, $a_1$, $\Lambda_0$) \\
\hline\hline
$A^{AA}=A^{BB}$ & $\frac{a_0}{\avr{E}}$ & $\frac{a_0}{\avr{E}}$ \\
\hline
$A^{AB}=A^{BA}$ & $\frac{a_1}{\avr{E}}$ & $\begin{array}{lcl} \frac{a_0}{\avr{E}}, & \textnormal{if} & E_1+E_2<\Lambda_0\avr{E} \\ \frac{a_1}{\avr{E}}, &\textnormal{if} & E_1+E_2>\Lambda_0\avr{E} \end{array}$ \\
\hline
$\begin{array}{c}\textnormal{long-time}\\ \textnormal{behaviour} \end{array}$ & $\begin{array}{c}\textnormal{growing}\\ \avr{E} \end{array}$ & $\begin{array}{c}\textnormal{growing or lowering}\\ \avr{E} \end{array}$
\end{tabular}
\caption{The two investigated cases with scaled modification parameters. The parameters $a_0$, $a_1$ and $\Lambda_0$ are positive.}
\label{tab:tab1MMBE}
\end{table}
In both cases, we choose the modification parameter in the terms $AA$ and $BB$ inversely proportional to the kinetic energy density $\avr{E}$, therefore the modification of the energy addition presents on every energy scale. That is, for two particles with average kinetic energy: $$\avr{E}\oplus \avr{E}=\avr{E}+\avr{E}+\frac{a}{\avr{E}}\avr{E}^2=(2+a)\avr{E}.$$ In case $I.)$, the constraint on the cross term $AB$ is modified similar to the terms $AA$ and $BB$, but with different pre-factor to $\avr{E}^{-1}$. In case $II.)$, the scaled-down modification parameters are the same below a scale $\Lambda_0$ and different above.\par
\begin{figure}
\centering
\subfloat[]{
\includegraphics[width=0.5\linewidth]{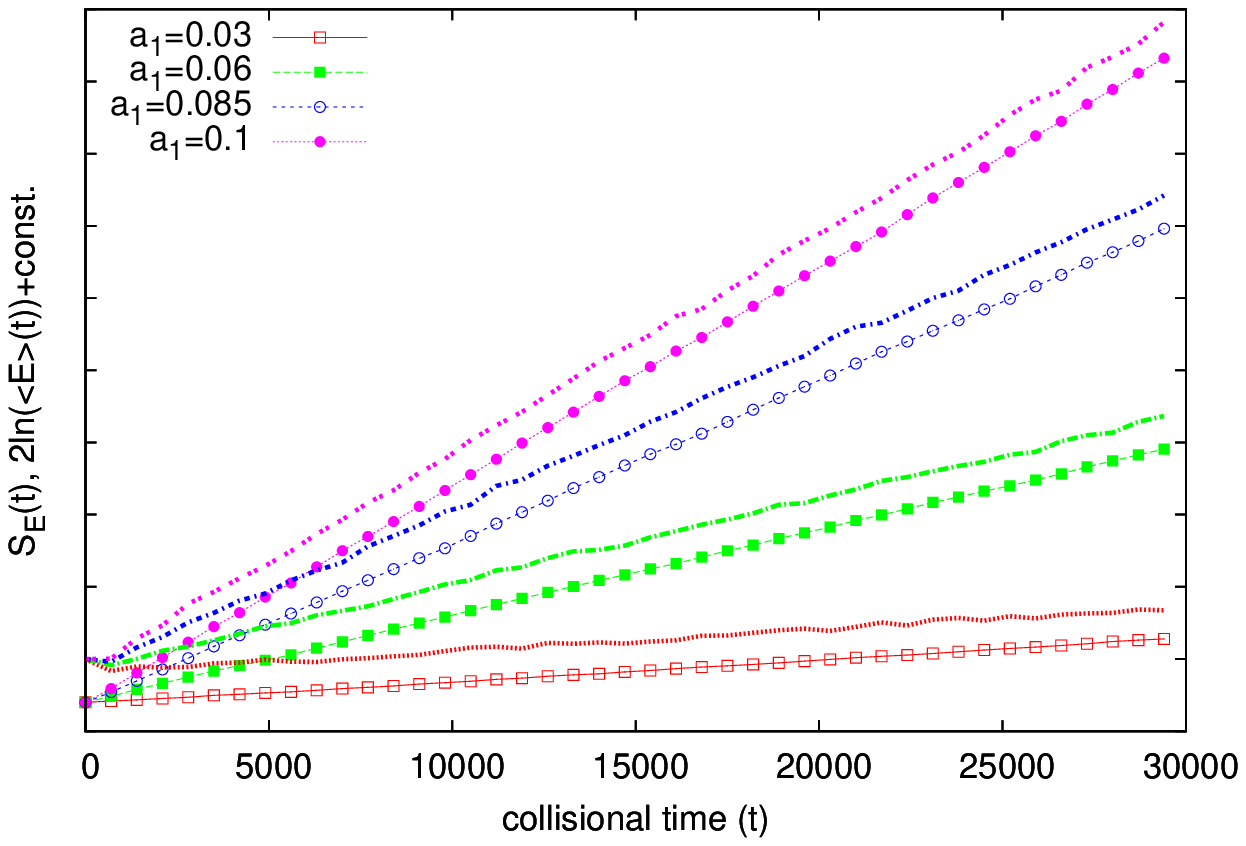}
\label{fig:entropy}
}
\subfloat[]{
\includegraphics[width=0.5\linewidth]{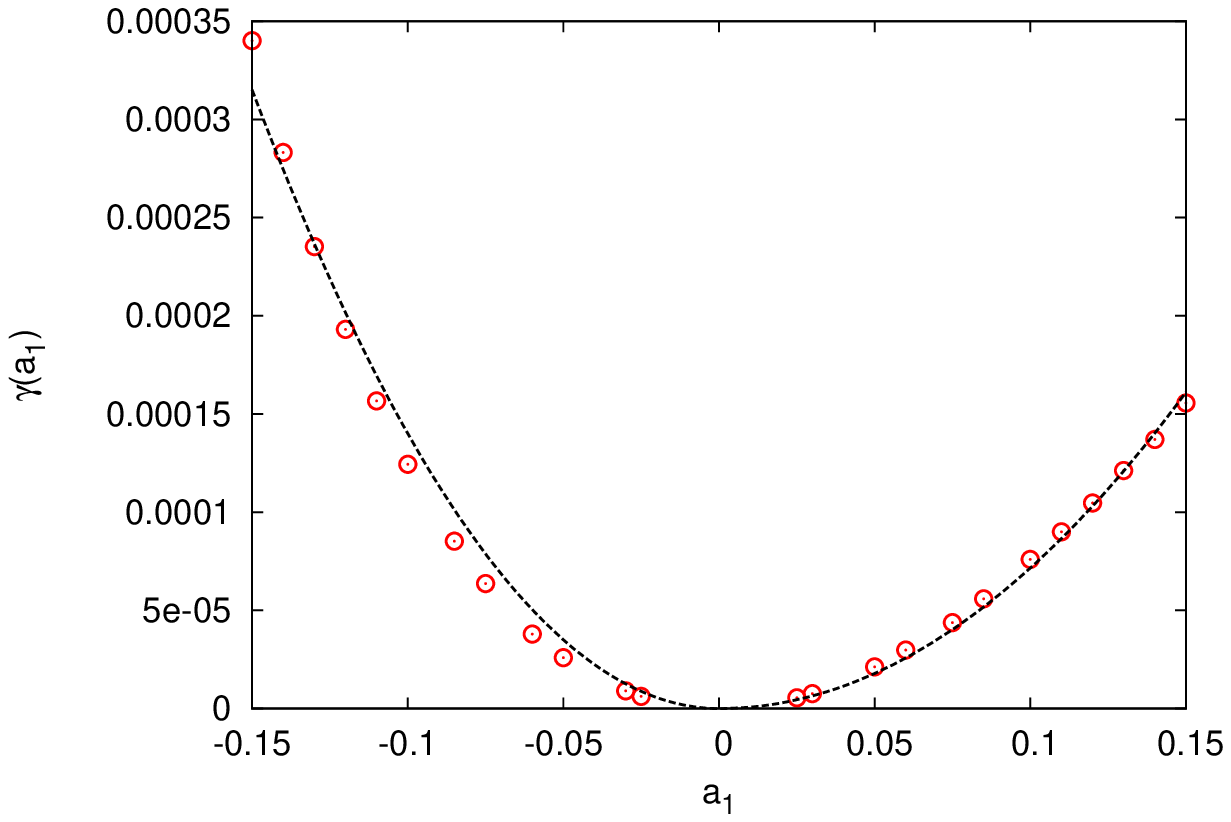}
\label{fig:gamma_vs_a}
}
\LNFIG
\caption{(a): Parametrized equation of states: $S_E(t)$, ${\ln\avr{E}(t)}$ for $a_0=0$, varying $a_1$. The entropy curves run in line with the logarithm of the total energy per particle after $t_{\mathrm{trans.}}$. (b): The exponent $\gamma$ for $a_0=0$ and different values of $a_1$. It approaches zero as ${\gamma \propto a_1^2}$.}
\end{figure}
\begin{figure}[!t]
\centering
\subfloat[]{
\includegraphics[width=0.5\linewidth]{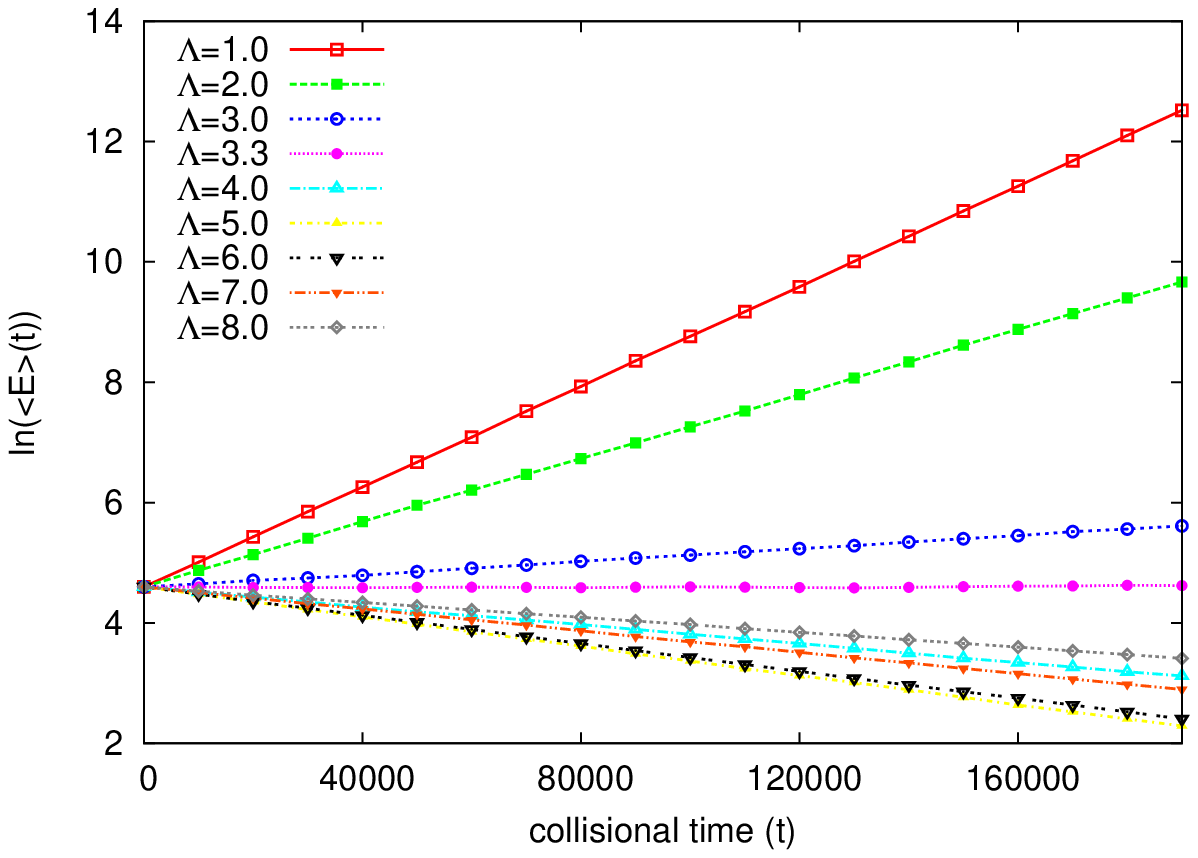}
\label{fig:en_saturation}
}
\subfloat[]{
\includegraphics[width=0.5\linewidth]{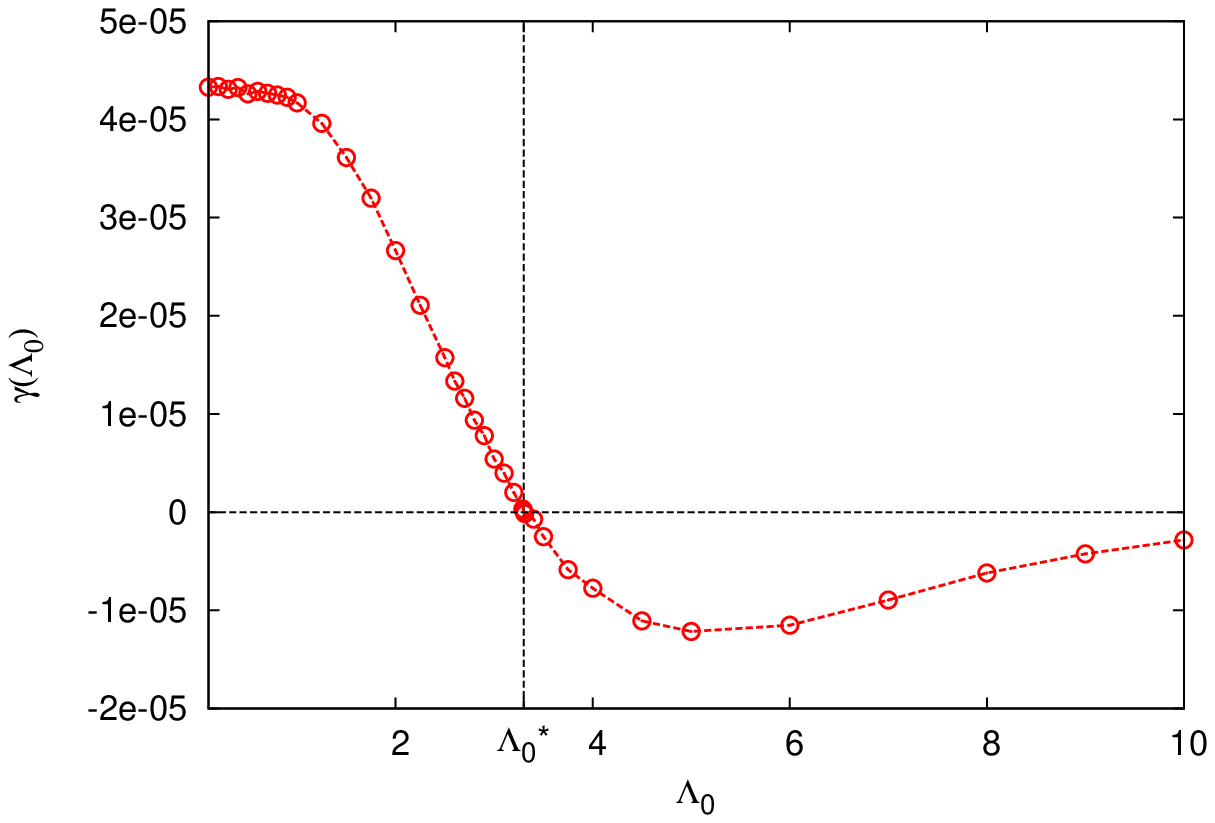}
\label{fig:gamma_vs_Lambda}
}
\LNFIG
\caption{(a): The running of $\ln\avr{E}(t)$ for various value of $\Lambda_0$. The total energy per particle can either be growing or lowering with time. With the fine-tuning of $\Lambda_0$, saturation occurs.\\ (b): The exponent $\gamma$ as a function of $\Lambda_0$. At its zero, the total energy per particle stands still.}
\end{figure}

\rem{Each one of the modifications $I.)$ and $II.)$ has an interesting feature. Both result in scaling density functions, insensitive to the initial conditions:
\begin{equation} \label{numscalingSOL}
f^{A,B}(E,t\rightarrow \infty) \sim \avr{E}^{-\frac{3}{2}}\psi(E/\avr{E}),
\end{equation}
in other words, the long-time evolution is defined by a one-parameter family of density functions. The $\psi$ shape-function is time-independent, the only time-dependence is due to $\avr{E}$. \par
In fact, the scaling behaviour (\ref{numscalingSOL}) can be observed after a transient time $t>t_\mathrm{trans.}$. $t_\mathrm{trans.}$ has the same order of magnitude as the relaxation time to the detailed balance state in one-component systems ($\mathcal{O}(10)$ collision per particle, in collision time). Since there is no parameter which could distinguish among the two components $A$ and $B$, we expect the same asymptotic density function, if a steady state evolves. We used two distinct states to prepare the initial conditions, namely a ''thermal'' one (Boltzmannian density) and the ''two fireball'' (one half of the particles moving into an assigned direction while the other half into the opposite direction). We did not experience any differences regarding the long-time behaviour for the various initial densities. Using the relation (\ref{numscalingSOL}) one can scale the densities belonging to different collisional times onto each other. However, the scaling is conspicuous on Fig. (\ref{fig:scale_dens_func}), it has another apparent feature, namely that the entropy is related to the total kinetic energy per particle as $S_E \sim \ln\avr{E}$, cf. Fig.~(\ref{fig:entropy}). \par
It turns out, that the scaling function $\psi(x)$ has power-law tail rather than an exponential one (when $x\rightarrow\infty$). The Tsallis-density function fits it quite well: ${~\psi(x) \sim \sqrt{x}(1+\tilde{A}x)^{-\frac{\tilde{B}}{\tilde{A}}}}$. This function is the detailed balance solution of the one-component MBE. An in-depth analysis would be needed to derive the fit parameters $\tilde{A}$ and $\tilde{B}$ starting from the given modification parameters $a_0$, $a_1$ and $\Lambda_0$, or reveal the numerically hardly visible bias from the Tsallis fitting function (Fig. (\ref{fig:dens_func_fit})).\par
The existence of the solution (\ref{numscalingSOL}) can be proven by using the algebraic identity $~E\oplus_{\avr{E}}^{\alpha\beta}E'= \avr{E}\left(\frac{E}{\avr{E}}\oplus_1^{\alpha\beta}\frac{E'}{\avr{E}}\right)$ valid for ${E\oplus_\avr{E}^{\alpha\beta} E'=E+E'+\frac{a^{\alpha\beta}}{\avr{E}}EE'}$. The details can be found in Appendix \ref{app2MMBE} in details. \par
The numerical experience indicates that such a family of density functions is asymptotically stable under time evolution.} Also, that in case $I.)$ $\avr{E}$ is always growing for long times ($a_0$, $a_1$ are positive constants), see Fig.~(\ref{fig:entropy}). Depending on the value of $\Lambda_0$, either the increase or decrease of the kinetic energy per particle can occur in case $II.)$ ($\Lambda_0$ is a positive constant), see Fig. (\ref{fig:en_saturation}). \par
The average kinetic energy per particle shows the following collisional-time dependence for the scaling solution described in Eq.~\ref{numscalingSOL}: \LNNL
\begin{align}
\avr{E}(t) &= \avr{E}_0 e^{\gamma (t-t_0)}, \label{avrEtimedepMMBE}
\end{align}
\rem{therefore one can quantify the growing or the lowering of the average kinetic energy by means of the exponent $\gamma$. The origin of this formula is discussed in Appendix \ref{app2MMBE}.} The connection between the collisional time and the laboratory time can be found in Appendix~\ref{app3MMBE}. We note here, that such scaling solutions occur in the context of the non-elastic Boltzmann equation (NEBE), referred as homogeneous cooling (heating) states \cite{ernst,bobylev1,bobylev2}. Both in NEBE and in MMBE models the total kinetic energy is not a conserved quantity -- the modelled systems are open.

\subsection{Heating and cooling of the pre-thermalized state}\label{heatingCoolingMMBE}
Now we turn to the interpretation of the scaling solution in Eq.~(\ref{numscalingSOL}), which apparently rules the long-time behaviour. This solution is characterized by the exponent $\gamma$ and the initial value of the average kinetic energy $\avr{E}_0$. The values of $a_0$, $a_1$ affect the shape function $\psi$ and also the value of $\gamma$. Depending on the modification parameters, $\gamma$ disappears smoothly: $\gamma \simeq C^\pm(a_1-a_0)^2$, when $~a_1-a_0 \rightarrow 0^\pm$, as it is demonstrated in Fig.~(\ref{fig:gamma_vs_a}). It is easy to see that for the $\kappa^{\text{th}}$ moments of $f(E,t>t_\mathrm{trans.})$ in the scaling regime $~\avr{E^\kappa} \sim \avr{E}^\kappa$ holds. Since the system is isotropic even in the momentum-space, the macroscopic behaviour can be described by these moments. As the time evolution starts following Eq.~(\ref{avrEtimedepMMBE}), we find (see Fig.~(\ref{fig:entropy})):\LNNL
\begin{align} \label{EoS}
S_E &= 2C_1\ln\avr{E} +C_2, \textnormal{ with}\\
C_1 &=\left(\int_0^\infty\!\!\mathrm{d}x\sqrt{x}\psi(x)\right) \ln\left(\int_0^\infty\!\!\mathrm{d}x x^\frac{3}{2}\psi(x)\right) \textnormal{and} \nonumber \\
C_2 &=-\int_0^\infty\!\!\mathrm{d}x \left(\sqrt{x}\psi(x)\ln(\sqrt{x}\psi(x))\right), \nonumber
\end{align}
where $C_1$ and $C_2$ are time- and phase-space-independent quantities. \par
\textit{That is, the system behaves like an ideal gas, which is heating up or cooling down depending on the sign of $\gamma$.}

\subsection{Saturation}\label{saturationMMBE}
The main difference between the investigated choices of the modification parameters (cf. Table~\ref{tab:tab1MMBE}) is the following. In case $I.)$, the section of the constraint surfaces $\mathcal{C}^{AA}(\avr{E})$ and $\mathcal{C}^{AB}(\avr{E})$ is empty (or at least its surface measure is zero), while \LNNL
\begin{align}
& \mathcal{C}^{AA}(\avr{E}) \rightarrow \mathcal{C}_0 \leftarrow \mathcal{C}^{AB}(\avr{E}),\,\,\text{for}\,\, \avr{E} \rightarrow \infty. \nonumber
\end{align}
Here, $\mathcal{C}_0$ is the constraint surface of the unmodified case $a_0=0$, $a_1=0$. In 3-dimension with $E \propto p^2$ dispersion $\mathcal{C}_0$ is a sphere. \par
In case $II.)$ with interaction threshold, there is a non-zero section of $\mathcal{C}^{AA}(\avr{E})$ and $\mathcal{C}^{AB}(\avr{E})$ for any time. While in case $I.)$ the system reaches a steady state without detailed balance, an equilibrium state develops in case $II.)$ if $\Lambda_0$ is fine-tuned, as it turned out by the numerical investigation. Since the energy cut-off was also scaled by the average kinetic energy $\avr{E}$, the scaling behaviour (\ref{numscalingSOL}) prevails. We depicted the running of $\avr{E}$ and $\gamma$ for various values of $\Lambda_0$ on Fig. (\ref{fig:en_saturation}). As it can be seen on Fig. (\ref{fig:gamma_vs_Lambda}), $\gamma(\Lambda_0)$ has a zero, in the present example ($a_0=0.15$, $a_{10}=0.25$) at $\Lambda_0^* \approx 3.3$. A qualitative explanation of this behaviour can be given if one takes notice of the fixed shape of the density function in the scaled variable $E/\avr{E}$. That is why the probability of such collisions, where the total kinetic energy grows (or decreases), is also constant in the scaling regime, being proportional to the integral of $\psi(E/\avr{E})$ on a definite domain in its variable. Therefore the probability depends on the modification parameters $a_0$, $a_1$, $\Lambda_0$ only, as all the quantities in the scaling regime do. Thus, $\Lambda_0$ prescribes how much the two distinct types of collisions featured by $a_0$ and $a_1$, respectively contribute to the probability of energy growing (decreasing) in the corresponding energy ranges $E_1+E_2 < \Lambda_0\avr{E}$ and $E_1+E_2 > \Lambda_0\avr{E}$. If the kinetic energy domain, which is responsible for the lowering effect, is large enough, then the energy change can be compensated statistically and the system equilibrates. \par
\begin{figure}[!t]
\centering
\includegraphics[width=.6\linewidth]{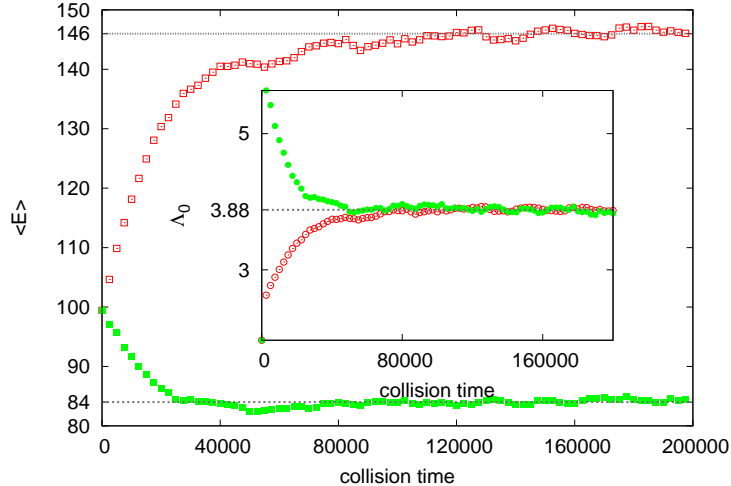}
\LNFIG
\caption{Feedback of the lowering or growing of $\Lambda_0$ as $\avr{E}$ varies. We used a simple smoothed step-function to mimic the feedback effect \FINAL{$\dot{\Lambda}_0 \propto \Lambda_0\cdot\text{sgn}(\dot{\avr{E}})$}. The system tends towards to a steady state for long times 
($a_0=0.11$, $a_1=0.21$, $\Lambda_0^*=3.88$.)}
\label{fig:saturFeedback}
\end{figure}

Studies on multi-component kinetic equations derived on first-principle basis show thermalization and develop equilibrium state for large times, see for example Ref.~\cite{cassing2} (although the particle number of a given species is not fixed in Ref.~\cite{cassing2}). In case $II.)$ it is possible to make the fixed point in the $\Lambda_0-\gamma$ phase-space attractive due to a dynamical feedback. Increasing $\Lambda_0$ when the energy is growing and lowering it if $\avr{E}$ is decreasing makes $\gamma(\Lambda_0^*)=0$ to be a stable fixed point of the time evolution. If the system relaxes to a scaling state fast enough when $\Lambda_0$ changes, then $\gamma(\Lambda_0)$ is indeed the allowed phase-space in the scaling regime. The result of this very simple feedback, \FINAL{$~{\dot{\Lambda}_0 \propto \Lambda_0\cdot\text{sgn}(\dot{\avr{E}})}$} can be seen in Fig.~(\ref{fig:saturFeedback}) from a numerical simulation. $\avr{E}$ and $\Lambda_0$ tend to a constant value, as it is expected.

\section{Conclusions}\label{conclusionsMMBE}
In this chapter we investigated whether an MMBE system tends to an equilibrium state or not. It is a non-trivial question, even with the conceptually simplest modification of the two-particle constraints (modifying the energy addition rule). As far as we know, though the modification of the BE due to the constraints is well discussed in the literature through several examples, there are no studies concerning modified, multi-component systems. The problem arises for all studied examples (either energetic or entropic reasoning for the modification of the collision integral) \cite{biro3,kaniadakis,biro4,rafelski}. The detailed balance state in the multi-component case is generally lacking because different conditions are to be satisfied for each piece of the collision integral to vanish. \par
Our conclusion is, that dealing with such kind of MMBE, equilibration is not guaranteed in general. The same quantity which is conserved in the one-component system with a non-additive energy composition rule, in the multi-component case describes an open system. In order to achieve a stationary state, one has to go beyond the simple kinetic treatment, and has to feedback the dynamics of the energy non-additivity to the MMBE. It is conceivable, that for a satisfactory description, one has to return to the microscopic description of the off-shell effects. \par
Although it is tempting to use such a simple modification to go beyond the on-shell particle picture used in kinetic theory, one should handle this non-self-consistent modification with care. \par To overcome the issue of the irregular equilibration properties, one can derive the kinetic theory on first principle basis: Starting with the Dyson-Schwinger equations of the two-point correlation function, a gradient expansion and the assumption of a near-equilibrium state leads to the Kadanoff-Baym equations. The quasi-particle approximation of the Green-functions (and in the simplest case, keeping the pole contribution only) gives the (quantum version) of the BE. \par
A possible generalization is to keep off-pole contributions. This would unavoidably lead to an effective modification of the kinetic picture but in a dynamical and self-consistent way, see for example Refs.~\cite{spicka2, spicka3, nonlocalMorawetz, beyondQPfermion} and also the references therein.
\clearpage
\chapter[Hydrodynamic behaviour of classical radiation patterns]{Hydrodynamic behaviour of \\ classical radiation patterns}\label{ellflow}

In this chapter, we present a simple phenomenological model built on the basis of semi-classical quasi-particles, aimed to describe the elliptic (or azimuthal) asymmetry factor, or simply the \textit{elliptic flow} ($v_2$) of the photonic and light hadronic yields observed in heavy-ion experiments. The quantity $v_2$ characterizes the second term in the Fourier series according to the $\varphi$ azimuthal angle: $2v_2\cos(2(\varphi-\psi))$, which we will discuss later in Sec.~\ref{v2EllFlow}. The widely accepted way to compute the particle yields in the final stage\footnote{When the produced particles are only streaming freely.} is the following: $i.)$ preparing the initial densities of the conserved quantities motivated by QFT calculations (like CGC), $ii.)$ let the conserved quantities evolve in time by solving the hydrodynamical equations numerically, $iii.)$ matching the final state energy-momentum and charge densities to those of a gas mixture composed by various hadronic species and $iv.)$ let them evolve using kinetic description (BE) until the interactions cease and the kinetic freeze-out happens. \par
Although hydrodynamics seems to be able to describe the expanding QGP in proton-nucleus and heavy-ion collisions -- see Refs.~\cite{hydro_bozek_1,hydro_bozek_2,hydro_bozek_3,hydro_gale_schenke_1,hydro_gale_schenke_2,hydro_shen,hydro_jiang,hydro_romatschke_1,hydro_wiedemann_1,hydro_wiedemann_2,hydro_wiedemann_3} and the references therein --, it is still not clear, how the final state of the system is effected by correlations other than the hydrodynamic ones. It is argued by several authors, that the correlations which are present also in the initial state, can survive the hydrodynamic evolution. It was found in Refs.~\cite{colorantennas, SU2dipole, lappi}, that quark bremsstrahlung coming from the initial state of QGP, could produce photons in comparable amount to those produced in the plasma phase. This eventually has a considerable contribution to the observables, like to the azimuthal asymmetry factor $v_2$. To understand these contributions is important in order to parametrize the hydrodynamic simulations realistically. \par
However, the fact that hydrodynamics has a strong predictive power does not imply that it is the only option to explain collective phenomena in such systems. There have been recent efforts to reproduce the flow patterns observed in RHIC and LHC using color scintillating antennas consisting of radiating gluons \cite{colorantennas,colorantennas2}. Other authors utilized phenomenological models of color-electric dipoles in order to account for angular correlations in high-energy processes \cite{dipole,dipole2,dipole3}. It is an ongoing debate though, whether a simple effective model, lacking hydrodynamics, could catch the flow-like behaviour or not. Unfortunately, it is rather complicated to explain the collective properties using microscopical models as a starting point. \par
Our goal here is to demonstrate, that the radiation originating from a dipole set-up is, in principle, able to match quantitatively the elliptic asymmetry factor $v_2$, measured in heavy-ion experiments. To do so, we discuss the yield of massless particles produced by a decelerating point-like charge in details in Sec.~\ref{deceleratingSourceEllFlow}. Then we compute the flow coefficient $v_2$ of a dipole composed of two, parallel displaced counter-decelerating charges in Sec.~\ref{v2EllFlow}. Then, a statistical ensemble of ordered radiators is used to approximate the situation, when a large number of microscopic sources are involved in Sec.~\ref{phaseShiftEllFlow}. Motivated by the formula in Eq.~(\ref{v2avr}), we fit our model to various experimental data in Sec.~\ref{fitsEllFlow}. Our analysis also involves discussing several open issues and their relevance for further, more realistic description. We also give a geometrical interpretation of the fitting parameters in the fitting formula Eq.~(\ref{v2fit}). Finally, we conclude speculating on how the model assumptions could be supported by mechanisms existing within the framework of the microscopic theory. The results of this chapter were mainly published in Ref.~\cite{v2paper}.

\section{Radiation produced by decelerating sources}\label{deceleratingSourceEllFlow}
According to classical electrodynamics, an accelerating point charge radiates. One can reinterpret this phenomenon quasi-classically as the emission of photons. It is straightforward to calculate the differential yield of emitted photons when the charge $e$ accelerates uniformly on a straight line. We do not go into the details of the derivation here, since we need the end result of the EM-analysis only.\footnote{The interested reader can find the detailed computation in Ref.~\cite{unruhgamma}.} The yield of photons emitted by a single point-like charge (or classically, the power density of EM-radiation per unit area) is determined by the formula:\LNNL
\begin{align}
Y_\text{sc} &:= \frac{\mathrm{d}^3N_\text{sc}}{k_\perp\mathrm{d}k_\perp\mathrm{d}\eta\mathrm{d}\psi} =2|\mathcal{A}_\text{sc}|^2, \nonumber \\
\mathcal{A}_\text{sc} &= \frac{e}{\sqrt{8\pi}}\int\!\!\mathrm{d}\xi e^{i\phi}\frac{\mathrm{d}}{\mathrm{d}\xi}\left(\frac{\epsilon\cdot u}{k\cdot u}\right), \label{yield0}
\end{align}
\begin{figure}[!t]
\centering
\includegraphics[width=0.5\linewidth]{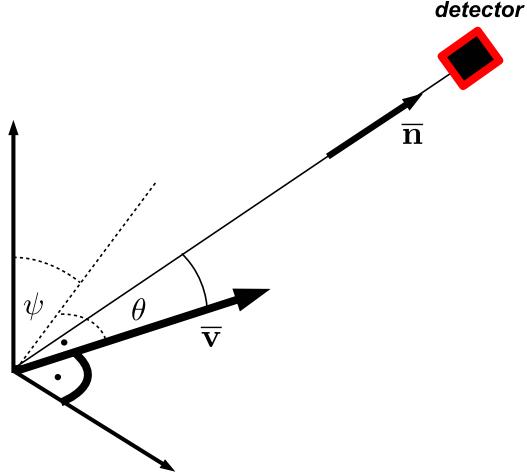}
\caption{\FINAL{Accelerating charge emits photons. The charge is moving on a straigt-line path appointed by $\ve{v}$. The photon detector is situated in the $\ve{n}$-direction. See the text for further explanations on the notations.}}
\label{fig:acc_charge}
\end{figure}
where the lower index ''sc'' stands for ''single charge''. The notations are the following. The azimuth angle $\psi$ is measured in the plane perpendicular to the trajectory, $\theta$ is the distortion angle measured from the tangent direction of the trajectory. \rem{We denoted the polarization vector by $\epsilon$.} The rapidity of the emitted photon, $\eta$, can be expressed as ${\text{tanh}\eta=\cos\theta}$. The phase $\phi$, the four-velocity $u$, the wave-vector $k$ and the rapidity $\xi$ in the direction of the motion read as
\[\begin{array}{lcl}
\phi= \omega(t-\ve{n}\cdot\ve{r}), & & k=\omega(1,\,\ve{n}), \\
u=\gamma_L(1,\,\ve{v}), & & \ve{v}=(v,\,0,\,0), \\
\xi=\text{arth}v, & & \gamma_L =\frac{1}{\sqrt{1-v^2}},
\end{array}\]
with $\ve{r}$ being the trajectory of the charge, whilst $\ve{n}$ points to the direction of the detector, \FINAL{see Fig.\! \ref{fig:acc_charge}}. We fixed the units as ${c=1}$. In case of uniform acceleration $g$ which lasts for a finite period of time, the formula of Eq.~(\ref{yield0}) simplifies -- using also the polarization vector ${\epsilon=\frac{1}{\omega}\frac{\partial k}{\partial\theta}}$:
\begin{equation}\label{yield}
Y_\text{sc}=\frac{\mathrm{d}^3N_\text{sc}}{k_\perp\mathrm{d}k_\perp\mathrm{d}\eta\mathrm{d}\psi}
=\frac{2\alpha_\mathrm{EM}}{\pi}\frac{1}{k_\perp^2} \left|
\, \int\limits_{w_1}^{w_2}\mathrm{d}w\frac{e^{iw\frac{k_\perp}{g}}}{(1+w^2)^\frac{3}{2}} \right|^2.
\end{equation}
Here the parameters $w_1$ and $w_2$ are in connection with the rapidity of the charge in the frame of the laboratory observer: \FINAL{$w_{1,2}=\text{sh}(\xi_{1,2}-\eta)$, $\xi_{1,2}=\xi_0+g\tau_{1,2}$, with initial rapidity value $\xi_0$}. The magnitude of the co-moving acceleration is $g$ and ${\alpha_\text{EM}=\frac{e^2}{4\pi\hbar}}$. For further details of the calculations of the emitted radiation, see Ref.~\cite{unruhgamma}. \par
\begin{figure}
\centering
\includegraphics[width=0.6\linewidth]{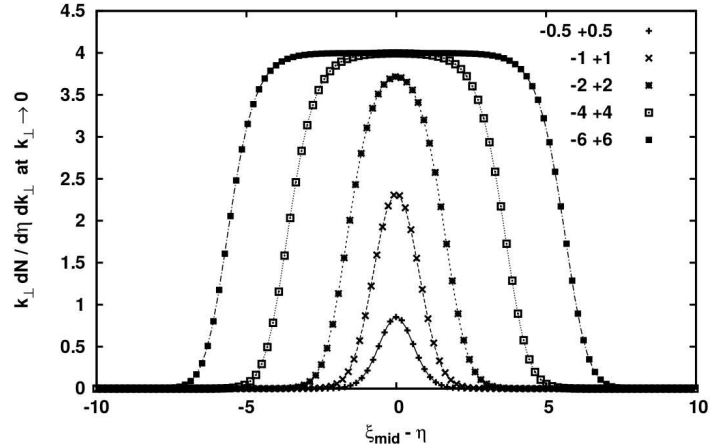}
\LNFIG
\caption{$k_\perp^2$ times the invariant photon yield at $k_\perp=0.01$ as a function of the rapidity ${\xi_\text{mid}-\eta}$, ${\xi_\text{mid}=\xi_0+g\frac{\tau_1+\tau_2}{2}}$. This numerical value we use as an approximation for the infrared limit. The different curves belong to varying proper time durations of the constant acceleration ($g = 1$) according to the legend (denoting $\tau_1$ and $\tau_2$ values). This figure is taken from Ref.~\cite{accelcharge}.}
\label{fig:tsb_zssz_zssch}
\end{figure}

It is noteworthy that Eq.~(\ref{yield}) in the $k_\perp\rightarrow 0$ limit reproduces a bell-shaped rapidity distribution, similar to Landau's hydrodynamical model, and also the plateau known from the Hwa--Bjorken scenario, depending on whether the accelerating motion of the charge covers a short or large range in rapidity -- as it is shown in Fig.~\ref{fig:tsb_zssz_zssch}. \par
Speculating further, we assume that in the case of light particles produced in a heavy-ion collision, a significant part of the yield comes from similar, deceleration induced radiation processes.
It is also worthwhile to mention that a gauge field theory which describes the radiation phenomena on the microscopic level, can be reformulated in the framework of hydrodynamics, as it was endeavoured in Refs. \cite{jackiw,jackiw2,jackiw3}.

\section{Elliptic flow coefficient}\label{v2EllFlow}
\begin{figure}[!t]
\centering
\includegraphics[width=0.75\linewidth]{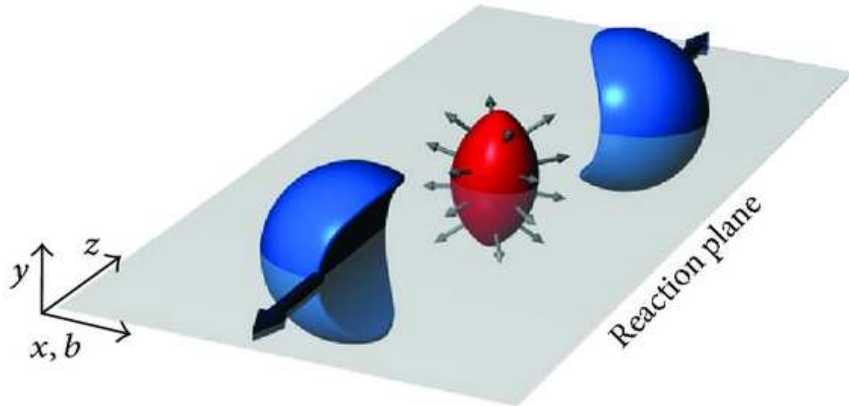}
\LNFIG
\caption{Schematic representation of a non-central heavy-ion collision. The distance of the trajectories of the two nuclei is denoted by $b$, the impact parameter. $x-z$ is the reaction plane, $x-y$ is the transverse plane of the collision. This figure is from \cite{introHydro2}.}
\label{fig:HICgeom}
\end{figure}
It is convenient to expand the particle yields into Fourier-series according to the azimuthal angle (measured in the so-called transverse plane, which is perpendicular to the reaction plane spanned by the pre-collisional trajectories of the nuclei, see Fig.~\ref{fig:HICgeom}):\LNNL
\begin{align}
Y &= Y_0\left(1+\sum_{n=1}^\infty 2v_n\cos(n(\varphi-\psi))\right),\label{yieldAzim} \\
v_n &= \frac{\intlim{\varphi}{0}{2\pi}Y\cos(n(\varphi-\psi))}{\intlim{\varphi}{0}{2\pi}Y},\label{v2def}
\end{align}
where $\psi$ is the azimuthal angle belonging to the reaction plane in the lab frame.\par
From the theoretical point of view, it is useful to check how various model assumptions are reflected in the various coefficients $v_n$. The experience shows that in a non-central event $Y_0$ and $v_2$ are far the most significant ($v_3$ is smaller with an order of magnitude compared to $v_2$). The coefficient $v_2$ is sensitive to the azimuthal asymmetry of the collision and it is smaller for more central events. Therefore it is called the elliptic asymmetry coefficient, or shortly the \textit{elliptic flow}.\par
We now turn to the analysis of the simplest structure which can cause a non-zero $v_2$. An elliptic asymmetry could stem from two decelerating point-like sources going into opposite directions on parallel paths. We calculate the emitted photon-equivalent radiation for the distance $d$ between the two sources, see Fig.~\ref{fig:dipole}. The yield is calculated by the coherent sum of the plane-wave-like amplitudes of the contributing sources:
\begin{eqnarray}\label{dipoleyield}
Y &\propto & |\mathcal{A}_1+\mathcal{A}_2|^2 =\nonumber \\
&=& |A_1e^{ik_\perp\frac{d}{2} \cos(\varphi-\psi)}+A_2e^{-ik_\perp\frac{d}{2} \cos(\varphi-\psi)}|^2
\nonumber \\
&=& |A_1|^2+|A_2|^2+2\mathrm{Re}(A_1A_2^* e^{ik_\perp d \cos(\varphi-\psi)}).
\end{eqnarray}
After Fourier-expansion -- which here is equivalent with the application of the so-called Jacobi--Anger expansion\footnote{$\cos(z\cos(\alpha))=J_0(z) +2\sum_{n=1}^\infty(-1)^nJ_{2n}(z)\cos(2n\alpha)$, \\ $\sin(z\cos(\alpha)=-2\sum_{n=1}^\infty(-1)^nJ_{2n-1}(z)\cos((2n-1)\alpha)$.} -- we get the following expression:\LNNL
\begin{align}
Y = &\underbrace{|A_1|^2+|A_2|^2+2\re{(A_1A_2^*)}J_0(k_\perp d)}_{=:Y_0} + \nonumber \\
+& 4\sum\limits_{n=1}^\infty \re{(A_1A_2^*)}(-1)^nJ_{2n}(k_\perp d)\cos(2n(\varphi-\psi)) +\nonumber \\
+& 4\sum\limits_{n=1}^\infty \im{(A_1A_2^*)}(-1)^nJ_{2n-1}(k_\perp d)\cos((2n-1)(\varphi-\psi)).
\end{align}
For arbitrary $n\geq 1$ integer the expansion coefficient $v_n$ reads as\LNNL
\begin{align}
v_{2n-1}&= (-1)^n\frac{2\im{(A_1A_2^*)}J_{2n-1}(k_\perp d)}{|A_1|^2+|A_2|^2+2\mathrm{Re}(A_1A_2^*)J_0(k_\perp d)},\label{dipolevOdd} \\
v_{2n}&= (-1)^n\frac{2\mathrm{Re}(A_1A_2^*)J_{2n}(k_\perp d)}{|A_1|^2+|A_2|^2+2\mathrm{Re}(A_1A_2^*)J_0(k_\perp d)}.\label{dipolevEven}
\end{align}
In the above formulae $J_n$ is the Bessel function of the first kind. Parametrizing the complex amplitudes as $A_1=Ae^{i\delta_0}$ and $A_2=\gamma Ae^{i(\delta_0+\delta)}$ with real $A$, $\gamma$ and $\delta_0$, where $\delta$ denotes the phase-shift and $\gamma$ is the ratio of $|A_2|/|A_1|$, we obtain the simplified expressions for the elliptic flow:
\begin{equation}
v_2 = \frac{-J_2(k_\perp d)\cos\delta}{\frac{1+\gamma^2}{2\gamma}+J_0(k_\perp d)\cos\delta }.
\label{v2final}
\end{equation}
We assumed here, that $A_1$ and $A_2$ are azimuthally symmetric, i.e. independent of the difference $\varphi-\psi$.\par

\section{Phase-shift averaged $v_2$}\label{phaseShiftEllFlow}
\begin{figure}[!t]
\centering
\subfloat[]{
\fbox{\includegraphics[width=0.5\linewidth]{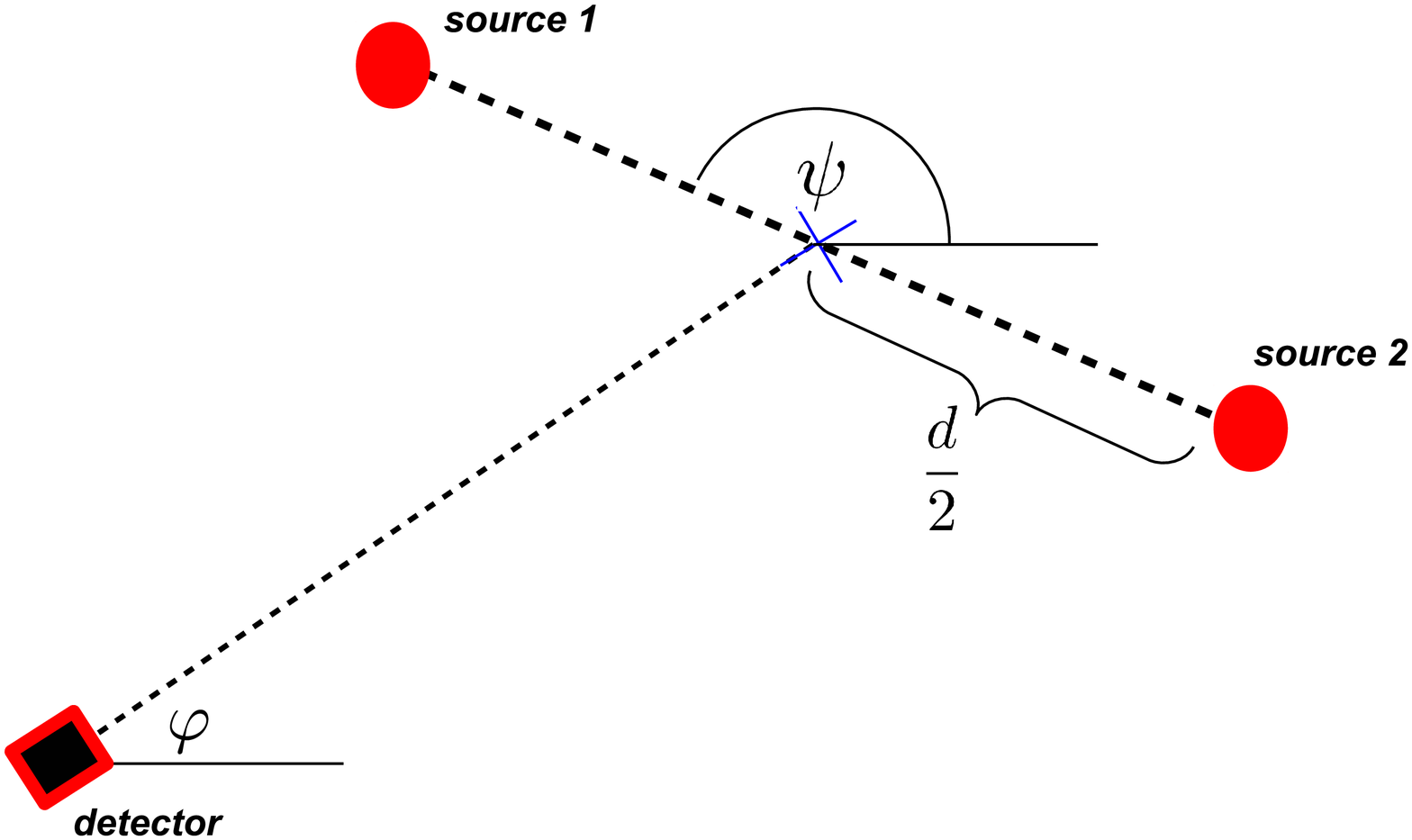}}
\label{fig:dipole}
}
\subfloat[]{
\fbox{\includegraphics[width=0.5\linewidth]{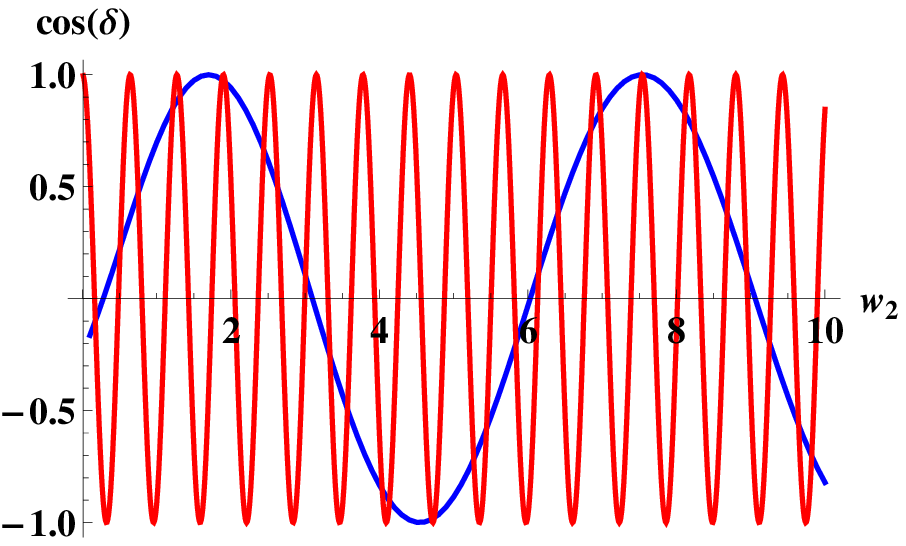}}
\label{fig:phaseshift}
}
\LNFIG
\caption{(a): Schematic representation of a single radiating pair of decelerating sources in the transverse plane. $\psi$ is the azimuthal orientation of the reaction plane in the lab frame. (b): Phase-shift factor $\cos\delta$ as a function of the final velocity $w_2$ in cases $\Delta=1.0$ (blue curve) and $\Delta=10.0$ (red curve).}
\end{figure}
In our picture $v_2$ depends on the phase-shift $\delta$, the dipole size $d$ and the strength asymmetry parameter $\gamma$. We assume that event-by-event $d$ and $\gamma$ might be well-determined, while $\delta$ fluctuates. The decelerating sources are strongly affected by the medium surrounding them, therefore, they radiate differently, depending on how long the interaction holds up. We consider one source decelerating from a velocity near $c$ to $0$ and the other one from $c$ to slightly above 0, in the opposite direction. The relevant amplitudes then read:
\begin{eqnarray}
\mathcal{A}_1 &\sim &
\int\limits_{-\infty}^{w_1=0}\mathrm{d}w\frac{e^{iw\Delta}}{(1+w^2)^\frac{3}{2}} =\nonumber \\
&=&\frac{\Delta}{2}\left[ 2K_1(\Delta) +i\pi \left(K_1(\Delta)-L_{-1}(\Delta)\right) \right],
\nonumber \\
\mathcal{A}_2 &\sim &
\int\limits_{\infty}^{w_2}\mathrm{d}w\frac{e^{iw\Delta}}{(1+w^2)^\frac{3}{2}}
= -\mathcal{A}_1^*+ \int\limits_0^{w_2}\mathrm{d}w\frac{e^{iw\Delta}}{(1+w^2)^\frac{3}{2}},
\end{eqnarray}
as it follows from Eq.~(\ref{yield}) with $\Delta=\frac{k_\perp}{g}$, $K_n$ and $L_\nu$ being the modified Bessel functions of second kind and the modified Struve function, respectively. The phase-shift factor, $\cos\delta=\frac{\mathrm{Re}(\mathcal{A}_1\mathcal{A}_2^*)}{|\mathcal{A}_1||\mathcal{A}_2|}$, can be evaluated numerically, the result is plotted in Fig.~\ref{fig:phaseshift}. It appears as a natural idea to average with respect to the phase difference variable, $\delta$, whenever $\cos\delta$ oscillates fast as a function of $w_2$ (cf. Fig.~\ref{fig:phaseshift}). \par
We can see, that for varying $w_2$, the phase shift $\delta$ quickly explores its all possible values. This observation enables to treat the ensemble of radiator pairs statistically, and to perform averaging on the phase-space parametrized by $\delta$. The uniform averaging respect to the phase-shift angle can be carried out analytically, resulting
\begin{equation}\label{v2avr}
\avr{v_2} = \frac{J_2(k_\perp d)}{J_0(k_\perp d)}\left(\frac{1}{\sqrt{1-\frac{4\gamma^2}{(1+\gamma^2)^2}J_0^2(k_\perp d)}}-1\right).
\end{equation}
We note here, that the odd $v_{2n-1}$ coefficients vanish after this averaging.

\section{Fits to experimental data}\label{fitsEllFlow}
Hereinafter we assume that the leading order contribution to the elliptic asymmetry comes from the yield produced by an ensemble of dipole-like structures discussed previously. We shall test our hypothesis on elliptic flow measurements in heavy-ion collisions at RHIC and LHC, where the fairly large number of dipoles ensures the validity of our working hypothesis, namely the uniform distribution of the phase-shift of the sources. We introduce an additional fit parameter, called $F$. This geometrical form factor is assumed to be independent of the transverse momentum $k_{\perp}$. Finally, the formula we use to fit the experimental data
\begin{equation}\label{v2fit}
\avr{v_2}_\mathrm{fit} \: = \: F \cdot \avr{v_2}\!\left(k_\perp d,\gamma\right),
\end{equation}
with $\avr{v_2}$ defined in Eq.~(\ref{v2avr}). \par
\begin{figure}[!h]
\centering
\LNFIG
\subfloat[Inclusive photon elliptic flow measured by \mbox{ALICE} group of LHC in Pb-Pb collisions at $\sqrt{s_{NN}}=2.76$ TeV for several centrality classes \cite{inclPhoton_v2_ALICE}. The continuous curves are the fitted ones.]{
\includegraphics[width=0.8\linewidth]{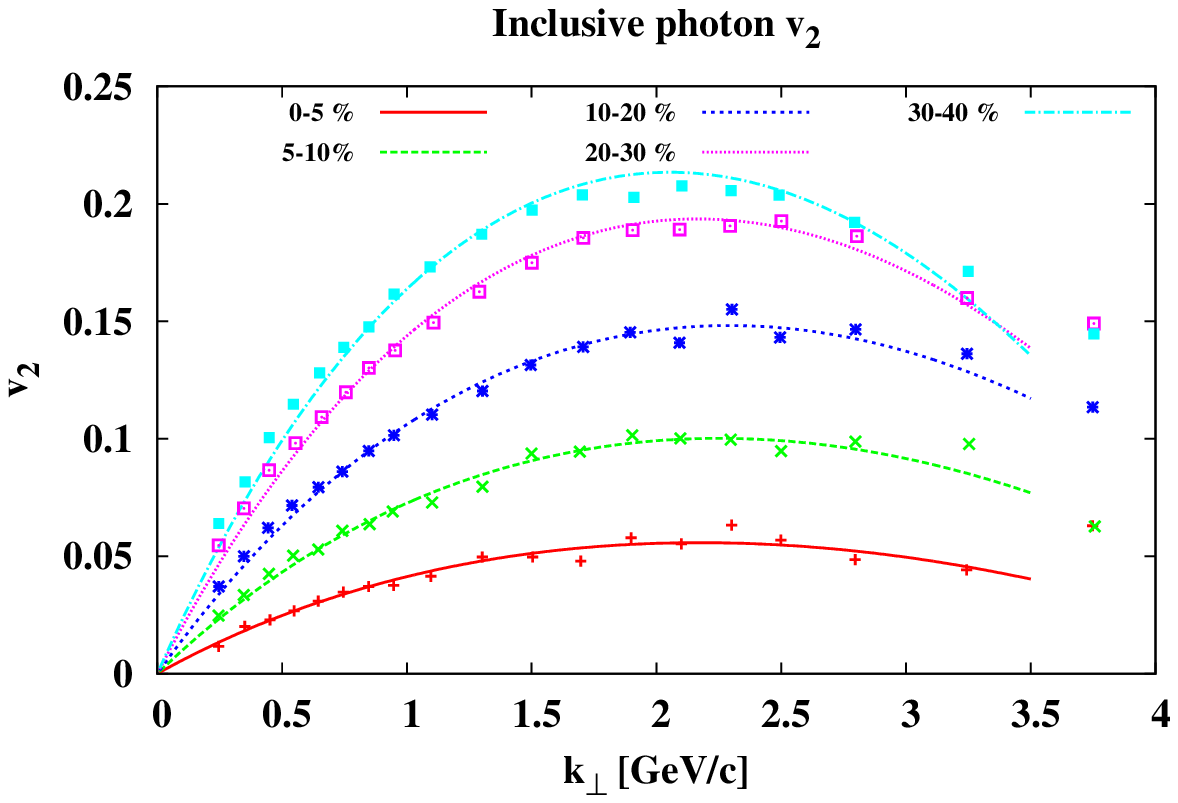}
\label{fig:photons}
}\\
\subfloat[Elliptic flow of charged pions measured by STAR of RHIC in Au-Au collisions at $\sqrt{s_{NN}}=62.4$ GeV (solid line) \cite{ideParts_v2_STAR}, and by PHENIX of RHIC in Au-Au collisions at $\sqrt{s_{NN}}=200$ GeV (dashed line) \cite{idParts_v2_PHENIX}. The continuous curves are the fitted ones.]{
\includegraphics[width=0.8\linewidth]{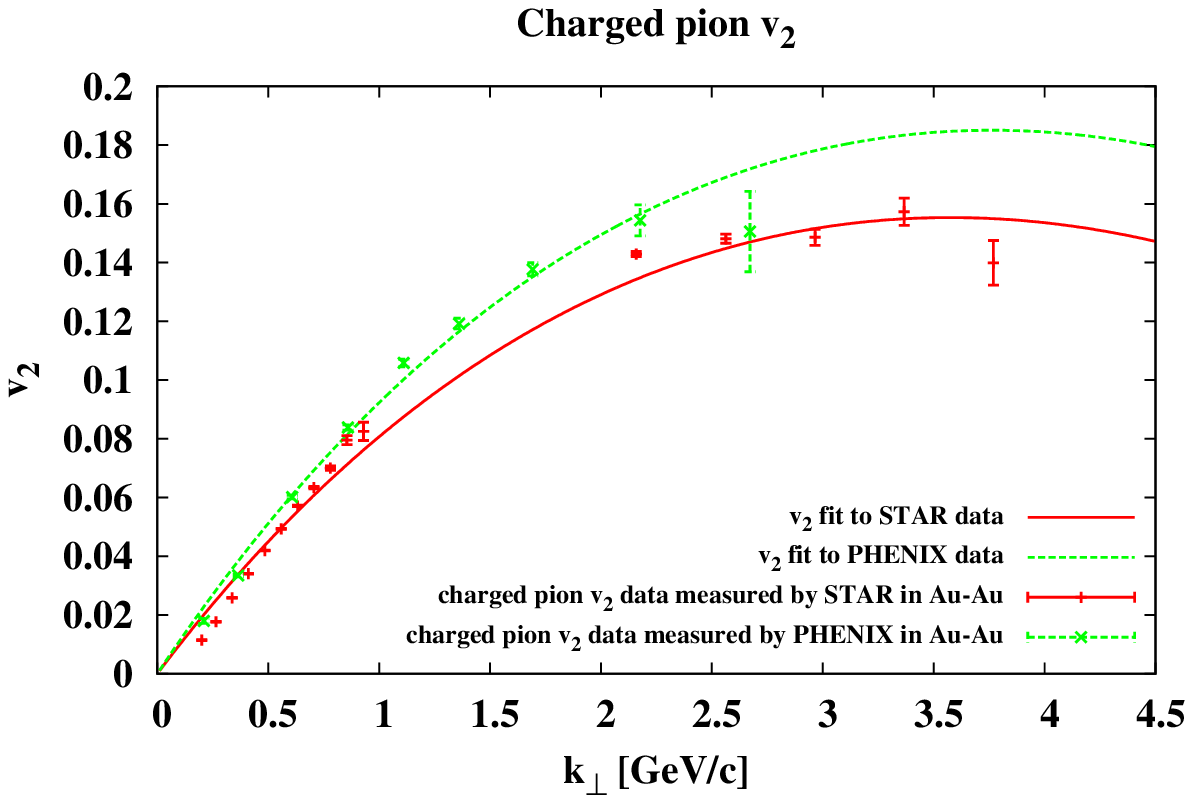}
\label{fig:pions}
}
\caption{}
\end{figure}

\begin{figure}[!t]
\centering
\LNFIG
\subfloat[Elliptic flow of charged hadrons measured by PHENIX in Au-Au collisions at $\sqrt{s_{NN}}=200$ GeV for several centrality classes \cite{idParts_v2_PHENIX2}. The continuous curves are the fitted ones.]{
\includegraphics[width=0.8\linewidth]{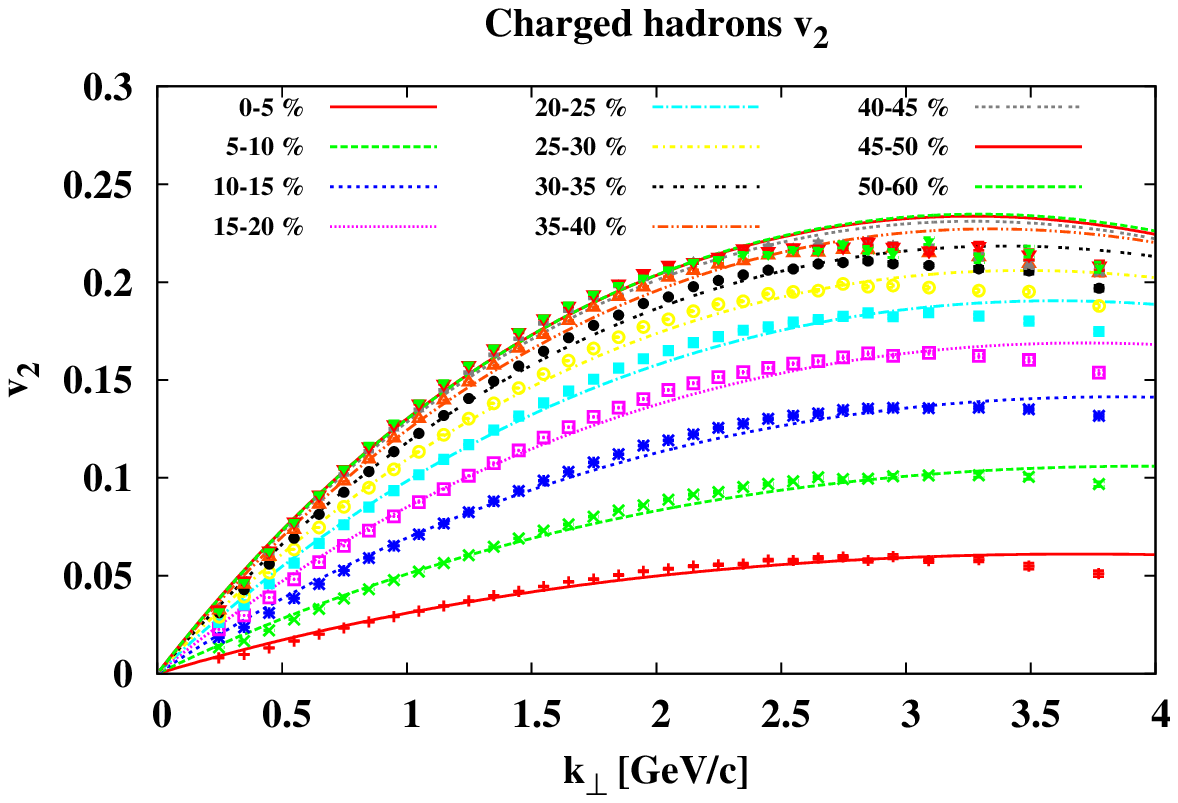}
\label{fig:chhadrons}
}\\
\subfloat[The form factor $F$ -- see Eq. (\ref{v2fit}) -- versus the impact parameter $b$ of the collision for inclusive photons and charged hadrons. The centrality--impact parameter relationship is taken from \cite{centr_vs_multip}. The continuous curves are the fitted ones.]{
\includegraphics[width=0.8\linewidth]{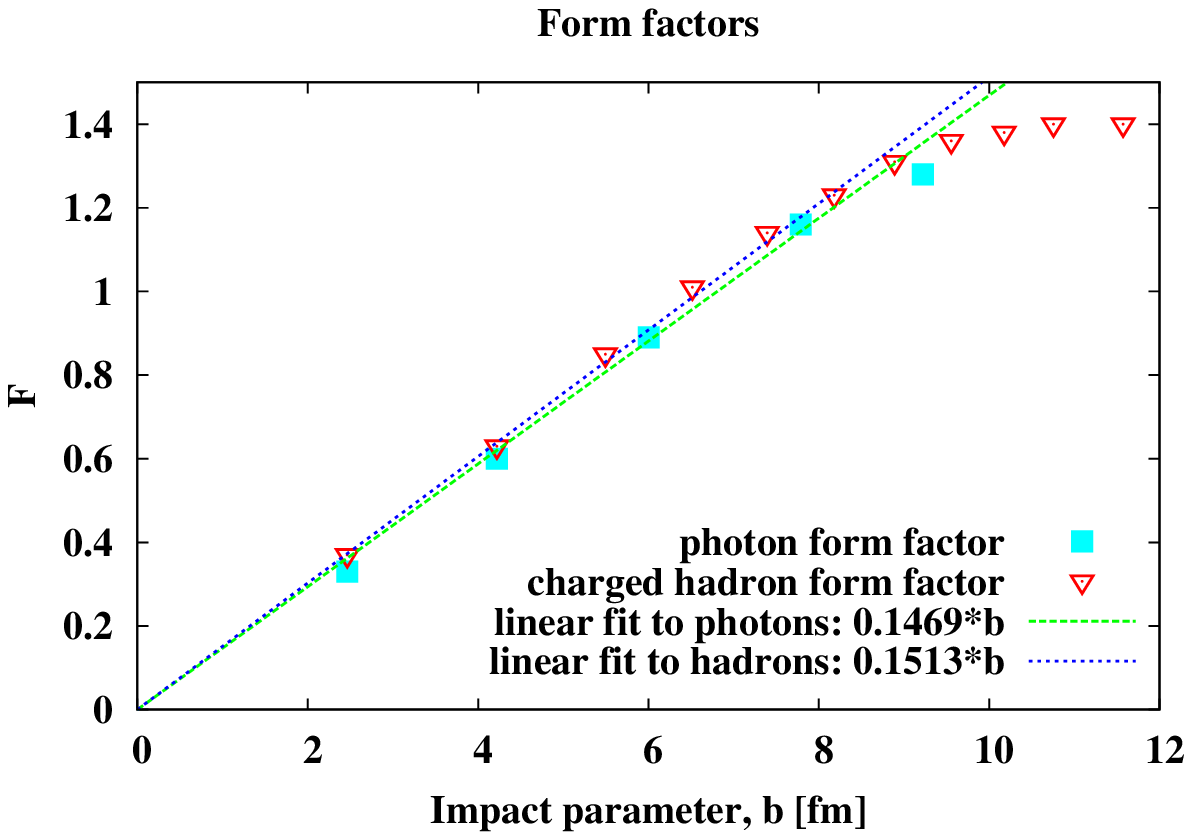}
\label{fig:formfactor}
}
\caption{}
\end{figure}
Comparison of measured and calculated $v_2$ as a function of $k_\perp$ are depicted on Figs.~\ref{fig:photons}, \ref{fig:pions} and \ref{fig:chhadrons}. For data resolved by centrality the fitted parameter, $d$, remains the same within 11\% for photons and 19\% for charged hadrons, cf. Table~\ref{tab:tab1}. Including more peripheral collisions, $F$ saturates somewhat below 1.5 (see Fig.\ref{fig:formfactor}). Since $F=1$ would mean that only the dipole term contributes to $v_2$, cf. Eq. (\ref{v2avr}), that suggests the need for further sources of elliptic flow, for example multipole contributions. In all cases $\gamma$ turns out to be very close to one, showing that symmetric dipole sources may dominate the radiation process. Interestingly, the $v_2$ values for charged pions on Fig.~\ref{fig:pions} and hadrons on Fig.~\ref{fig:chhadrons} fit as well as for the emitted photons on Fig.~\ref{fig:photons}. \rem{Above $2-3$GeV/c in transverse momentum, our fitting formula seems to overestimate the experimental data systematically. In this momentum region the ''hard physics'' of the QCD starts to overcome the low-momentum particles which thought to reflect the bulk properties of the SIM. See for example Refs.~\cite{RHICwhiteP1, RHICwhiteP2} and the references therein.\footnote{We note, that this overestimation effect can be partially the result of the subtraction of non-flow contributions during the analysis made by the various research collaboration.} }\par
\begin{table}[!h]\centering
\begin{tabular}{c|c|c|c|c}
\hline
Cent. [\%] & $F$ & $d$ [fm] & $\gamma$ & $\chi^2$ \\
\hline\hline
\multicolumn{5}{c}{ALICE photon $v_2$ fit parameters}\\ \hline
0$-$5& 0.33& 0.10& 1.00& 0.36\\
5$-$10& 0.60& 0.10 & 1.00& 0.36\\
10$-$20& 0.89& 0.09 & 1.00& 0.58\\
20$-$30& 1.16& 0.10 & 1.00& 0.82\\
30$-$40& 1.28& 0.10 & 1.00& 1.60\\
\hline
\multicolumn{5}{c}{PHENIX charged hadron $v_2$ fit parameters} \\ \hline
0$-$5& 0.37& 0.06& 1.00& 4.46\\
5$-$10& 0.63& 0.05 & 1.00& 4.31\\
10$-$15& 0.85& 0.06 & 1.00& 5.25\\
15$-$20& 1.01& 0.06 & 1.00& 6.09\\
20$-$25& 1.14& 0.06 & 1.00& 6.88\\
25$-$30& 1.23& 0.06 & 1.00& 7.24\\
30$-$35& 1.31& 0.06 & 1.00& 7.55\\
35$-$40& 1.36& 0.06 & 1.00& 7.91\\
40$-$45& 1.38& 0.07 & 1.00& 8.19\\
45$-$50& 1.40& 0.07 & 1.00& 7.53\\
50$-$60& 1.40& 0.07 & 1.00& 7.57\\
\hline
\multicolumn{5}{c}{STAR charged pion $v_2$ fit parameters} \\ \hline
& 0.93& 0.06& 1.00& 4.48\\
\hline
\multicolumn{5}{c}{PHENIX charged pion $v_2$ fit parameters} \\ \hline
& 1.11& 0.06& 1.00& 0.15
\end{tabular}
\LNFIG
\caption{List of fitted parameters according to the formula in Eq.~(\ref{v2fit}). $\chi^2$ values contain the published measurement errors and also the error of fitting: \FINAL{$\chi^2 = \frac{1}{N}\sum_{i=1}^N\frac{(v_2^i-\avr{v_2}_\text{fit}^i)^2}{\Delta^2+\sum_j\left(\frac{\partial\avr{v_2}_\text{fit}}{\partial p_j}\delta p_j\right)^2}$, where $N$ denotes the number of data records, $\Delta$ measures the systematical errors and $p_j$ stands for the parmeters of the fitting function.}}
\label{tab:tab1}
\end{table}
We briefly list recent literature studies about how various stages of the heavy-ion collision could contribute to the azimuthal asymmetry of the flow in order to support the phenomenological picture we sketched above. We focus on the role of dipole-like structures being revealed at the early-time stage of the HIC.\par
\begin{enumerate}[\it i.)]
\item \textit{Strong electromagnetic (EM) fields.} In non-central collisions, the magnitude of the magnetic field due to the geometrical asymmetry of the system could reach $\sim 5 m^2_\pi$ for a short time of $0.2$fm/c \cite{HIJING_EM,HSD_EM}. The pure EM-effect (caused by the coupling of charged quasi-particles and the EM-field) is, however, not significant at the level of global observables, as it is suggested by hadron string dynamics simulations \cite{HSD_EM}, or contributes to higher order asymmetries only (quadrupole electric moment) \cite{HIJING_EM}. Note, that these simulations are based on transport models using quasi-particles and improved, but essentially perturbative cross sections.\par
In Ref. \cite{mclerran_skokov} the authors use an order-of-magnitude estimation, leading order in perturbation theory, for the gluon-photon coupling in order to argue that the direct photon flow maybe affected at RHIC, but unlikely at LHC.\par
\item \textit{QCD in magnetic field.} Lattice Monte-Carlo simulations suggest, that QCD at high temperature is paramagnetic, see Ref.~\cite{latticeQCD}. Therefore a ''squeezing'' of the plasma could occur, elongating it in the direction of external magnetic field, which, in case of non-central collisions, points perpendicular to the reaction plane. Charge separation of quarks in the direction of the external magnetic field due to the fluctuation of the topological charge (known as the chiral magnetic effect, CME) can also contribute to the asymmetry of the plasma, as it is indicated by lattice results \cite{latticeQCD2}. These effects are not yet incorporated in simulations based on quasi-particles, like in Refs.~\cite{HIJING_EM,HSD_EM}.
\item \textit{Radiation of non-Abelian plasma.} The classical limit of non-Abelian fields generated by ultra-relativistic sources is analysed in Ref.~\cite{SU2dipole}. It is shown, that dipole-like structures will emerge with the same geometric properties like their EM-versions. These could be important, when the initial state of the matter -- like the color glass condensate (CGC) -- is melting down and converting to QGP, while a considerable amount of quark-antiquark pairs are produced. This happens probably when dipoles are smaller than 1 fm, accompanied by fast oscillation of the sources.\par
It is pointed out in Ref.~\cite{Goloviznin}, that the bremsstrahlung of quarks on the surface of the QGP, pulled back by the confining force, could produce photon radiation in comparable amount to those produced in the plasma phase.
\end{enumerate}
The typical value of the effective dipole size $d$ is about $0.06$ fm from the hadronic fits and about $0.1$ fm from photon data according to our investigation. This is rather small compared to the size of heavy-ion fireballs. It may hint to subhadronic sources of this part of radiation. We mention here, that other authors pointed out the quark-level origin of the flow independently of our present analysis \cite{lacey}. \par

\section{The interpretation of the form factor $F$}\label{formFactorEllFlow}
\begin{figure}
\centering
\subfloat[]{
\fbox{\includegraphics[width=0.6\linewidth]{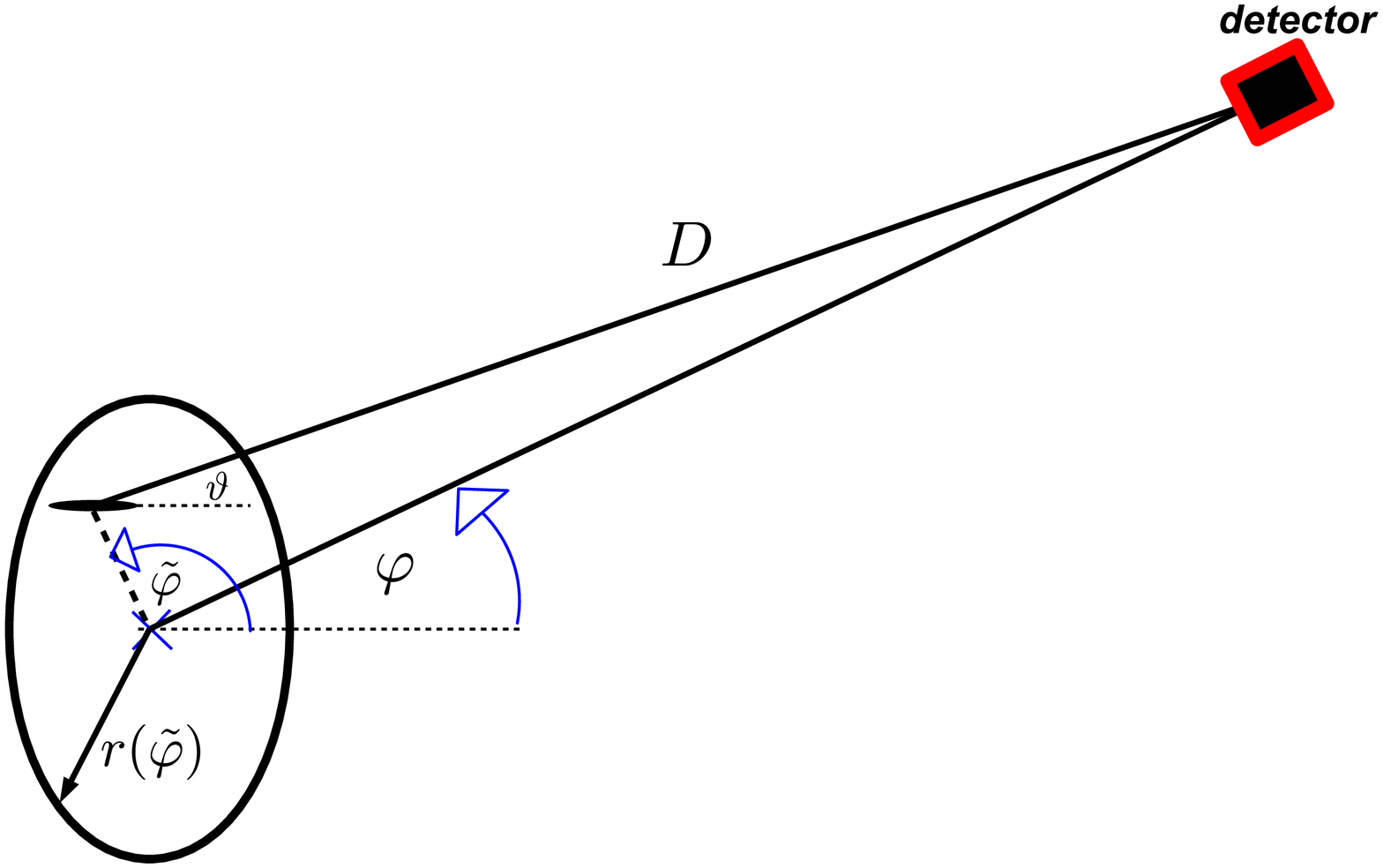}}
\label{fig:geomAvr}
}
\subfloat[]{
\fbox{\includegraphics[width=0.35\linewidth]{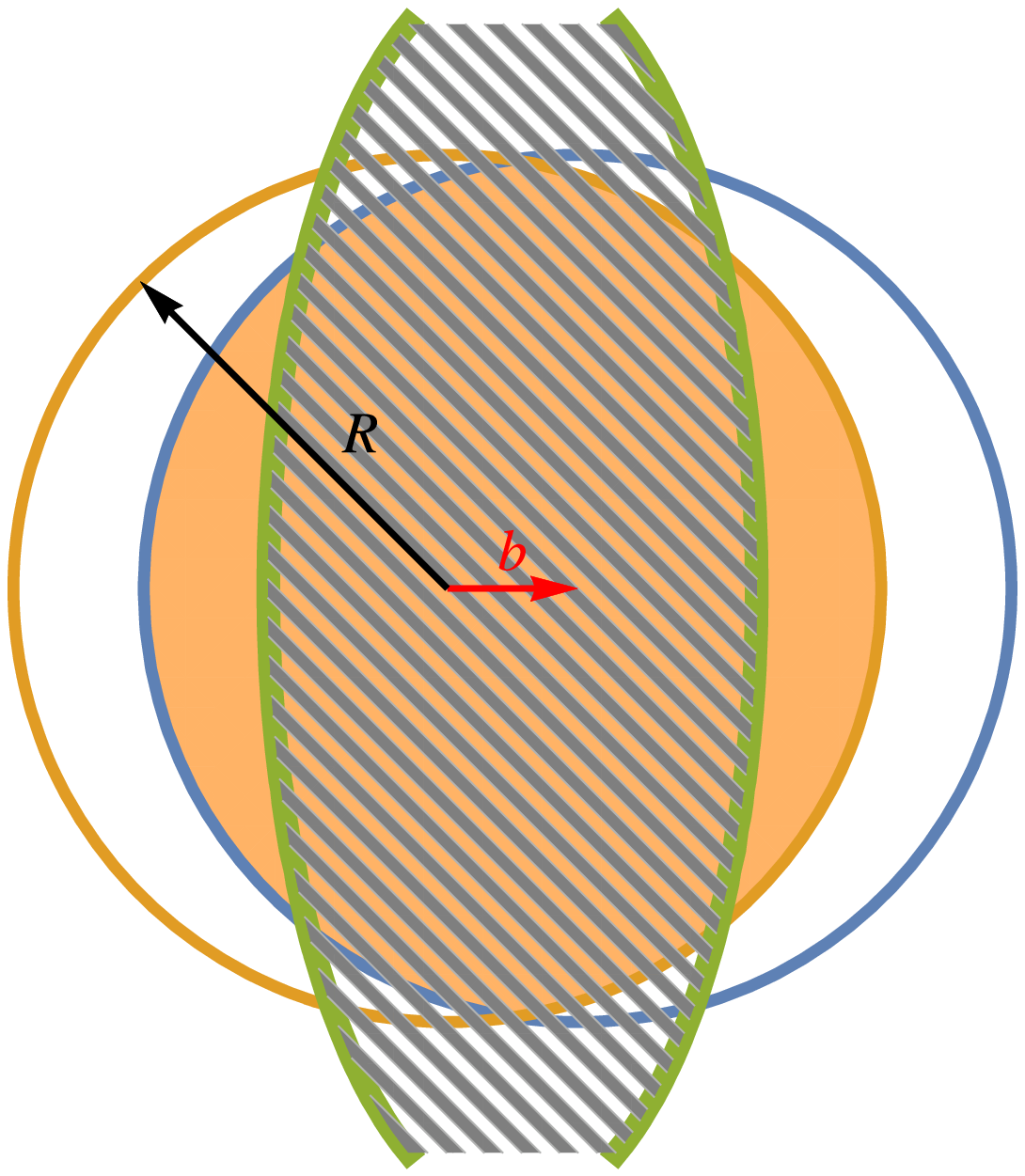}}
\label{fig:HICcartoon}
}
\LNFIG
\caption{(a): Geometrical average over a small domain of radiator-pairs. In the transverse plane, $\varphi$ is the direction of the detector from the center of the domain. $\tilde{\varphi}$ is the direction of a small radiator-pair in the domain. For a given elementary radiator from the domain, the detector is in the angle $\vartheta$. $r(\tilde{\varphi})$ is the contour of the domain. (b): Geometric cartoon of a HIC in the transverse plane -- the two nuclei are indicated by the two circles, which are shifted by the impact parameter $b$ compared to each other. The intersection of the two circles (orange) has equal area with the ellipse (meshed), which is the effective source-size of the radiation.}
\end{figure}

At this point, the physical interpretation of the form factor $F$ is due. Since we wish to keep $F$ momentum independent, a simple geometric cartoon of a HIC can be suggested. We perform a geometrical averaging over an ordered ensemble of radiator pairs. This ensemble is described by the profile function $r(\tilde{\varphi})$, where $\tilde{\varphi}$ is the polar angle measured around its center, cf. Fig.~\ref{fig:geomAvr}. The radiation of an elementary dipole-like radiator at $\tilde{\varphi}$ reaches the detector from a slightly different direction $\vartheta(\tilde{\varphi})$. Now, utilizing the fact that the characteristic size of the domain of the radiator-ensemble is much smaller than its distance $D$ from the detector, ${r\ll D}$, it is also approximately true, that ${\pi-\tilde{\varphi} +\vartheta\approx \tilde{\varphi}-\varphi}$. This can be seen on Fig.~\ref{fig:geomAvr}, as the triangle, which consists of a given elementary radiator at $\tilde{\varphi}$, the center and the detector, is approximately isosceles. \par
In the light of the above mentioned simplifications, the geometrical averaging after Fourier-expansion of the yield in Eq.~(\ref{dipoleyield}) is straightforward:\LNNL
\begin{align}
\overline{Y} &= \intlim{\tilde{\varphi}}{0}{2\pi}\underbrace{Y_0\left(1+2v_2\cos(2\vartheta(\tilde{\varphi}))\right)}_{=Y}\frac{r^2(\tilde{\varphi})}{2} =\underbrace{Y_0\intlim{\tilde{\varphi}}{0}{2\pi}\frac{r^2(\tilde{\varphi})}{2}}_{=:Y_0T=:\overline{Y}_0} +\nonumber \\
& +2Y_0v_2\cos(2\varphi)\underbrace{\intlim{\tilde{\varphi}}{0}{2\pi}\frac{r^2(\tilde{\varphi})}{2}\cos(4\tilde{\varphi})}_{=:TF} +2Y_0v_2\sin(2\varphi)\underbrace{\intlim{\tilde{\varphi}}{0}{2\pi}\frac{r^2(\tilde{\varphi})}{2}\sin(4\tilde{\varphi})}_{=0} =\nonumber \\
&= \overline{Y}_0\left(1+ 2Fv_2\cos(2\varphi)\right) =:\overline{Y}_0\left(1+ 2\overline{v}_2\cos(2\varphi)\right).
\end{align}
The resulted ${\overline{v}_2=Fv_2}$ is exactly what our goal was: keeping the original form of the elliptic flow coefficient multiplied by a momentum-independent factor. We take the simplest shape with non-zero $F$, an ellipse with half-axes $A$ and $B$:\LNNL
\begin{align}
F &= \frac{\intlim{\tilde{\varphi}}{0}{2\pi}r^2(\tilde{\varphi})\cos(4\tilde{\varphi})}{\intlim{\tilde{\varphi}}{0}{2\pi}r^2(\tilde{\varphi})} =\left(\frac{A-B}{A+B}\right)^2.
\end{align}
Let us consider the two colliding nuclei as circular disks (squeezed due to Lorentz contraction) with radius $R$, displaced by impact parameter $b$ between the centres. There are several ways to attach an ellipse to the geometry of the collision. We intend to match one to the almond-shaped intersection of the two nuclei, with the requirement of equal area. This area can be expressed as\LNNL
\begin{align}
T_\text{int} &= 2R^2\text{arccos}\frac{b}{2R} -bR\sqrt{1-\frac{b^2}{4R^2}} \approx \pi R^2 -2Rb +\mathcal{O}(b^3) 
\end{align}
An ellipse with half-axes ${A=R+\sqrt{\frac{2Rb}{\pi}}}$ and ${B=R-\sqrt{\frac{2Rb}{\pi}}}$ has equal area to $T_\text{int}$ in the leading order of $b$. This approximation gives the maximal impact parameter value ${b_\text{max}=\frac{\pi}{2}R}$, when $B$ approaches zero. Thus, the geometric form factor has the following $b$-dependence:\LNNL
\begin{align}
F(b) &= \frac{2b}{\pi R}.
\end{align}
In fact, experimentally $F(b)$ turns out to be linear in a wide range of impact parameter values: ${F(b)\approx \alpha b \approx \frac{0.15b}{\text{fm}}}$, see Fig. \ref{fig:formfactor}. Comparing the numerical values, we get ${R=\frac{2}{\pi\alpha}\approx 4.24\text{fm}}$. This is an effective size of the source of the dipole-like radiation\footnote{We should keep in mind, that our approximation is meaningful for ${F\leq 1}$ only, i.e. for ${b\precapprox\frac{1}{\alpha}\approx 6.67\text{fm}.}$}. It is smaller than the typical size of a Pb-nucleus by a factor of 1.5. This finding warns against a collective source extending in the whole media, but does not exclude hydrodynamic evolution.

\section{Conclusions and outlook}
It seems that dipole-like structures coupled to the initial geometric asymmetry of heavy-ion collisions are quite natural in a wide scale of models concerning the early time evolution of the hot nuclear matter. We suggest that these domains could be the sources of intense photon and/or gluon radiation having similar geometric properties to its EM counterparts. The orientation of these dipoles may be ordered by EM effects like the mentioned squeezing of the QCD plasma and CME, triggered by the early-time intense fields present in non-central events. Therefore, the cumulative effect of small but not necessarily coherent radiators may affect the macroscopic observables, contributing significantly to the azimuthally asymmetric component of the flow. It is indeed convincing, that such an initial-state effect could be important besides the ones caused by the collective motion. \par
An other important aspect of the issue of the azimuthal asymmetries is in what extent are those evolved on the microscopic level of the dynamics or caused by collective behaviour. There is an ongoing debate in the literature about the contributions of initial and final state asymmetries to the elliptic flow \cite{Goloviznin,lappi,schenke,rybicki}. Emphasizing only a few examples, it is observed, that in proton-nucleon collisions the CGC-correlations could be directly visible in the measurable particle spectrum. Using classical Yang-Mills simulations for p-Pb collisions, in the first half fm/c after CGC was initiated, significant build-up of contributions to $v_2$ and $v_3$ was observed \cite{lappi}. These momentum space anisotropies are not correlated with the final state global asymmetries described as collective flow behaviour. In Pb-Pb collisions, the early-time contributions are relatively small, supporting the role of collective effects. In this case the sources are uncorrelated, localized color field domains, resulting the gluon spectrum to be isotropic \cite{schenke}. EM effects also could play a role in the final state. The directed flow of charged pions could be a result of a spectator induced splitting, as it is demonstrated in Refs.~\cite{rybicki,rybicki2}. \par
Concluding this chapter, we emphasize the necessity of exploring how microscopic causes can lead to macroscopic anisotropies. Especially in large systems, where collective (flow-like) effects may take place, it is important to distinguish those from the amplified sum of individual subhadronic causes.\par
A few comments are in order on how the investigation we presented here could be extended and continued in the future. It would be necessary to investigate -- beyond numerical simulations --, what is the microscopic mechanism that makes the dipol-like sources of radiation ordered. This radiation patterns emerge from the underlying field theoretical description of the gluonic matter, the so-called color glass condensate (CGC). Further investigations are needed to understand the dynamics of this structures. It would be interesting to find out, whether it is possible to describe the phenomenology of structures with higher asymmetry contributing to $v_3$, $v_4$, etc.\! as well.\par
A different question is how the intense electromagnetic (EM) fields -- which can be present in a non-central collision -- influence the radiation patterns. It is possible, that there is an EM-CGC coupling through the EM-charged quark sources of the gluon field. It would be interesting to quantify the strength of such an effect.
\clearpage
\chapter[Effective field theory for transport coefficient estimation]{Effective field theory for\\ transport coefficient estimation}\label{etaOs}

In this chapter, we establish an effective field theory framework to tackle both the thermodynamic and transport properties of a medium without any conserved charges -- besides the energy-momentum. We demonstrate, how the generalization of the quasi-particle picture leads to a non-trivial, yet simply tractable description of fluids with a broad range of transport properties, i.e.\! small and also large viscosities.\par
Thermodynamic and transport properties of physical systems give a basis for the comparison of the theoretical predictions to the physical reality. Despite the diversity of models, concepts like conductivity, viscosity, densities of energy and entropy etc.\ allow us to phenomenologically access a wider range of physical systems from cold atomic gases through fluids at room temperature to the hot and dense matter created in heavy-ion collisions. These macroscopic observables are, however, in a very complicated relationship with the microscopic quantities (i.e.\ the fundamental degrees of freedom) of a given theory. There are numerous examples in the literature illustrating this elaborate issue, see for example Refs.~\cite{thermo_gluondamping, thermo_gluonplasma, thermo_quasipart0, thermo_quasipart1, thermo_quasipart2, thermo_quasipart3, thermo_selfcons1, thermo_selfcons2, thermo_QCDlattice} for the analysis concerning thermodynamical quantities. Furthermore see Refs.~\cite{Jeon, transport_YMfrg, transport_NJL3, transport_2PI1, transport_largeN1, transport_largeN2, transport_eff2, transport_SU3lattice, transport_2PI2, transport_eff1} for transport coefficients obtained from quantum field theory (QFT), functional renormalization group (FRG) or lattice calculations, and see Refs.~\cite{transport_QMD, transport_NJL4, transport_eff3, transport_eff4, transport_eff5} for kinetic theory or quasi-particle (QP) approaches. Our goal is to parametrize the system in the language of microscopic quantities, which yet can be phenomenologically meaningful, possible measurable ones. Then we try to link those quantities to the macroscopic physics as straightforward as it is possible.\par
\begin{figure}
\centering
\subfloat[]{
\includegraphics[width=0.38\linewidth]{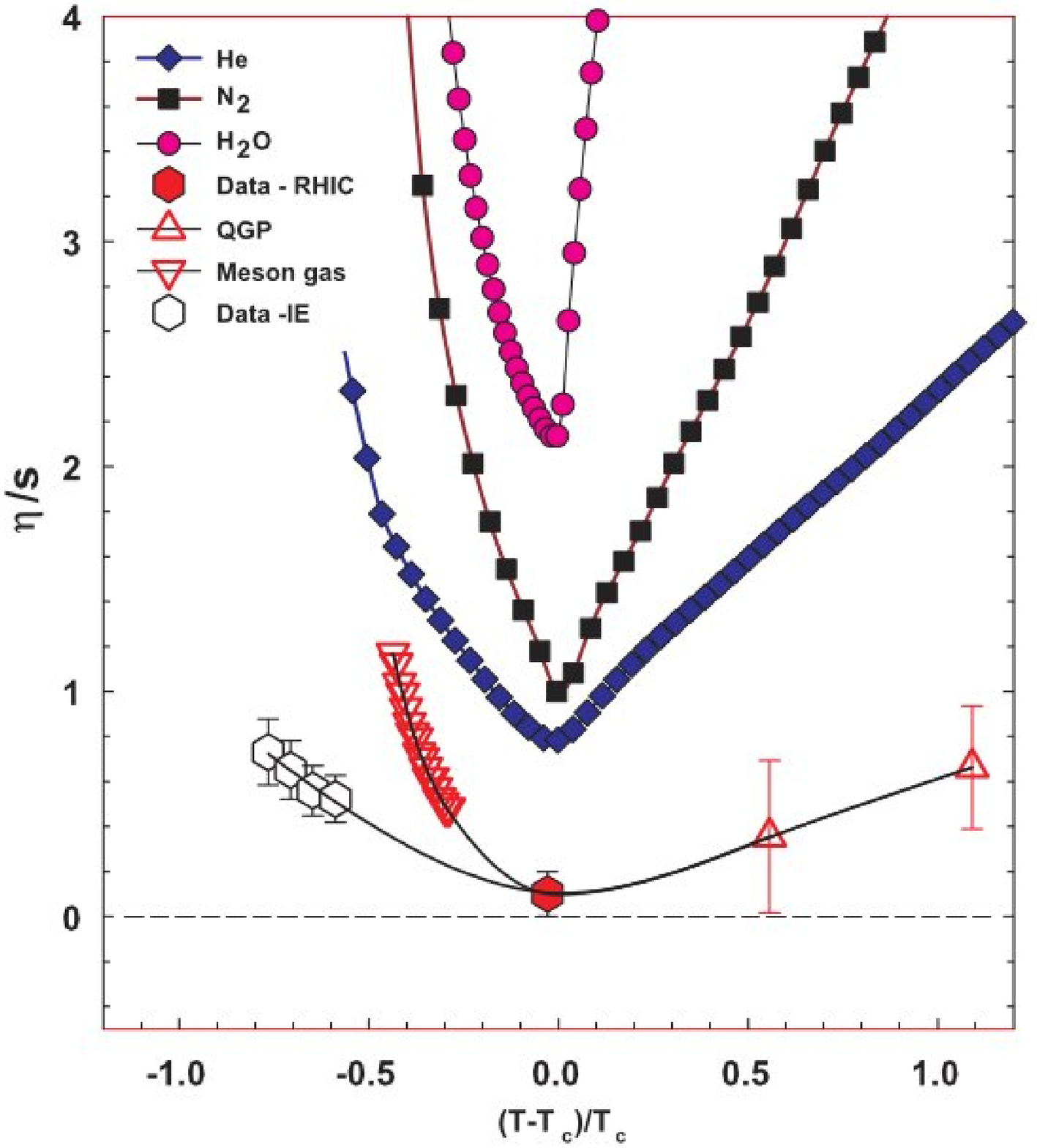}
}
\subfloat[]{
\includegraphics[width=0.42\linewidth]{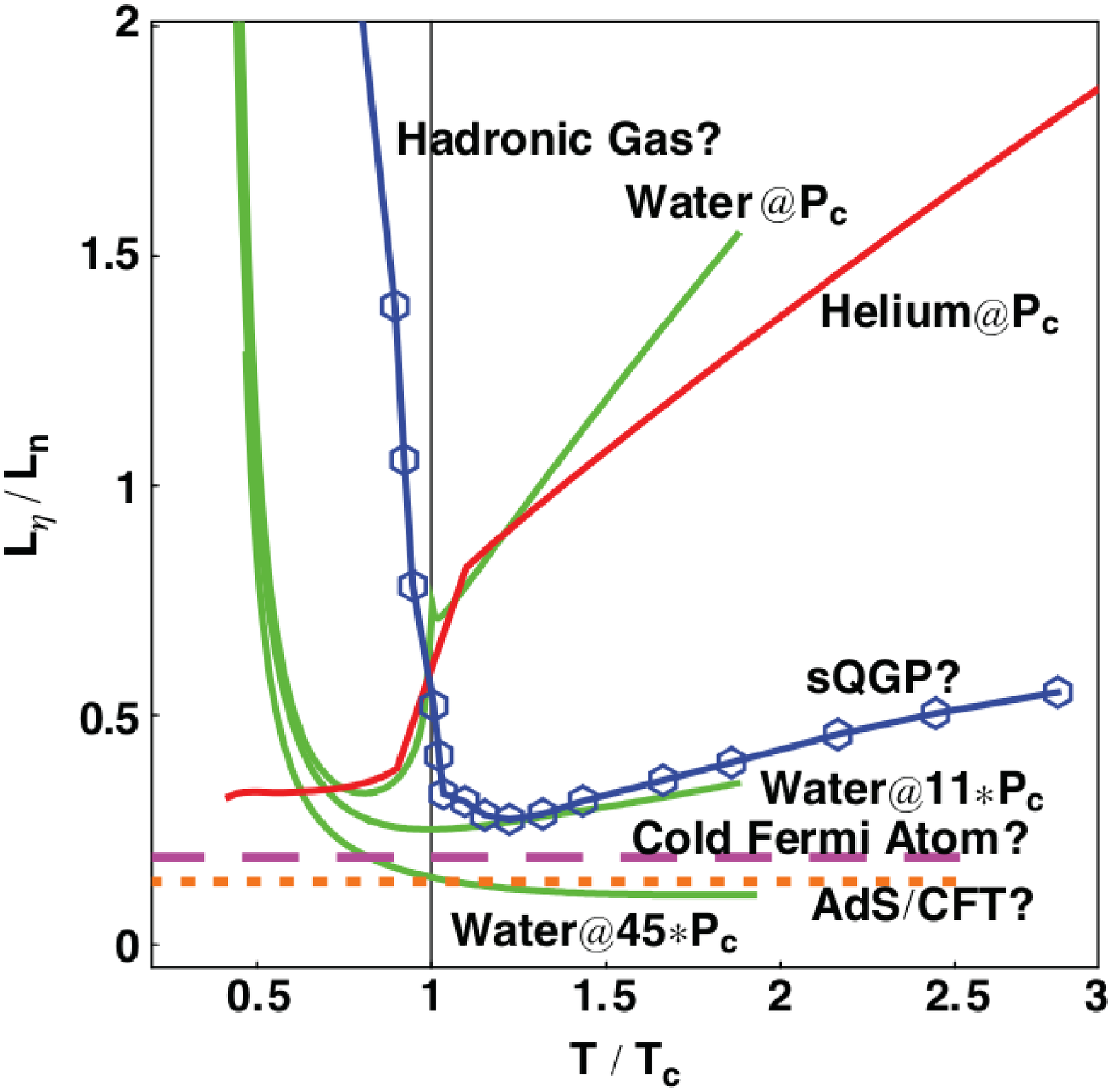}
}
\LNFIG
\caption{(a): Fig.\! is taken from Ref.~\cite{transport_exp1}. It shows the characteristic behaviour of the fluidity measure $\eta/s$ for several material near the critical temperature. \\(b): Fig.\! is taken from Ref.~\cite{transport_exp2}. Another fluidity measure, but the characteristic behaviour -- i.e.\! the minima near $T_c$ -- remains. The definition $~{\frac{L_\eta}{L_n}=\frac{n^{1/3}\eta}{c_s\rho}}$ measures the ratio of the minimal wavelength of a sound wave is needed to propagate, versus an inner length scale, derived from the density: $n^{-1/3}$. The mass density $\rho$ and the sound velocity $c_s$ ensures that the ratio is dimensionless.}
\label{fig:fluidityMesEtaOs}
\end{figure}

Interestingly enough, the ratio of the shear viscosity $\eta$ to the entropy density $s$ has qualitatively the same temperature dependence in several systems, showing in general a fluid-like behaviour. Near to the critical endpoint of the liquid-gas phase-transition, the fluidity measure $\eta/s$ achieves its minimal value \cite{transport_fermigas, transport_graphene, transport_exp1}, indicating that these materials are most fluent near to their critical state. The presence of increased fluidity is also supported by the analysis of other quantities aimed to measure the fluidity of the medium \cite{transport_exp2}, see Fig.~\ref{fig:fluidityMesEtaOs}. \par
The structure of this chapter is the following. After a short introduction on the spectral density of states in Section~\ref{spectFuncEtaOs}, we summarize first the concept of quasi-particles and its limitations in effective modelling in Section \ref{exQP}. We introduce thermodynamic notions through the energy-momentum tensor in Section \ref{thermo}. The issue of thermodynamic consistency is briefly discussed. In Section \ref{linRes}. the transport coefficients in linear response are elaborated using Kubo's formula. After a short discussion on the lower bound of the ratio $\eta/s$ in the extended quasi-particle picture in Section \ref{lowerB}., we turn to analyse physically motivated examples in Section \ref{examples}. \par
Most of the results presented here have been published in Ref.~\cite{etaOspaper}.

\section{The spectral density of states}\label{spectFuncEtaOs}
Simply put, we are interested in what makes matter more fluent. Our goal here is to analyse the transport coefficient $\eta$ and the thermodynamic quantities in the framework of an effective field theory. We are aiming to quantify how the robust properties of microscopically meaningful quantities relate to the qualitative behaviour of macroscopic observables. We are going to use the spectral density of states or spectral function for this purpose, as it is meaningful even on the level of the fundamental theory. The spectral function $\rho_{x,y}$ is the response of the theory at the space-time point $y$ to a small, local perturbation occurred at $x$. In the momentum space, it characterizes the density of the quantum states in the energy $\omega$ if all other quantum numbers (including the momentum $\ve{p}$) kept fixed. Roughly speaking, $\rho_{\omega,\ve{p}}\mathrm{d}\omega$ quantifies the probability of the creation of an excitation with momentum $\ve{p}$ and energy within the interval ${[\omega,\omega+\mathrm{d}\omega]}$. Since it can be seen as a response function, when one is intended to ''scan'' the system's reaction -- for example in scattering experiments -- a spectral function is the result\footnote{Note, that realistically this is a spectral function of a composite operator. However, since our discussion is about effective modelling, it is not a surprise, that our effective DoF are represented as a composite operator on the level of the fundamental theory.}.\par
\begin{figure}
\centering
\includegraphics[width=0.75\linewidth]{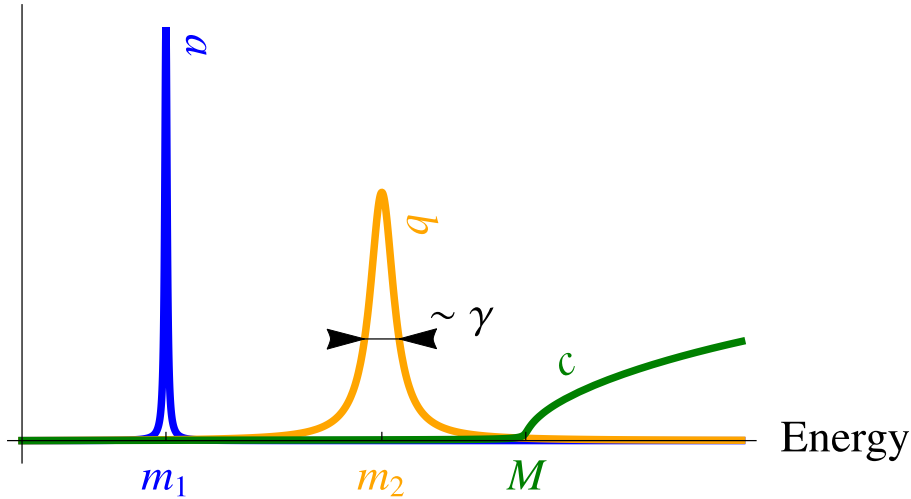}
\LNFIG
\caption{Robust features of a generic spectral function (color online). \textit{QP-behavior}: practically infinitely long lifetime (blue, $a$), \textit{Broad peak} with lifetime $\sim 1/\gamma$ (orange, $b$), \textit{Continuum of multi-particle states} with threshold $M$ (green, $c$)}
\label{fig:spectralfunctions}
\end{figure}

A physically important characterization of $\rho_{\omega,\ve{p}}$ is whether it has a narrow-peak structure or not (see Fig.~\ref{fig:spectralfunctions}). If so, the behaviour of the system is dominated by (quasi-)particles, with inverse lifetime proportional to the half-width of the peak. The dispersion relation $\omega(\ve{p})$ is determined by the position of the peak. Kinetic description and perturbation theory work usually well in this case. On the other hand for wide peak(s) or in the presence of a relevant continuum contribution, the situation is more intricate. The continuum contribution to $\rho_{\omega,\ve{p}}$ signals, that multi-particle states are significant. Such spectra are resulted by non-perturbative methods, for example the resummation of the infrared (IR) contributions of the perturbation theory \cite{BNresum1, BNresum2, BNresum3} or FRG calculations. Typically, the phenomenology of such systems cannot be described in terms of conventional quasi-particles with long lifetime. \par

\section{Extended quasi-particles}\label{exQP}
We focus on the transition between hydrodynamical and kinetic regimes. For this purpose, generalization of the notions of the QP-description is needed. From phenomenological point of view, quasi-particles are objects with infinite (or with very long) lifetime, usually well-localized in space. Resonances and other short-living yet particle-like entities are also often referred to as quasi-particles, confusingly.\par
From the side of QFT, particles are the asymptotic states of the theory in question. This definition, however, does not cover finite lifetime particle-like intermediate states often appearing in particle physics experiments. In effective modelling, one possibility is to associate a new field degree of freedom to every observed particle-like object. But non-physical symmetries could be generated via this resonance--field correspondence, it is not obvious how to avoid the double-counting of thermodynamic degrees of freedom \cite{SpectrFuncThermo}. \par
Finite lifetime bound-states and resonances are more natural to appear via interaction among some elementary fields. It is very unlikely though to guess the underlying fundamental structures when constructing an effective theory, due to the lack of basic understanding -- the reason we needed effective description in the first place. When the width of the peaky structures in the spectral function\footnote{The width can be interpreted as the inverse life-time of a QP excitation.} is large, we must not rely on perturbative treatment any more: in this case an effective field theory approach may help to re-define the fundamental structures. \par
Let us consider a scalar (spin-0) operator $\varphi$ bearing all the physical degrees of freedom we are interested in. We call $\varphi$ an \textit{extended quasi-particle (EQP)} if its equation of motion is linear in $\varphi$. An equivalent statement is that the action is a quadratic functional of $\varphi$: \LNNL
\begin{align}
S[\varphi] &=\frac{1}{2}\int_x\int_y\varphi_x\mathcal{K}_{x-y}\varphi_y, \label{EQPactionEtaOs}
\end{align}
and therefore the equation of motion (EoM) reads as \LNNL
\begin{align}
\int_y\mathcal{K}_{x-y}\varphi_y &=0. \label{EQPeomEtaOs}
\end{align}
If so, all correlation functions are determined by the single two-point function ${\rho_x=\avr{[\varphi_x,\varphi_0]}}$ -- which we call the \textit{spectral function} from now on --, through Wick's theorem and causality. \par
In other words, we use wave-packet-like modes instead of plane waves. The idea of using a suitable basis of quantization, chosen to the actual problem, is widely used e.g.\ in solid state physics: like Cooper-pairs in superconductivity or atomic orbits and Wannier-functions in the description of crystals \cite{wannier1}. Choosing the appropriate degrees of freedom, the theory of strongly interacting elementary objects can become a weakly (or non-) interacting theory of composite ones. \par
The action (\ref{EQPactionEtaOs}) of the EQP-description is non-local in the sense, that the two field operators are inserted in different space-time points. There are several known examples in the literature for theories with non-local quadratic action, including pure gauge theories, low-energy effective theories of particles etc., see for example Refs.~\cite{nonlocal_scalar1, nonlocal_scalar2, nonlocal_resummation, nonlocal_gauge1, nonlocal_gauge2, nonlocal_condmat1}. \par
We view the non-local EQP-action as the leading order (or the relevant part) of an IR-resummed theory. Our goal is to describe the physics near to a critical point, where a second order phase transition occurs. We assume that the relevant field operator remains unchanged, but, since the long-range correlations may also play a major role, we allow the appearance of derivative terms to arbitrary order. These are the physical criteria cumulated in the non-local quadratic action. The quasi-particle nature is reflected in the linearity of the EoM (\ref{EQPeomEtaOs}) , i.e.\ any linear combination of solutions also satisfies the field equation.\par
We stress here the main advantage of the quadratic nature of the description: the integrability. This allows us to calculate thermodynamic observables using two-point functions. Also in the case of transport coefficients, where higher correlators are needed, the knowledge of two-point functions is sufficient for the linear response calculations, since $\avr{\varphi\varphi\varphi\varphi} \sim \sum\avr{\varphi\varphi}\avr{\varphi\varphi}$, where the summation runs over all the possible pairings of the field operators $\varphi$.

\section{Equilibrium thermodynamics with EQP}\label{thermo}
Thermodynamic quantities (if no conserved charges are present) can be originated from either the energy density $\varepsilon$ or from the free energy density $f$. In both cases the averaging is performed over spatially translational invariant field configurations. Despite the lack of a well-defined canonical formalism, in non-local theories, $e^{-\beta T^{00}}$ serves as the usual Boltzmann statistical operator. With the time-evolution operator $e^{itT^{00}}$ the Kubo-Martin-Schwinger-relation (KMS) still holds. The field theoretical calculations of correlation functions in thermal equilibrium and also the response functions in the linear response approximation (this will come later in Sec.~\ref{linRes}) can be carried out conveniently in a common framework, if one treats the theory in the so-called Keldysh-formalism. We do not wish to break, however, the continuity of the discussion which focuses on the phenomenological aspects mainly. Therefore, we recommend the interested reader to study Appendix~\ref{propagators}, \ref{enmomTensor} and Ref.~\cite{SpectrFuncThermo} for further details and a basic introduction into the formalism. The line of thought of this chapter is intended to be comprehensible without referring to those formal details. We recall the important results independent from this discussion, whenever it is needed.\par 
Due to the quadratic form of the action (\ref{EQPactionEtaOs}), thermodynamic quantities can be expressed\footnote{We subtract the vacuum contribution to the thermodynamic quantities, so that $\varepsilon(T=0)=0$ and $P(T=0)=0$.} using the spectral function $\rho(\omega, |\ve{p}|)$ and the Fourier-transformed kernel $K(\omega, |\ve{p}|)$, see Ref.~\cite{SpectrFuncThermo} or App.~\ref{enmomTensor} for a detailed computation:
\begin{eqnarray}
f &=& -\intpos{p}n(\omega/T)\intlim{\tilde{\omega}}{}{\omega}\frac{\partial K(\tilde{\omega},|\ve{p}|)}{\partial\tilde{\omega}}\rho(\tilde{\omega},|\ve{p}|) =-P,\label{pressure0} \\
\varepsilon &=& \intpos{p}\omega\frac{\partial K(\omega,|\ve{p}|)}{\partial\omega}\rho(\omega,|\ve{p}|) n(\omega/T). \label{energy0}
\end{eqnarray}
Here we used the notation $\intpos{p}\equiv\int\!\!\frac{\mathrm{d}^3\ve{p}}{(2\pi)^3}\int\limits_{0}^{\infty}\!\!\frac{\mathrm{d}\omega}{2\pi}$ for phase-space integration with respect to the four-momentum $p=(\omega,\ve{p})$, restricted to positive frequencies. Note, that $\rho$ and $K$ are not independent: $\rho=-2\mathrm{Im}G(\omega+i0^+,\ve{p})$ and $K(\omega,\ve{p})=\mathrm{Re} G^{-1}(\omega+i0^+,\ve{p})$ with $G(\omega,\ve{p})=\intlim{\tilde{\omega}}{-\infty}{\infty}\frac{\rho(\tilde{\omega},\ve{p})}{\omega-\tilde{\omega}}$.

\subsection{Thermodynamic consistency}\label{thermoConsist}
We wish to also include systems with temperature-dependent parameters into the EQP description. The consistency of Eq.~(\ref{pressure0}) and (\ref{energy0}) is fulfilled, however, only if $\rho$ and $K$ are temperature-independent. We mean by consistency, that the relations $s=\frac{\partial P}{\partial T}$ and $sT=\varepsilon+P$ hold, and therefore $\varepsilon=T^2\frac{\partial (P/T)}{\partial T}$, as a consequence. \par
To overcome this issue and also keeping the simplicity of the EQP-picture, we let $\phi:=\avr{\varphi}$ be non-zero, homogeneous and temperature-dependent. This is equivalent with a non-trivial, temperature-dependent ''bag constant'' -- for a similar argument, see Ref.~\cite{thermo_quasipart0}. The correlators are shifted, thus $\varepsilon =\varepsilon_{\phi\equiv 0} +B$, $P=P_{\phi\equiv 0}-B$, with the temperature-dependent quantity $B$ (referring to the ''background''). This procedure leaves the entropy formula $s=\frac{\varepsilon+P}{T}$ unchanged (for further details see Appendix \ref{sourceTerm}). That is, the thermodynamic consistency is fulfilled using the same entropy formula, with temperature-dependent spectral function ${~\rho(\omega,\ve{p},\{m_i(T)\})}$. The $T$-dependence of the background field is not arbitrary, its effect precisely cancels the extra terms coming from the temperature-dependent parameters $m_i$: \LNNL
\begin{align}
&\frac{\partial B}{\partial T}=\sum_i\frac{\partial m_i}{\partial T}\frac{\partial P_{\phi\equiv 0}(T,\{m_i(T)\})}{\partial m_i}.
\end{align}

\subsection{Microcausality}\label{mCausality}
Microcausality (or also often referred as locality) means that there is no correlation between two space-time points separated by a space-like interval. Since in our description all measurable quantities can be expressed by the spectral function, $\rho(x-y)\equiv 0$ is required for spatially separated space-time points $x$ and $y$. \par
In case of self-consistent approaches or perturbative calculations, microcausality is guaranteed by construction, as it originates from the non-interacting theory and the space-time-local interaction vertices. In effective theories, this is not necessarily true. In order to guarantee microcausality, we choose the Fourier-transform of $\rho$ as $${\rho(\omega,\ve{p})=\theta(\omega^2-\ve{p}^2)\mathrm{sign}(\omega)\overline{\rho}(\omega^2-\ve{p}^2)},$$ which simplifies Eq.~(\ref{pressure0}), (\ref{energy0}):
\begin{eqnarray}
P &=&\intlim{p}{0}{\infty}\frac{\partial K}{\partial p}\overline{\rho}(p)T^4 \chi_P(p/T), \label{pressure1} \\
\varepsilon &=&\intlim{p}{0}{\infty}\frac{\partial K}{\partial p}\overline{\rho}(p)T^4 \chi_\varepsilon(p/T), \label{energy1}
\end{eqnarray}
with the notation $p^2=\omega^2-\ve{p}^2$. The thermodynamic weight-functions are:
\begin{eqnarray}
\chi_P(x)&=& \frac{x^3}{4\pi^3}\intlim{y}{1}{\infty}y\sqrt{y^2-1}\cdot n(xy) \approx \frac{x^2}{4\pi^3}K_2(x), \\
\chi_\varepsilon(x)&=& \frac{x^4}{4\pi^3}\intlim{y}{1}{\infty}y^2\sqrt{y^2-1}\cdot n(xy) \approx \nonumber \\
&\approx & \frac{x^3}{4\pi^3}K_1(x)+\frac{3x^2}{4\pi^3}K_2(x),
\end{eqnarray}
where $n$ is the Bose--Einstein distribution, $~{K_1,\, K_2,\, \dots}$ are modified Bessel functions appearing in the limit of the Boltzmannian approximation, when $n(x)\approx e^{-x}$. $T^4\chi_P(x)$ and $T^4\chi_\varepsilon(x)$ respectively are the densities of pressure and energy of an ideal gas, with temperature $T$ and particle mass $xT$. It is apparent, that the combination $\frac{\partial K(p)}{\partial p}\overline{\rho}(p)$ acts as a mass-distribution (i.e.\ normalizable\footnote{$\int_0^\infty\!\!\mathrm{d}p\frac{\partial K}{\partial p}\overline{\rho}(p)$ has to be finite, otherwise the high-temperature Stefan-Boltzmann-limit of the thermodynamics does not exist.}), therefore our quasi-particle description of the thermal observables can be interpreted as a mass-distributed ideal gas \cite{thermo_quasipart1, thermo_quasipart2}.\par
Note here, that the temperature-dependence of $\rho$ can break the manifest Lorentz-covariance through the temperature-dependent parameters, which are thought to be measured in the frame assigned to the heat bath.

\section{Shear viscosity in linear response}\label{linRes}
Hydrodynamics describes the collective motion of fluids with given material properties, based on the analysis of the energy-momentum conservation during the motion. The relaxation time of the system after a macroscopic perturbation is measured by the hydrodynamic transport coefficients. In case of a given transverse wave with wavenumber $\ve{k}$ perpendicular to the local flow velocity $\ve{v}$, its relaxation to the equilibrium configuration is controlled by $\eta/s$. Expressed with the energy-momentum tensor: $~{\pi_\perp^{\mu\nu}=\pi_{\perp,0}^{\mu\nu}+\delta \pi_\perp^{\mu\nu}}$, the fluctuation part decays as $~{\delta \pi_\perp^{\mu\nu}(t)=e^{-\frac{\eta}{s}\frac{\ve{k}^2t}{T}}\delta \pi_\perp^{\mu\nu}(0)}$, where $T$ is the local temperature, see Ref.~\cite{hydroLectures} for further details. \par
Transport coefficients can also be interpreted from the kinetic theory point of view. The shear viscosity $\eta$ is the diffusion coefficient of momentum transfer perpendicular to the local velocity of the fluid. In case of a gas of particles $\eta \sim v\lambda\rho$ with $v$ being the root mean square particle velocity, $\rho$ is its mass density and $\lambda$ is the mean free path of gas particles (see for example Sec. 3.4 of~\cite{liboff}). Typically speaking, $\eta$ is large (compared to some internal scale) in gases (or in fluids where kinetic description is acceptable) compared to ordinary liquids. From the kinetic point of view, it means that the mean free path is significantly smaller in liquids. \par
A possible way to connect these two regimes is to define the transport coefficients in the linear response approximation. This allows us to go beyond the quasi-particle picture used in the kinetic theory and discuss the transport properties translated to those of a continuous medium, represented by its energy-momentum tensor.

\subsection{Linear response in relativistic hydrodynamics}\label{linResHydro}
Let us take a small perturbation in the action: ${~\delta S=\int_x h_xA_x}$, where $A$ is a measurable quantity (Hermitian operator) and $h$ is a scalar function. Our goal is to express the change in the expectation value of $B$ up to first-order in $h$ as ${\delta\avr{B_x}=\int_yi\mathcal{G}^{ra}_{BA}(x-y)h_y}$. Kubo's formula characterizes the response function, supposing the system relaxes to thermal equilibrium in which it was before the perturbation occurred: $${i\mathcal{G}^{ra}_{BA}(x-y)=\theta_{x^0-y^0}\avr{[B_x,A_y]}=\theta_{x^0-y^0}\rho_{BA}(x-y)},$$ where $\avr{.}$ refers to averaging over configurations in thermal equilibrium. \par
Now we summarize the results regarding the linear response of the energy-momentum tensor. The response to a long-wavelength perturbation in the rest frame of the fluid defines the hydrodynamic transport coefficients. Following the derivation of Ref.~\cite{linres_hydro}, we add an extra term to the action, depending on the flow velocity $u$ explicitly\footnote{Let us write $~{\delta S=\int_{-\infty}^t\delta H}$. The perturbation of the Hamiltonian, $\delta H$, also can be thought of as an additional source term to the grand-canonical density operator, when the thermal averaging is performed: $~{\sim e^{-\beta(T^{00}-\delta H)}}$. In this case, $\delta c$ causes heat current, while $u$ is the source of momentum-current.}:\LNNL
\rem{
\FINAL{
\begin{align}
\delta S &= \intlim{^3\ve{x}}{}{} (U_\mu-\overline{u}_\mu) T^{0\mu}.
\end{align}}
}
\rem{
In the above formula $~{\overline{u}=(1,0,0,0)}$ is the velocity in the rest frame of the fluid, while $U$ denotes the combined source of both flow and heat perturbations due to the inhomogeneous temperature field. We denote the total source as \FINAL{$~{U=(1+\delta c)u}$}. The energy-perturbation $\delta{c}$ can be expressed as $\delta c=\frac{\partial\ln \varepsilon}{\partial T}\delta T$. Taking the time-derivative of $\delta S$, then using the energy-momentum conservation $~{\partial_\nu T^{\mu\nu}=0}$ one arrives at
\begin{align}
\partial_0\delta S &= \intlim{^3\ve{x}}{}{} [\left(\partial_0 U_\mu\right)T^{0\mu}+(U_\mu-\overline{u}_\mu)\underbrace{\partial_0 T^{0\mu}}_{=-\partial_i T^{i\mu}}], \\
\delta S&=\intlim{\tau}{-\infty}{t}\intlim{^3\ve{x}}{}{} T^{\mu\nu}\partial_{[\mu}U_{\nu]}. \label{hydro_deltaS}
\end{align}}
We also introduced the notation ${\partial_{[\mu}U_{\nu]}=(\partial_\mu U_\nu +\partial_\nu U_\mu)/2}$ for the symmetrized derivative. \par
Now, plugging Eq.~(\ref{hydro_deltaS}) into Kubo's formula, after tensorial decomposition of the energy-momentum tensor the shear viscosity coefficient can be identified as:\LNNL
\begin{align}
\delta\avr{T^{ij}} &\overset{\text{rest frame}}{=} \underbrace{\lim\limits_{\omega\rightarrow 0}\frac{\frac{1}{5}\avr{[T^{lm},T_{lm}]}(\omega,\ve{k}=0)}{\omega}}_{=:\,\eta}\left(\partial^{[i}u^{j]}-\frac{1}{3}\delta^{ij}\partial_ku^k\right). \label{hydro_Kubo}
\end{align}
The detailed calculation of $\eta$ -- and of other transport coefficients also -- can be found in Appendix \ref{linresAppEtaOs}.

\subsection{Shear viscosity with EQP}
We intend to get the transport coefficients using a field theory framework. In case of the shear viscosity we are interested in the linear response to a small perturbation in the energy-momentum tensor $T^{\mu\nu}$. For that response function we need the spectral function $\rho_{TT}$, which we give in Appendix \ref{visco} in detail, since this gives the shear viscosity in the limit of long-wavelength, as we have seen in Eq.~(\ref{hydro_Kubo}). With EQP, one gets the following expression for $\eta$:\LNNL
\begin{align}
\eta &= \lim\limits_{\omega\rightarrow 0}\frac{\rho_{(T^\dagger)^{12}T^{12}}(\omega,\ve{k}=0)}{\omega} = \\
&=\intpos{p}\left(\frac{p^1p^2}{\omega}\frac{\partial K(\omega,|\ve{p}|)}{\partial\omega}\rho(\omega,|\ve{p}|)\right)^2\left(-\frac{\partial n(\omega/T)}{\partial\omega}\right). \label{eta0}
\end{align}
This particularly simple form of $\eta$ is a result of the quadratic nature of the EQP-description. There are, however, several examples for calculations done in interacting theories resulting formulae with similar structure \cite{Jeon, transport_YMfrg, transport_NJL3, transport_2PI1, transport_largeN1, transport_largeN2, transport_eff2}. \par
Contrary to Eqs.~(\ref{pressure0}) and (\ref{energy0}), this result cannot be interpreted simply as the sum of viscosities in a mass-distributed gas-mixture. We will see later, that Eq.~(\ref{eta0}) can cover phenomenology beyond the relaxation time approximation. Furthermore, due to the integrable nature of the EQP-action, it is symmetry-preserving, and there is no need of further operator-improvement (e.g.\ by the resummation of vertex corrections as it would be necessary in the 2PI approximation, see for example Ref.~\cite{Jeon}). This means that the four-point function $\avr{TT}$ is exact, because in the EQP description, any higher correlation functions can be expressed by the two-point ones, i.e.\! by $\rho$. \par
As a matter of thermodynamic consistency, it turns out, that for a homogeneous and temperature-dependent background Eq.~(\ref{eta0}) is unchanged. For the details of the calculations with non-zero background see Appendix \ref{sourceTerm}.

\section{Non-universal lower bound for $\eta/s$}\label{lowerB}
In the previous sections we have derived quite simple expressions for the entropy density and the shear viscosity in Eqs.~(\ref{pressure0}, \ref{energy0}) and (\ref{eta0}). Using dimensionless quantities, the entropy density over $T^3$ reads as\LNNL
\begin{align}\label{entropy1}
\sigma :=\frac{s}{T^3} =\intlim{p}{0}{\infty}g(p,T)\chi_s(p/T),
\end{align}
where ${g(p,T)=\frac{\partial K}{\partial p}\overline{\rho}}$, while the thermodynamic weight is\LNNL
\begin{align}
\chi_s(x) =\chi_{\varepsilon}(x)+\chi_P(x) \approx \frac{x^3}{4\pi^3}K_3(x).
\end{align}
The expression for the shear viscosity contains the very same function $g$:\LNNL
\begin{align}\label{eta1}
\eta =\intlim{p}{0}{\infty}g^2(p,T)T^4\lambda_\eta(p/T),
\end{align}
with the weight function\LNNL
\begin{align}
\lambda_\eta(x) =\frac{1}{4\pi^3}\frac{x^5}{15}\intlim{y}{1}{\infty}(-n'(xy))(y^2-1)^{5/2} \approx \frac{x^2}{4\pi^3}K_3(x).
\end{align}
Now we focus on the fluidity measure $\eta/s$, the relaxation coefficient of transversal hydrodynamical perturbations, as it was mentioned earlier. There is a great interest in theoretical physics, whether a universal lower bound for $\eta/s$ exists. It has been theorized in Ref.~\cite{KSSpaper}, that this lower bound is $\frac{1}{4\pi}$ in certain conformal field theories with holographic dual. Further investigation showed the possibility of violating this universal value of the lower bound even in the framework of the AdS/CFT duality \cite{KSSviolation1, KSSviolation2} and also in effective theories \cite{nonUnivLowerBnd, nonUnivLowerBnd2, nonUnivLowerBnd3, CohensArgument}. Although we do not expect any universal result in the framework of EQP, the question is still valid. In fact, we are able to give an answer within our description. The following variational problem is to be solved:\LNNL
\begin{align}
\frac{\delta}{\delta g}(\eta[g]-\alpha s[g]) =0,
\end{align}
with the ($p$-independent) Lagrange's multiplier $\alpha$. Our strategy is to minimize $\eta$, while the value of $s$ is kept fixed. Since $s$ is a linear functional of $g$ whilst $\eta$ is quadratic, the solution for the minimizing function is\LNNL
\begin{align}
g^*(p,T) = \frac{\alpha}{2T}\frac{\chi_s(p/T)}{\lambda_\eta(p/T)}. \label{gstar}
\end{align}
Keeping the value of $s$ fixed, we are able to compute $\eta^*$, the lowest possible value of the shear viscosity in the EQP description is then given by thermodynamic quantities:\LNNL
\begin{align}
\eta^* =\frac{\alpha^2}{4T^2}\intlim{p}{0}{\infty} \frac{\chi_s^2(p/T)}{\lambda_\eta^2(p/T)}T^4\lambda_\eta(p/T)= \frac{1}{\intlim{y}{0}{\infty}\frac{\chi_s^2(y)}{\lambda_\eta(y)}}\frac{s^2}{T^3}.
\end{align}
Therefore the lower bound for $\eta/s$ is\LNNL
\begin{align}
\frac{\eta}{s} \geq \frac{\eta^*}{s}= \frac{1}{\intlim{y}{0}{\infty}\frac{\chi_s^2(y)}{\lambda_\eta(y)}} \sigma =:\frac{\sigma}{I}\approx \frac{1}{48}\sigma.
\end{align}
A speciality of this minimal-$\eta/s$ system is that all the thermodynamic and transport quantities are controlled by $\eta/s=\frac{\sigma}{I}$. The two types of averages we considered in this chapter are proportional to $\sigma$ or $\sigma^2$, despite a constant tensorial factor:
\begin{eqnarray}
\frac{\avr{T^{\mu\nu}}}{T^4}\sim\sigma, & \mathrm{ and } & \frac{\eta}{T^3},\,\frac{\zeta}{T^3},\,\frac{\kappa}{T^3}\sim \sigma^2, \nonumber
\end{eqnarray}
$\zeta$ and $\kappa$ being the bulk viscosity and the heat conductivity, respectively -- see Appendix \ref{linresAppEtaOs}.

\subsection{Some properties of the kernel function $g^*$}
It is an interesting question, what kind of spectral functions -- if any -- can realize the kernel in Eq.~(\ref{gstar}), at least in the Boltzmannian approximation:\! ${g^*(p)=\frac{\sigma}{I}\frac{p}{T^2}}$. We do not address to solve the inverse problem ${g^* \rightarrow \overline{\rho}^*}$ here. Instead, we aim to reproduce $g^*$ as the asymptotic behaviour
of an ansatz, namely\LNNL
\begin{align}
\overline{\rho}(p) &= \frac{\gamma Z(p,\gamma)}{p(p^2+\gamma^2)}= \frac{Z_0(\gamma)\gamma p e^{-\zeta(\alpha\gamma p)}}{p^2+\gamma^2},
\end{align}
where we assume $Z$ to be analytic for any $p$ on the upper-half complex plane\footnote{If $Z(p)$ has poles, we expect that the residue of those behave like $\mathcal{O}(\gamma^{-1})$ for large values of $\gamma$.}. Also, $\zeta$ is non-negative for any real values of its argument and it is ${\zeta(x)\approx |x|+\text{const.}+\mathcal{O}(x^{-1})}$ for $x$ with large absolute value. With these assumptions\footnote{Here we are not going to deal with the question of the existence of such a $Z(p)$ further.}, the kernel function is the following:\LNNL
\begin{align}
g(p) &= K(p)\overline{\rho}(p) =\frac{\gamma\left(2-p\frac{Z'(p)}{Z(p)}\right)}{p^2+\gamma^2} =\frac{\gamma^2\zeta'(\alpha\gamma p)}{p^2+\gamma^2}\alpha p \overset{\gamma\rightarrow\infty}{\longrightarrow} \alpha p =g^*(p),
\end{align}
with $\alpha=\frac{\sigma}{IT^2}$. This result also holds for large $\gamma$, if ${p<\gamma}$, otherwise possible corrections in the order of $\mathcal{O}(\gamma^{-1})$ can enter. Taking the sum rule:\LNNL
\begin{align}
\frac{1}{\pi}\intlim{p}{0}{\infty} \frac{\gamma Z(p)}{p^2+\gamma^2} =\frac{Z_0(\gamma)}{\pi\alpha^3}\intlim{x}{0}{\infty} \frac{x^2e^{-\zeta(x)}}{\frac{x^2}{\alpha^2}+\gamma^4} \overset{Z_0(\gamma)= z_0\gamma^4}{\underset{\gamma\rightarrow\infty}{\longrightarrow}} \frac{z_0}{\pi\alpha^3}\intlim{x}{0}{\infty}x^2e^{-\zeta(x)} \overset{!}{=}1,
\end{align}
therefore the $\gamma$-dependence of $Z_0$ is fixed. \FINAL{Now, we observe the behaviour of the normalized combination $pg^*(p)$ for $p\rightarrow 0$ by scaling $p=c/\gamma$ ($c$ is a positive constant) and also for $p>0$}:\LNNL
\begin{align}
pg^*(p) =\frac{z_0\gamma^5p^2e^{-\zeta(\alpha\gamma p)}}{p^2+\gamma^2} \overset{\gamma\rightarrow\infty}{\longrightarrow}\left\{
\begin{array}{lcl}
\text{for } p\neq \frac{c}{\gamma}, & & 0 \\
\text{for } p=\frac{c}{\gamma}, & & z_0c^2e^{-\zeta(\alpha c)}\gamma \rightarrow\infty
\end{array}
\right.
\end{align}
That is, in the limit ${\gamma\rightarrow\infty}$, when $g^*$ is realized, this ansatz predicts the accumulation of the accessible states at ${p=0}$. Interestingly, the density of states $\overline{\rho}^*$ of this momentum-space condensate is independent of $\alpha$ in the leading order of the large-$\gamma$ expansion.

\section{Examples to liquid--gas crossover}\label{examples}
In this section we turn to analyse several counterexamples. The main objective here is to demonstrate the changes in $\eta/s$ while the spectral function interpolates between quasi-particle-like behaviour with narrow peaks and cases with significant continuum contribution.

\begin{figure}
\centering
\includegraphics[width=0.75\linewidth]{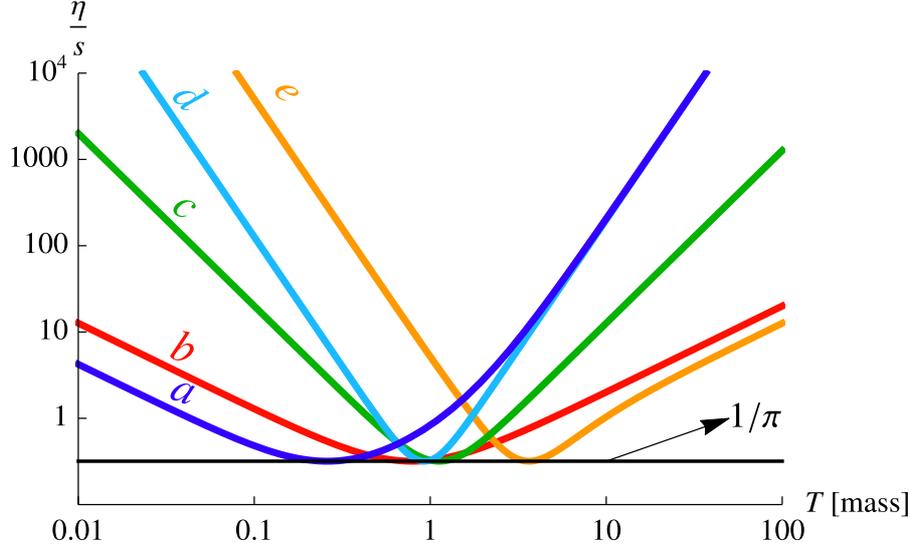}
\LNFIG
\caption{$\eta/s$ versus temperature $T$, provided by the Lorentzian ansatz Eq.~(\ref{BWansatz}). The minimal value is universally $1/\pi$. The plotted lines belong to various choices of $\gamma$: ${\sim (T^2+T_0^2)^{-1}}$ (blue, $a$), constant (red, $b$), $\sim T^3$ (green, $c$), $\sim T^{-2}$ (light blue, $d$), ${\sim T^{2+\frac{1}{\epsilon}}}(T_0^\frac{1}{\epsilon}+T^\frac{1}{\epsilon})^{-1}$ with $\epsilon=0.5$ (yellow, $e$).}
\label{fig:lorentzian}
\end{figure}
\subsection{Lorentzian quasi-particle peak}\label{exLorentzian}
First we consider a Lorentzian ansatz. It is consistent with the Dyson resummation in the special case when the self-energy equals to ${~\gamma^2-2\gamma\omega\text{i}}$:
\begin{equation}\label{BWansatz}
\rho_L(\omega,\ve{p})=\frac{4\gamma\omega}{(\omega^2-\ve{p}^2-\gamma^2)^2+4\gamma^2\omega^2}
\end{equation}
The sum-rule ${\frac{1}{2\pi}\intlim{\omega}{-\infty}{\infty}\omega\rho_L(\omega,\ve{p})=1}$ is fulfilled, moreover, ${\rho_L(\omega,\ve{p})\overset{\gamma\rightarrow 0}{\rightarrow} 2\pi\delta(\omega^2-\ve{p}^2)}$. Although this spectral function is unrealistic as a result of a first-principle calculation, it can be interpreted as a toy-model for quasi-particles with finite lifetime $\frac{1}{\gamma}$. Interestingly, this ansatz is microcausal without any restriction\footnote{Its Fourier-transform is not Lorentz-invariant, but closely related to the free-particle limit $\gamma=0$: $\rho(x)=e^{-\gamma t}\rho_{\gamma=0}(x)$.}. Using Eqs.~(\ref{pressure0}) and (\ref{energy0}) we get:
\begin{equation}
s_L=\frac{1}{4\pi^3}\intlim{\omega}{0}{\infty}2\pi\omega^3\left(-\frac{1}{\omega}\mathrm{ln}(1-e^{-\frac{\omega}{T}})+\frac{1}{T}\frac{1}{e^{\frac{\omega}{T}}-1}\right)= \frac{2\pi^2}{45}T^3,
\end{equation}
where $2\pi\omega^3$ before the parenthesis equals to $\int\!\!\mathrm{d}^3\ve{p}\omega\frac{\partial K_L}{\partial\omega}\rho_L$. It coincides, apparently, with the entropy of the ideal Bose gas. For the shear viscosity we evaluate Eq.~(\ref{eta0}):
\begin{equation}
\eta_L= \frac{1}{60\pi^2}\frac{1}{T}\intlim{\omega}{0}{\infty}\left(5\gamma\omega^2+\frac{\omega^4}{\gamma}\right)\frac{1}{\mathrm{ch}\frac{\omega}{T}-1}= \frac{1}{18}\gamma T^2+\frac{2\pi^2}{225}\frac{T^4}{\gamma}.
\end{equation}
Besides the expected $\sim\gamma^{-1}$ term, a linear one appears. The fluidity measure $\eta/s$ reads as:
\begin{equation}\label{etaLor}
\frac{\eta_L}{s_L}= \frac{5}{4\pi^2}\frac{\gamma}{T}+\frac{1}{5}\frac{T}{\gamma}.
\end{equation}
A peculiar property of this expression, that regardless of the temperature-dependence of $\gamma$, it has the minimal value $\left.\frac{\eta_L}{s_L}\right|_{T^*}=\frac{1}{\pi}$ (Fig.~\ref{fig:lorentzian}). The position of the minimum satisfies the equation ${\gamma(T^*)=\frac{2\pi}{5}T^*}$, for constant $\gamma$ this is $T^*=\frac{5\gamma}{2\pi}$. \par
It is worthwhile to mention, that Eq.~(\ref{etaLor}) is clearly beyond the relaxation time approximation as it has a contribution proportional to the inverse of the quasi-particle lifetime $\sim\gamma$.

\subsubsection*{The long lifetime limit $m\gg\gamma$}\label{exQPapprox}
In the quasi-particle limit with finite mass $m\gg\gamma$ the Eqs.~(\ref{pressure1}), (\ref{energy1}) and (\ref{eta1}) with the Dirac-delta-approximating spectral function ${\overline{\rho}(p)=2\pi\delta_\gamma(p^2-m^2)}$ result in the following simple expressions:
\begin{eqnarray}
s_{QP} &=& \frac{m^3}{2\pi^2}K_3(m/T), \label{sQP} \\
\eta_{QP} &=& \frac{1}{2\pi^2}\frac{m^2T^2}{\gamma}K_3(m/T). \label{etaQP}
\end{eqnarray}
Here, the width of the peak appears in $\eta$ only, due to the regularization of the square of the Dirac-delta: ${\delta_\gamma^2(p^2-m^2) \overset{m\gg\gamma}{\approx} \frac{2\pi}{\gamma}\delta(p^2-m^2)}$. The $\eta$ over $s$ ratio is the following:\LNNL
\begin{align}
\frac{\eta_{QP}}{s_{QP}} = \frac{T^2}{\gamma m}. \label{etaOsQP}
\end{align}
\begin{figure}
\centering
\includegraphics[width=0.6\linewidth]{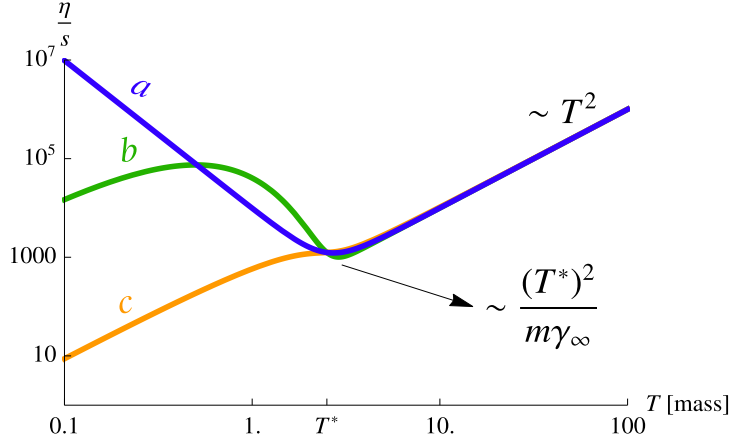}
\LNFIG
\caption{$\eta/s$ in QP-approximation for various $\gamma(T)$ with sudden change at $T^*$. Different low-temperature behaviour of $\eta/s$ are depicted for different $\gamma$-characteristics in the transition region.\\ Exponential relaxation with local minimum and maximum: power-law relaxation with diverging result when $T\rightarrow 0$: ${\sim T^{2-\frac{1}{\tilde{\epsilon}}}((T^*)^{\frac{1}{\tilde{\epsilon}}}+T^{\frac{1}{\tilde{\epsilon}}})}$, $\tilde{\epsilon}=0.2$ (blue, $a$), ${\sim T^2(1+\text{tanh}((T-T^*)/\epsilon))^{-1}}$, $\epsilon=0.5m$ (green, $b$), power-law relaxation with inflexion in $T^*$: ${\sim T^2(1+2/\pi\cdot\text{arctan}((T-T^*)/\epsilon))^{-1}}$ $\epsilon=0.9m$ (yellow, $c$). The value of the minimum is: ${\left.\eta/s\right|_\text{min}=(T^*)^2(m\gamma_\infty)^{-1}+\mathcal{O}(\epsilon)}$, $\gamma_\infty=0.01m$.}
\label{fig:QPapprox}
\end{figure}
Let us assume, that for some reason, the particle-lifetime changes significantly around $T=T^*$, but $\gamma\ll m$ still holds (Fig.~\ref{fig:QPapprox}). We parametrize the width as ${\gamma(T) =\gamma_\infty\theta_\epsilon(T-T^*)}$, where $\gamma_\infty$ is its value when $T\gg T^*$ and $\epsilon$ ($\tilde{\epsilon}$) is the size of the transition region in energy dimensions (or in dimensionless units) and ${m\varepsilon,\,\tilde{\varepsilon}\ll T^*}$ holds. We further assume that $m$ does not change significantly. In case of sharp change in $\gamma$, $\eta/s$ has a well-defined minimum at $T^*+\mathcal{O}(\epsilon)$. Depending on how the transition region is localized, the low-temperature limit of $\eta/s$ could be different:
\begin{itemize}
\item $i)$ When $\gamma$ goes to 0 in an exponential manner, $\eta/s$ reaches zero as $\sim T^2$.
\item $ii)$ If the transition in $\gamma$ is power-law-like: ${\sim (1+(T^*/T)^\frac{1}{\tilde{\epsilon}})^{-1}}$, the ratio is either divergent in $T=0$ or zero:
\begin{equation}
\frac{\eta}{s}\sim \left\{\begin{array}{lll}
T^2, &\mathrm{when} & T\gg T^*, \\
T^{2-\frac{1}{\tilde{\epsilon}}}, &\mathrm{when} & T\ll T^*.
\end{array}\right.
\end{equation}
\end{itemize}
The minimal value is ${\left.\frac{\eta}{s}\right|_{T^*}\approx\frac{2(T^*)^2}{m\gamma_\infty}}$. \par
A physically realistic situation is when $\gamma\sim T$ and $m \approx\text{const.}$ for $T>m$. In this case the fluidity measure is proportional to $T$ at high temperature. \newpage

\subsection{Quasi-particle and its continuum tail}\label{exQPtail}
\begin{figure}
\centering
\includegraphics[width=0.6\linewidth]{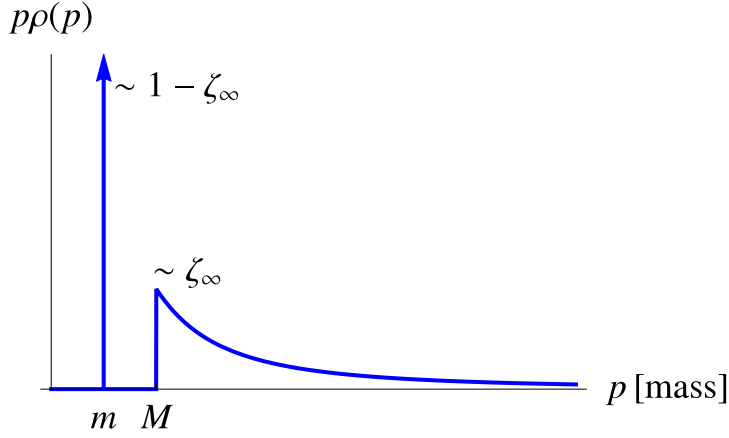}
\LNFIG
\caption{Schematic plot of the spectral function ${p\rho(p)}$. In the calculations of Sec.~\ref{exQPtail}, we used this depicted approximation, i.e.\! we took the QP-peak as a Dirac-delta, and assumed very sharp change near ${p\approx M}$.}
\label{fig:QPtailSchematic}
\end{figure}
We move towards to more general situations and parametrize the retarded propagator\footnote{For the precise notations regarding the correlation functions we use throughout this chapter, see Appendix~\ref{propagators}.} with momentum dependent self-energy: ${m^2(p)-p\gamma(p)\text{i}}$ and wave-function renormalization $Z(p)$\footnote{$Z$ can lead to an anomalous dimension, i.e.\ changing the asymptotic behaviour of the propagator $G \sim p^{-2+\delta}$, in case of $Z \sim p^\delta$ for large $p$.}:
\begin{equation}
G^{ra}(p)=\frac{Z(p)}{p^2-m^2(p)+ip\gamma(p)}.
\end{equation}
We assume $\gamma(p)$ and $Z(p)$ to be analytic functions and keep $m$ constant. The kernel function then reads as follows:\LNNL
\begin{align}
g(p) &= \frac{\partial K}{\partial p}\overline{\rho}(p) =\frac{\left(2p-(p^2-m^2)\frac{Z'(p)}{Z(p)}\right)p\gamma(p)}{(p^2-m^2)^2+p^2\gamma^2(p)} =\\
&=: g_\text{peak}(p)-\frac{p\gamma(p)(p^2-m^2)\frac{Z'(p)}{Z(p)}}{(p^2-m^2)^2+p^2\gamma^2(p)}=g_\text{peak}(p)+g_\text{cont}(p), \nonumber
\end{align}
where we separated the Lorentzian \textit{peak contribution}. The remaining \textit{continuum} part bears the same pole structure as $g_\text{peak}$ but with ${p^2-m^2}$ in the nominator also, and therefore disappears in the $\gamma\rightarrow 0$ limit. \rem{We are aimed to ''engineer'' a spectral function plotted in Fig.~\ref{fig:QPtailSchematic}. So we will use $Z(p)$ to suppress the spectral density for every $p<M$ but $p\approx m$. That is, the continuum part of the kernel $g_\text{cont}$ contributes only for momenta above the threshold $M$ }\par
Keeping in mind, that we are interested in going beyond the QP-spectrum in a parametrically controlled way, we link $Z$ and $\gamma$ together. For $Z=1$ and $\gamma=0$ we expect the particle excitation to be restored with mass $m$ and with infinite lifetime. Therefore we force $Z<1$ and $\gamma>0$ to happen simultaneously by setting ${\gamma\overset{!}{=}\frac{\Gamma}{\zeta_\infty}(1-Z(p))=:\frac{\Gamma}{\zeta_\infty}\zeta(p)}$ with a constant $\Gamma$ with energy-dimension. To ensure, that $Z<1$ is restricted to ${p>M>m}$, we put ${\zeta(p)=\zeta_\infty\theta_\epsilon(p-M)}$, where $0<\zeta_\infty<1$ and $\epsilon$ encodes how sudden the change from 0 to $\zeta_\infty$ is. We sketched the spectral function $\overline{\rho}$ in Fig.~\ref{fig:QPtailSchematic}. If ${M\gg\epsilon}$, the integrals in Eqs.~(\ref{entropy1}) and (\ref{eta1}) pick the $p\approx M$ contributions only, resulting in\LNNL
\begin{align}
s \approx & s_{QP}(m,T) + \frac{\frac{\zeta_\infty}{1-\zeta_\infty}\Gamma M(M^2-m^2)}{(M^2-m^2)^2+\Gamma^2M^2}s_{QP}(M,T), \\
\eta \approx &\eta_{QP}(m,T,\gamma_p) + \nonumber \\
&+ \frac{\frac{\zeta^2_\infty}{1-\zeta_\infty}\Gamma^2M^2(M^2-m^2)\left(4\epsilon M+\frac{M^2-m^2}{1-\zeta_\infty}\right)}{\left[(M^2-m^2)^2+\Gamma^2M^2\right]^2}\eta_{QP}(M,T,\epsilon),
\end{align}
with $s_{QP}$, $\eta_{QP}$ defined by Eqs.~(\ref{sQP}) and (\ref{etaQP}), respectively. ${\gamma_p=\Gamma\theta_\epsilon(m-M)\ll\Gamma}$ and $\epsilon$ are present to regularize the $\delta^2$-like parts in the viscosity integral. Writing out $\eta$ over $s$ explicitly:\LNNL
\begin{align}
\frac{\eta}{s} &\approx \frac{\eta_{QP}(m,T,\gamma_p)}{s_{QP}(m,T)}\frac{1+A^2(m,M,\Gamma,\zeta_\infty)\frac{\eta_{QP}(M,T,\epsilon)}{\eta_{QP}(m,T,\gamma)}}{1+A(m,M,\Gamma,\zeta_\infty)\frac{s_{QP}(M,T)}{s_{QP}(m,T)}} =\nonumber \\
&= \frac{T^2}{m\gamma_p}\frac{1+A^2(m,M,\Gamma,\zeta_\infty)\frac{M^2}{m^2}\frac{\gamma_p}{\epsilon}\frac{K_3(M/T)}{K_3(m/T)}}{1+A(m,M,\Gamma,\zeta_\infty)\frac{M^3}{m^3}\frac{K_3(M/T)}{K_3(m/T)}}, \,\, \text{with}\\
&A(m,M,\Gamma,\zeta_\infty) = \frac{\frac{\zeta_\infty}{1-\zeta_\infty}\Gamma M(M^2-m^2)}{(M^2-m^2)^2+\Gamma^2M^2} \nonumber.
\end{align}
The above expression results in a reduced value of ${\eta/s}$ compared to ${\eta_{QP}(m,T,\gamma_p)/s_{QP}(m,T)}$ whenever ${\zeta_\infty<(1+\frac{1}{2}\frac{m}{M}\frac{\Gamma}{\epsilon}\theta_\epsilon(m-M))^{-1}}$ holds.
The ratio ${r=\frac{\eta/s}{\eta_{QP}/s_{QP}}}$ has a minimal value ${2\frac{m}{M}\frac{\gamma_p}{\epsilon}\left(\sqrt{1+\frac{M}{m}\frac{\epsilon}{\gamma_p}}-1\right)}$ for $T\gg M$. \par
\begin{figure}
\centering
\includegraphics[width=0.75\linewidth]{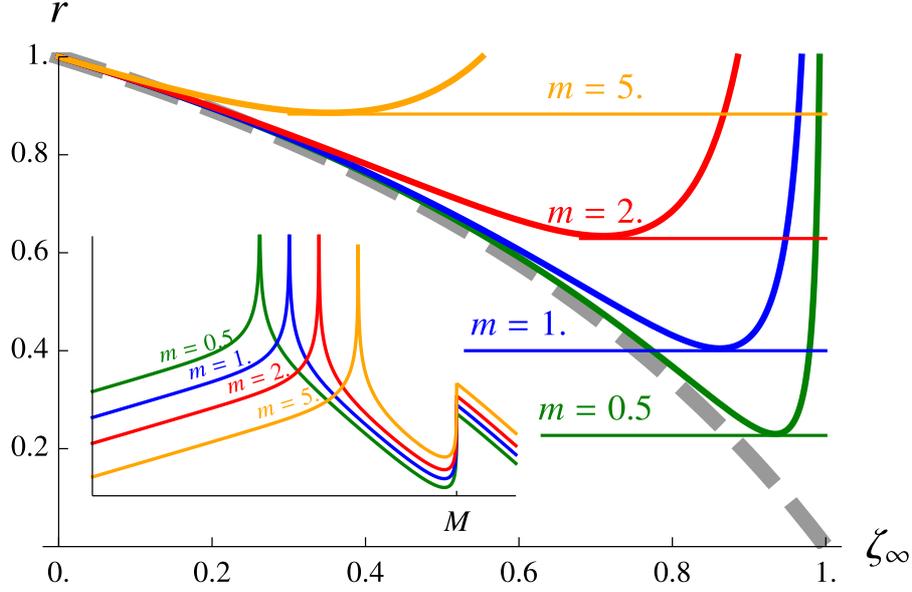}
\LNFIG
\caption{The ratio $r=\frac{\eta/s}{\eta_{QP}/s_{QP}}$ versus $\zeta_\infty$. Fixing $\frac{\gamma_p}{m}=0.05$, $M=50.0m$, $\epsilon=0.015m$ and $T=100m$, graphs with various values of $m$ (0.5, 1.0, 2.0, 5.0) are plotted, so that ${m,\,M\gg\epsilon}$ holds. For given $m$, $M$ and $\gamma_p$ the sum rule provides: ${\Gamma\approx \frac{\pi}{4}M}$ for $m\ll M$. The dashed line indicates the limiting case $m=0$: ${\left(1+c\frac{\zeta_\infty}{1-\zeta_\infty}\right)^{-1}}$, with ${c=\frac{\pi}{4}(1+\frac{\pi^2}{16})^{-1}}$. An illustration of the corresponding spectral functions are inset on a double-logarithmic plot.}
\label{fig:QPtail}
\end{figure}

Since $Z(p)$ and $\gamma(p)$ are momentum-dependent, and the sum rule ${\frac{2}{\pi}\intlim{p}{0}{\infty}p\overline{\rho}(p)=1}$ imposes a constraint on the parameters. $\Gamma$ happens to be proportional to $M$. For increasing $\zeta_\infty$ its value drops considerably, see Fig.~\ref{fig:QPtail} for examples. 
Consequently, the fluidity measure is modified by the ''continuum'' parameters $M$ and $\epsilon$. Its minimal value is the following:\LNNL
\begin{align}
\left.\frac{\eta}{s}\right|_\text{min.} &\overset{\frac{M}{T}\ll 1}{\approx} \frac{2T^2}{M\epsilon}\left(\sqrt{\frac{M\epsilon}{m\gamma_p}+1}-1\right) =\nonumber \\
&\overset{\epsilon =\gamma_p}{=} \frac{T^2}{\gamma_p m}\underbrace{\frac{2m}{M}\left(\sqrt{\frac{M}{m}+1}-1\right)}_{\leq 1} \overset{\frac{m}{M}\ll 1}{\approx} \frac{2T^2}{\gamma_p \sqrt{mM}}.
\end{align}
\FINAL{The $m=0$ limiting case is also worth noting:
\begin{align}
r &=\frac{\eta_{QP}/s_{QP}}{T^2/(m\gamma_p)} =\left\{\begin{array}{l} \gamma_p=\epsilon \\ \Gamma=\frac{\pi}{4}M \\ m=0\end{array}\right\}= \frac{1}{1+c\frac{\zeta_\infty}{1-\zeta_\infty}\frac{M^3}{8T^3}K_3(M/T)} \overset{M\ll T}{=} \frac{1}{1+c\frac{\zeta_\infty}{1-\zeta_\infty}},
\end{align}
with $c=\frac{\pi}{4}\left(1+\frac{\pi^2}{16}\right)^{-1}$.}

\subsection{Beyond the QP-pole}\label{exCutSigma}
\begin{figure}[!t]
\centering
\includegraphics[width=0.7\linewidth]{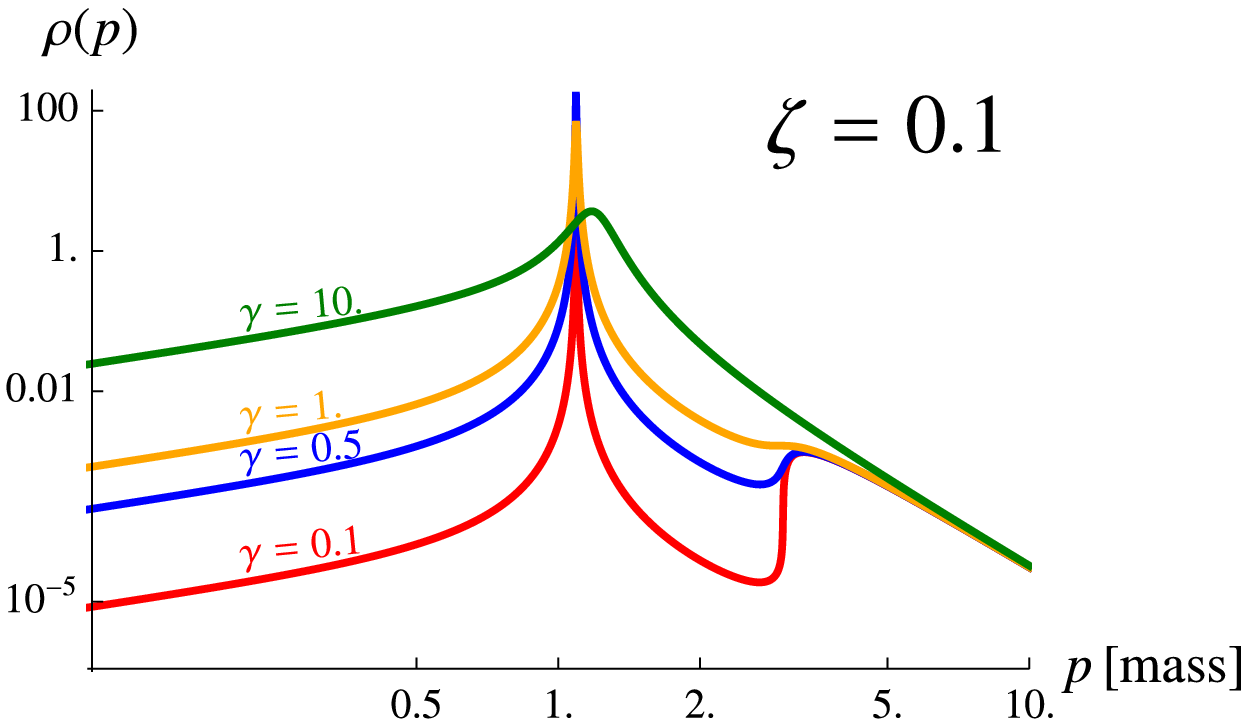}
\includegraphics[width=0.7\linewidth]{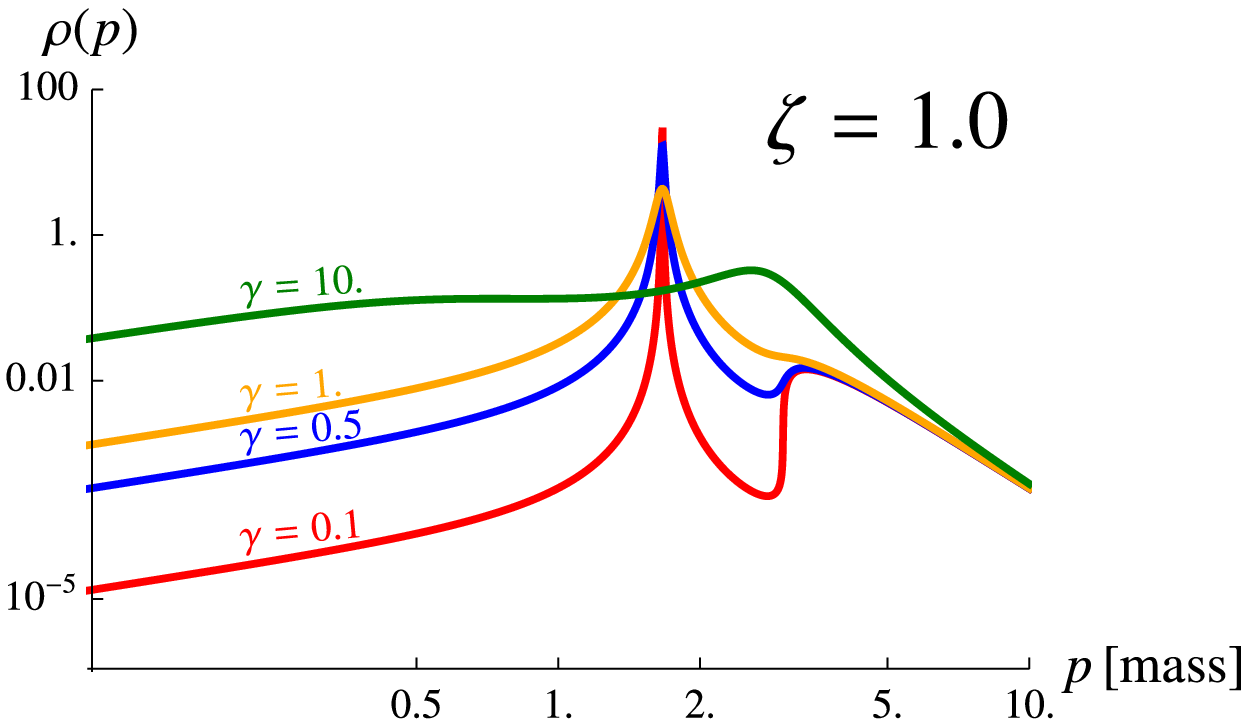}
\includegraphics[width=0.7\linewidth]{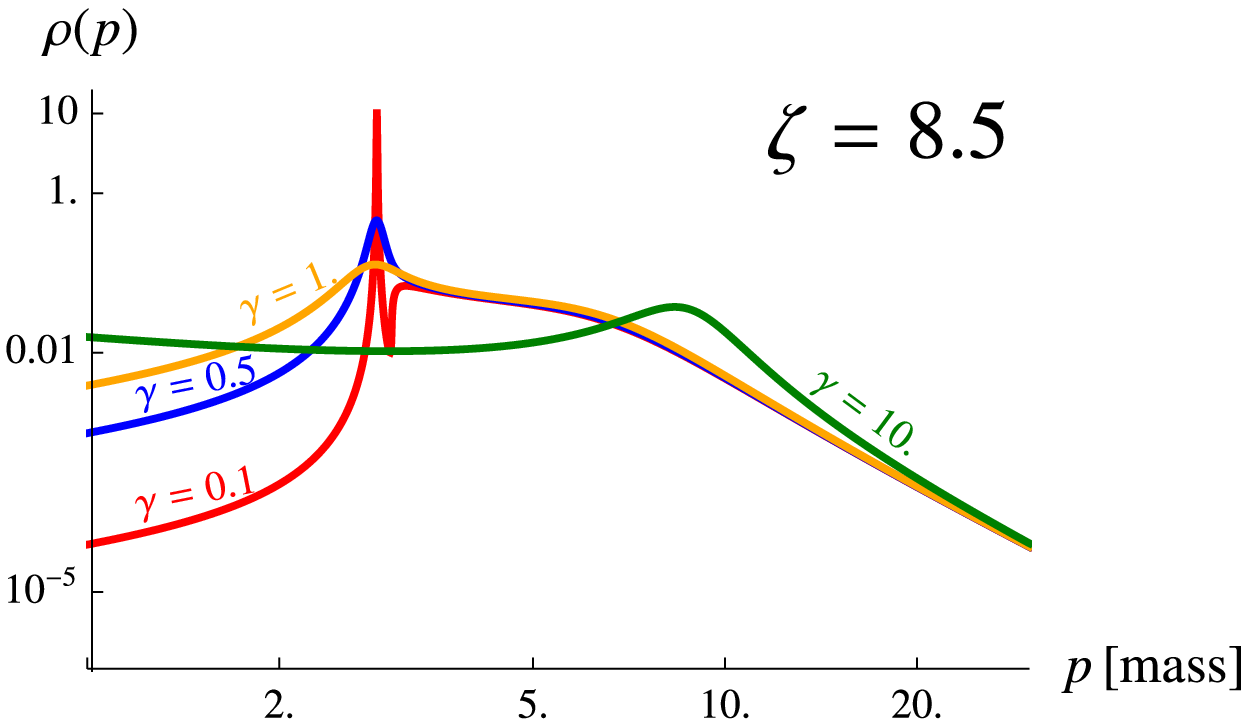}
\LNFIG
\caption{Spectral density $\overline{\rho}(p)$ for various values of ${~\gamma=0.1-10.0}$, with fixed parameters $m=1.0$, $M=3.0$ (in the dimension of mass) and ${~\zeta=0.1,\,1.0,\,8.5}$ (in the dimension of mass square). After the pole-part and the continuum ''melted'' into each other ($\gamma\approx 1.0$), the further increase of $\gamma$ shifts the remaining hump-like structure towards higher momenta.}
\label{fig:cutSigmaRho}
\end{figure}

As we have seen, if $G^{ra}$ has only pole singularities, those control the overall behaviour of the theory inevitably. Mimicking the features of the multi-particle contribution using the QP-tail is inadequate in the sense, that its effect is suppressed by the imaginary part of the pole position: the width of the QP-peak. In more realistic situations, i.e.\ in interacting QFTs, the propagator has branch cuts beside its poles. Branch cuts are generated even in one-loop order in perturbation theory, corresponding to the opening of multi-particle scattering channels. For example, in a theory with the lowest mass excitation $m$, the continuum contribution of the spectrum starts at $M=2m$ (at zero temperature, if {~1-to-2} decay or {~2-to-2} scattering is allowed at tree-level). To take into account these cut contributions, we parametrize the inverse retarded propagator and the spectral function as follows:\LNNL
\begin{align}
(G^{ra})^{-1} &= p^2-m^2-\Sigma_s, \label{invGCut} \\
\overline{\rho} &= \frac{\text{Im}\Sigma_s}{(p^2-m^2-\text{Re}\Sigma_s)^2+(\text{Im}\Sigma_s)^2}. \label{rhoCut}
\end{align}
At zero temperature, we assume $\Sigma_s$ to have a branch cut along the real line, starting at $p=M$. At finite temperature we expect the near-$M$ behaviour of $\im\Sigma_s$ smoothens. We use an ansatz, that shows this kind of behaviour. It is motivated by the self-energy correction of a cubic scalar model (see for eample Sec.~24.1.1 of Ref.~\cite{SchwartzQFT}) and by the IR-safe resummation discussed in Ref.~\cite{GenBoltzmannEqu}:\LNNL
\begin{align}
\im\Sigma_s(p) &=\zeta \pi\frac{\sqrt{\sqrt{\left(1-\frac{M^2}{p^2}\right)^2+4\frac{\gamma^4}{M^4}}+1-\frac{M^2}{p^2}}}{\sqrt{\sqrt{1+4\frac{\gamma^4}{M^4}}+1}} \\
&\overset{\gamma\rightarrow 0}{\longrightarrow} \zeta\pi\theta(p-M)\sqrt{1-\frac{M^2}{p^2}},\nonumber\label{cutImSigma}
\end{align}\LNNL
\begin{align}
\re\Sigma_s(p) &=\frac{1}{\pi}\mathcal{P}\!\!\!\!\!\!\! \intlim{q}{0}{\infty}\frac{2q\im\Sigma_s(q)}{p^2-q^2} \\
&\overset{\gamma\rightarrow 0}{\longrightarrow} \left\{ \begin{array}{lll}
2\zeta\sqrt{\frac{M^2}{p^2}-1}\cdot\arcsin\left(\frac{p}{M}\right), & \,\, & p<M,\\
-2\zeta\sqrt{1-\frac{M^2}{p^2}}\cdot\text{ln}\left(\frac{p}{M}-\sqrt{\frac{p^2}{M^2}-1}\right), &\,\, & M<p.
\end{array}\right. \label{cutReSigma}
\end{align}
\begin{figure}[!t]
\centering
\subfloat[]{
\includegraphics[width=0.5\linewidth]{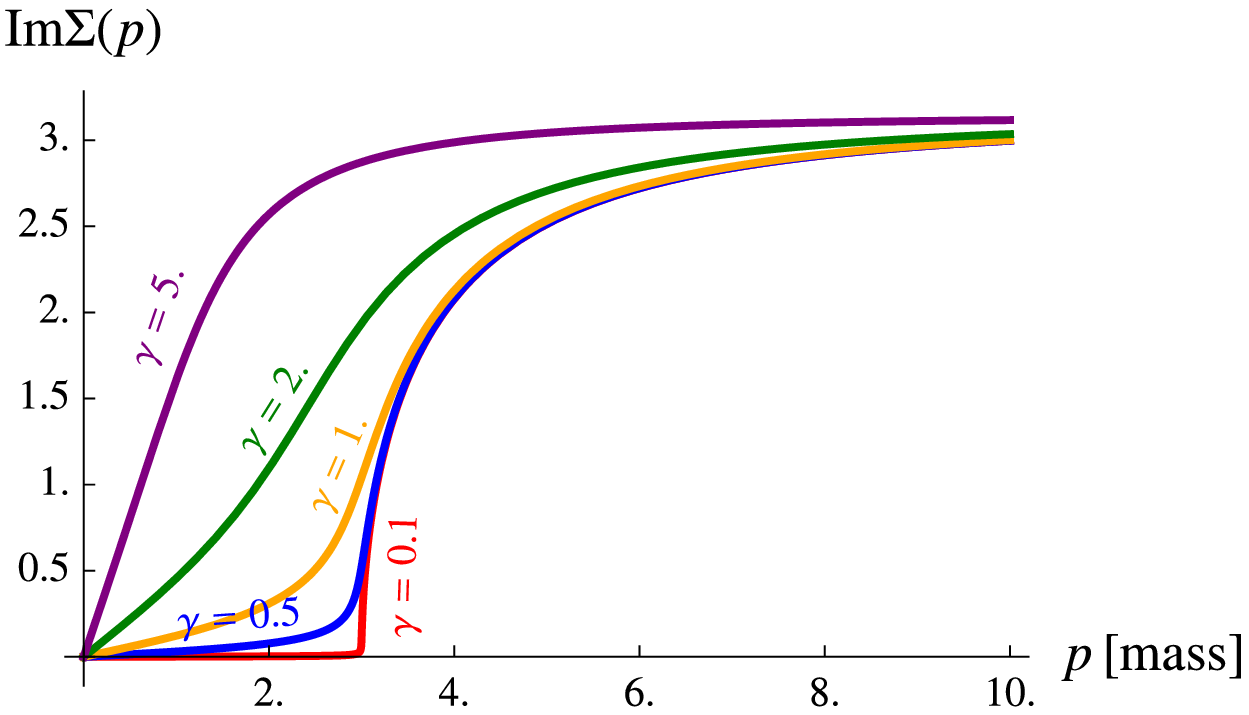}
}
\subfloat[]{
\includegraphics[width=0.5\linewidth]{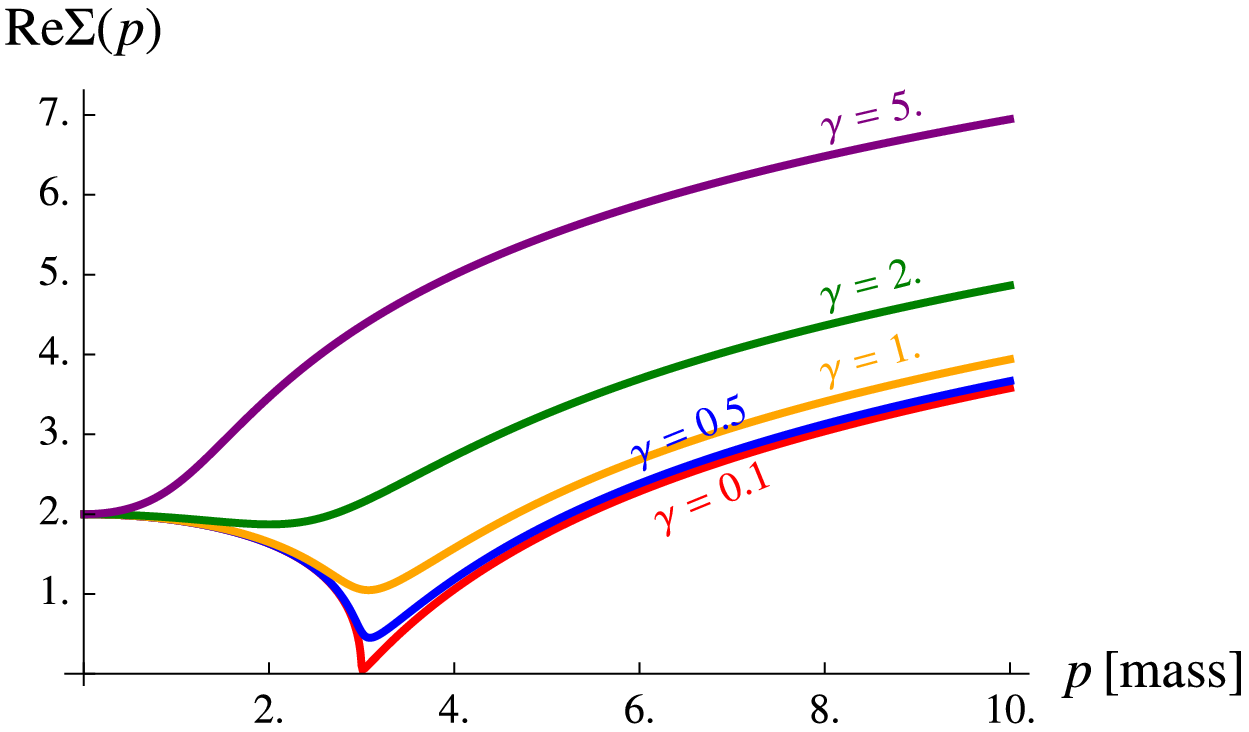}
}
\LNFIG
\caption{Imaginary and real parts of the self energy $\Sigma_s$ on the real line, with parameters $M=3.0$ (in the dimension of mass) and $\zeta=1.0$ (in the dimension of mass square).}
\label{fig:Sigma}
\end{figure}
The Kramers-Kronig relation is used to evaluate $\re\Sigma_s$ for any values of $\gamma$ numerically. We plotted the self-energy $\Sigma_s$ and the spectral density $\overline{\rho}$ on Figs. \ref{fig:Sigma} and \ref{fig:cutSigmaRho} for illustration. The limit $\gamma=0$ is also given analytically in Eq.~(\ref{cutReSigma}). \par
\begin{figure}
\centering
\includegraphics[width=0.75\linewidth]{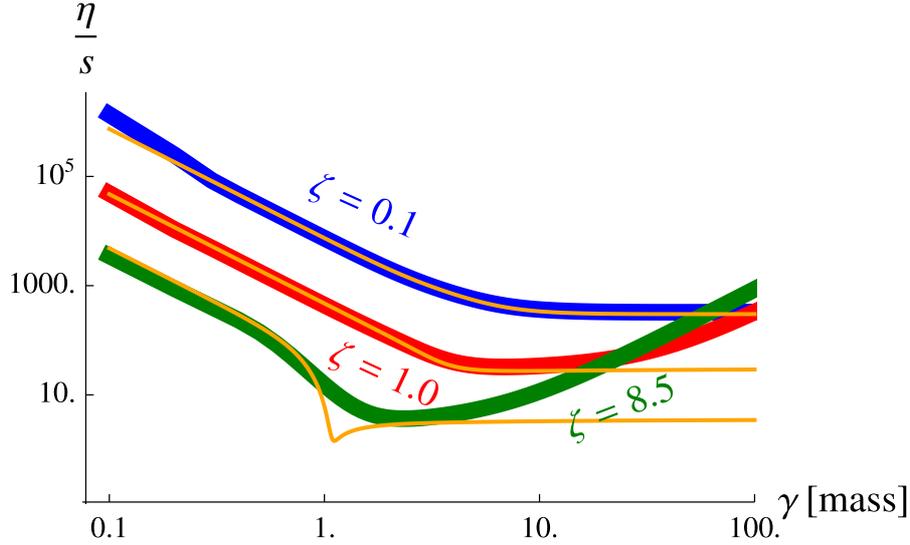}
\LNFIG
\caption{The fluidity measure $\eta/s$ computed with a realistic ansatz for the self energy $\Sigma_s$. The ratio $\eta/s$ reduces as the continuum contribution is more and more pronounced by the increase of $\zeta$. The $\gamma$-dependence shows a minimal value of $\eta/s$, far from the region where the QP-peak and the continuum part are well distinguishable. The QP-pole approximation of Eq.~(\ref{QPpoleAppr}) is indicated by the thin curves. The values of other fixed parameters are $m=1.0$, $M=3.0$ and $T=10.0$.}
\label{fig:etaOsSigma}
\end{figure}

Formulae in Eqs.~(\ref{entropy1})~and~(\ref{eta1}) are used to evaluate the fluidity measure $\eta/s$. The numerical results are depicted on Fig.~\ref{fig:etaOsSigma} for various values of $\gamma$ with fixed $\zeta$. The main conclusion here is, that the increase of the weight of the continuum in $\overline{\rho}$ by increasing the value of $\zeta$, the ratio $\eta/s$ decreases. As for the $\gamma$-dependence of the fluidity measure, we find a power-law-like decay ending in a minimum. This decrease of $\eta/s$ seems to be connected to the ''melting'' of the QP-peak and the multiparticle continuum in $\overline{\rho}$. Leaving this region of the parameter space, i.e.\ further enhancing $\gamma$, the ratio saturates, than starts to slowly increase, see Fig.~\ref{fig:etaOsSigma}. This effect is similar to those of we observed in the case of the Lorentzian, i.e.\! ${\eta/s\sim\gamma}$ for large enough $\gamma$, cf.\ Eq.~(\ref{etaLor}).
\subsection{The QP-region}\label{exQPinCutSigma}
We are interested in how far the QP-approximation remains reliable. Therefore, we briefly overview the key steps performing such a calculation. Starting from Eqs.~(\ref{invGCut}),~(\ref{rhoCut}) we force a real QP-peak into the spectrum $\overline{\rho}$, i.e.\! a Dirac-delta peak at ${p=M_*<M}$. The model self-energy $\Sigma_s$ in Eqs.~(\ref{cutImSigma}),~(\ref{cutReSigma}) does the job, when the limit ${\gamma\rightarrow 0}$ is taken. The position of the peak is assigned by $M_*$, the pole-mass, which satisfies the following equation:\LNNL
\begin{align}
M_*^2-m^2-\re\Sigma_s(M_*)&=0.\label{QPpoleMass}
\end{align}
This immediately makes the sum rule into a constraint, which connects $M$, $M_*$, $m$ and $\zeta$:\LNNL
\begin{align}
\frac{2M_*}{2M_*-\frac{\partial\re\Sigma_s(M_*)}{\partial p}}+\frac{1}{\pi}\intlim{p}{M}{\infty}\frac{p(-\im\Sigma_s)}{(p^2-m^2-\re\Sigma_s)^2+(\im\Sigma_s)^2} &=1.\label{QPsumrule}
\end{align}
Expanding the viscosity kernel function $g^2(p)$ up to first order in ${1/\im\Sigma_s(M_*)}$ provides $\eta$ in the QP-approximation. Furthermore, we take the QP entropy density $s_{QP(T,M_*)}$ to calculate $\eta/s$, as it is discussed in details in App.~\ref{QPsumruleAnal}. Thus, the resulted formula is:\LNNL
\begin{align}
\left.\frac{\eta}{s}\right|_{\text{QP-pole}} &= \frac{T^2}{2M_*}\frac{2M_*-\frac{\partial\re\Sigma_s(M_*)}{\partial p}}{\im\Sigma_s(M_*)}, \label{QPpoleAppr}
\end{align}
The contribution of the QP-peak is shown on Fig.~\ref{fig:etaOsSigma} (indicated by the thin curves), for comparison. It seems to catch the power-law-like fall of $\eta/s$ correctly. This approximation of $\eta$ and $s$, however, becomes worse and worse with increasing the value of $\zeta$, as it is expected. \par
The resulted pole mass from Eq.~(\ref{QPpoleMass}) for large $M$ is ${M_*^2=m^2+2\zeta+\mathcal{O}(\zeta/M^2)}$, and the corrections vanish fast, as it is demonstrated in Appendix \ref{QPsumruleAnal}. We use Eq.~(\ref{cutImSigma}) with small $\gamma$ to regularize the divergent behaviour of ${1/\im\Sigma_s(M_*)}$ in the QP viscosity. With the help of the leading order result for $M_*$, Eq.~(\ref{QPpoleAppr}) can be simplified further:\LNNL
\begin{align}
\left.\frac{\eta}{s}\right|_{\text{QP-pole}} &\overset{M\gg m,\,M_*}{\sim}\frac{1}{M_*} \sim \frac{1}{\sqrt{m^2+2\zeta}}.
\end{align}
That is, the main effect in the regime dominated by the contribution of the QP-peak is the dynamical shifting of the pole mass $M_*$.

\subsection{On phase transition} \label{nearCEP}
Hitherto we investigated systems whose thermodynamical quantities were continuous functions of the temperature. We argued, that our framework may tackle the phenomenology in the crossover-region, near to a possible critical end point (CEP), where the long-range correlations play an important role. Let us now make here few remarks on the issue of phase transition. \par
As we mentioned earlier, it was observed in a wide range of materials with a CEP in their phase diagrams, that $\eta/s$ shows a considerable reduction near the critical temperature $T_c$. We note here two jointly present effects, both which can contribute to the behaviour of $\eta/s$ as a function of the temperature near to $T_c$. The dimensionless entropy density $s/T^3$ changes more and more sharply approaching the critical temperature. It saturates to the Stefan--Boltzmann-limit for high $T$ and vanishes by lowering the temperature. Therefore, depending on the details of the transition, $T^3/s$ could show significantly different behaviour below and above $T_c$ -- even possibly diverge for $T\rightarrow 0$. That in itself is enough to develop a minimum for $\eta/s$ , even if $\eta/T^3$ is monotonous. Crossing a 1$^\text{st}$ order type phase boundary, the value of $T^3/s$ jumps, whilst for a 2$^\text{nd}$ order transition it turns back. \par
Besides, $\eta/T^3$ may also behave differently as a function of temperature above and below a characteristic temperature value $T^*$. We refer to Eq.~(\ref{etaLor}) as a simple example. Although it depends smoothly on temperature for constant $\gamma$, a jump or turning back of the slope is conceivable, whenever the temperature dependence of $\gamma$ changes passing the critical temperature. The value of $T^*$ characterizing this transition point is expected to be close to the critical temperature of the system, ${T_c/T^*\approx\mathcal{O}(1)}$. In case of the Lorentzian shape with constant $\gamma$, this temperature value is in the order of $\gamma$, namely ${T^*=\frac{5\gamma}{2\pi}}$. \par
In fluids, it is observed that $\eta$ acts like a susceptibility and diverges weakly as the correlation length $\xi$ goes to infinity. The critical exponent of the shear viscosity is reported to be very small compared to those of the correlation length \cite{etaOsCEP1, etaOsCEP2, etaOsCEP3}. We can use Eq.~(\ref{etaLor}) again, with the tentative identification ${\gamma\sim\xi^{-1}}$, where $\xi$ is the correlation length (since $\gamma$ is also the mass parameter in the example of Sec.~\ref{exLorentzian}). This would result in a critical behaviour ${\eta/s\sim |T-T_c|^{-\nu}}$, i.e. the critical exponents of $\xi$ and of $\eta/s$ would be the same\footnote{The correlation length behaves as ${\xi\sim|T-T_c|^{-\nu}}$ near to the CEP. The exponent ${\nu>0}$ is typically in the order of one, in the mean field (Ginzburg--Landau) case ${\nu=\frac{1}{2}}$.}. This value of the critical exponent is way too high compared to the experimental findings. It is worth to emphasize though, that our approach is based on the Gaussian approximation of the generating functional. Therefore it is not expected to describe the phenomenology in the CEP, where the fluctuations of the order parameter are huge.

\section{Conclusions}\label{ConclEtaOs}
\begin{figure}
\centering
\fbox{\includegraphics[width=0.75\linewidth]{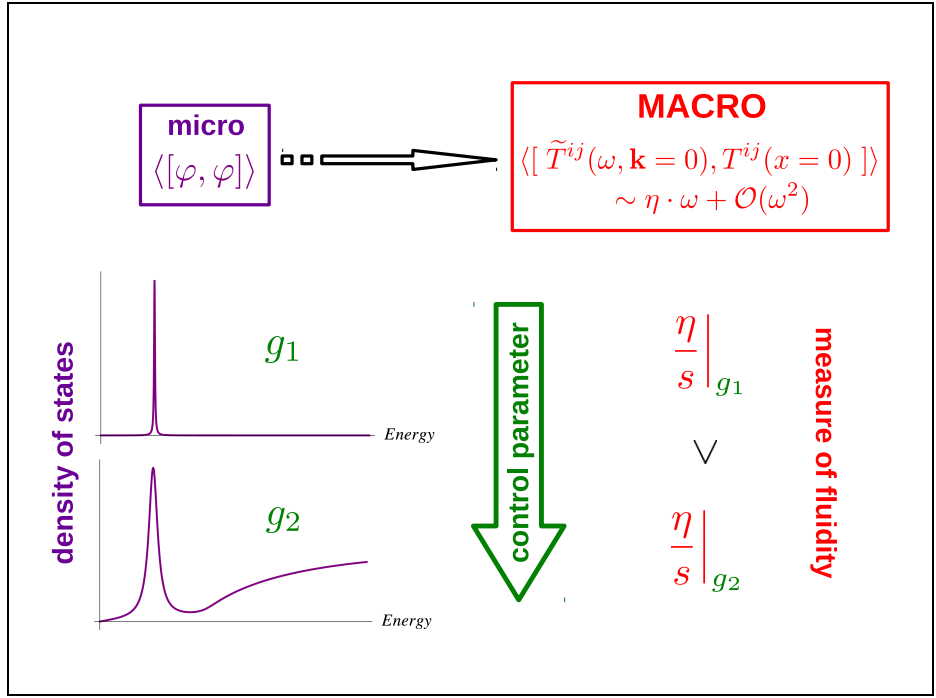}}
\LNFIG
\caption{Summarizing the phenomenological findings of the chapter}
\label{fig:conclEtaOs}
\end{figure}

In this chapter, we investigated how the robust properties of the spectral density of states $\rho$ of a QFT define the value of the fluidity measure $\eta/s$ in the framework of extended quasi-particles. Without other conserved charges, this ratio characterizes the relaxation to thermal equilibrium after a small shear stress is applied. We worked out formulae for the thermal quantities and also for the transport coefficients in the linear response regime regarding an approximation scheme parametrized solely by $\rho$. This scheme is able to incorporate finite lifetime effects and multi-particle correlations caused by interaction. \par
Parametrizing $\rho(p)$ by microscopically meaningful quantities, like the inverse lifetime and mass of quasi-particle excitations (position of the pole singularity of $\rho(p)$), and multi-particle threshold (position of the branch point of $\rho(p)$) we analysed the fluidity measure $\eta/s$. Our main finding is, that the more non-quasi-particle-like $\rho$ is, the more fluent the medium it describes. More precisely, we tuned the parameters of the spectral function $\rho$ in such a way, that the strength or residuum of the quasi-particle peak became less and less pronounced, and we observed the reduction of $\eta/s$, see Fig.~\ref{fig:conclEtaOs}. All-in-all, the particularly simple formula of Eq.~(\ref{etaLor}) has proven to be very insightful, especially in the light of the more complicated examples, since it seems to be showing all the key features we have explored during the analysis done in Sec.\ \ref{exQPtail}~and~\ref{exCutSigma}. \par
Our result supports the observations of other authors. The weakening of $\eta$ is also observed in resummed perturbation theory of the quartic interacting scalar model \cite{Jeon}, and also supported by numerical evidences in case of hadronic matter, when one takes into account a continuum of Hagedorn-states besides the hadronic resonances \cite{transport_greiner}. \par
We pointed out, that in our framework there is a lower bound for $\eta/s$, which is proportional to the entropy density over $T^3$. As long as one can constrain the thermodynamic quantities, our approach provides a restriction to the transport. \par
Moreover, the approximation of the transport coefficients is feasible based on the detailed knowledge about the thermal observables. Supposing that one knows all the independent thermodynamic quantities as a function of some control parameter (e.g.\ temperature), there is room for a model with as many parameters as the number of the independent thermal observables. Fitting the formulae to the known data set, the parameters $\alpha_i(T)$ in ${g(p,\{\alpha_i(T)\})=\frac{\partial K}{\partial p}\overline{\rho}}$ can be fixed. Therefore the viscosity in the framework of EQP is determined, using $g^2(p)$ and the formula (\ref{eta1}). There are available data from lattice Monte-Carlo simulations describing observables in thermal equilibrium in QCD, and also from condensed matter systems and other field theories. However, it is still challenging to extract the transport coefficients. The estimation based on thermal observables can be a good guideline here.\par

\section{Outlook}\label{OutlookEtaOs}
There are numerous ways to further develop this effective fluid description presented in this chapter. The most straightforward way is to incorporate conserved quantities in order to have non-zero chemical potential. Since more independent thermodynamical quantities mean more conserved charges (besides the energy-momentum density), the formulae given here need to be generalized. The first straightforward step into this direction is to consider the cases of the charged scalar field and the Dirac-field. It would be also interesting to see, how the lower bound on $\eta/s$ changes when the chemical potential corresponding to the charge density comes into play. These subjects, however, are left to be discussed in future publications.\par
Also, we need to see how general the phenomenon of the interaction induced increase of fluidity is, by means of analysing various interacting QFTs. A good candidate for this matter is the Bloch-Nordsieck theory, i.e. the QED in quenched approximation -- when the effects of fermion loops are neglected. Since this model is exactly solvable, despite the usual IR-sensitivity of gauge theories, it is an ideal landscape for studying the shear viscosity, which is also sensitive for the details of the infrared physics.\par
Another interesting issue is to quantify the fluidity properties of a QFT, when interaction with another field degree of freedom is switched on in a controlled way. For example, let us speculate on the coupled system of scalar fields $\varphi$ and $\chi$ given by the following Lagrangian: \LNNL
\begin{align}
\mathcal{L}(\varphi,\chi) &=\mathcal{L}_\mathrm{env.}(\chi)-\frac{1}{2}\varphi\Box\varphi-\frac{g_1}{2}\varphi\chi^2 -\frac{g_2}{4}\varphi^2\chi^2.
\end{align}
We refer to the part $\mathcal{L}_\mathrm{env.}$ which depends on solely $\chi$ as ''environment'', and neglect the contribution of $\varphi$-loops to its correlation functions -- i.e.\! we use quenched approximation. The correlation functions of the ''test particle'' field $\varphi$ are, however, modified. Now, the question is how thermodynamic and transport quantities can be determined consistently, as a function of ${(g_1,\,g_2)}$. For $g_1=0$ -- when only 2-to-2 scattering processes are allowed at tree-level -- resummation of the vertex corrections to $\avr{\varphi^2\chi^2}$, which are needed to both the entropy density and the viscosity, may be possible.\par
We mention another approach: using FRG techniques with $\mathcal{L}$ as a starting point, an effective action could be established, and as such, contributions beyond the EQP-action can be identified. \par
These kind of models may also be important to tackle the in-medium energy loss of partonic jets in the QGP -- as an above-mentioned ''test particle'' is moving through the medium encoded in the sector $\chi$.
\clearpage
\chapter{Summary and conclusions}\label{thStatements}

In this thesis we discussed three separate analysis of various phenomenological aspects of heavy-ion collisions. In Section~\ref{introduction}, we summarized the key problems and unanswered questions regarding the standard model of HIC. We also pointed out the inevitability of using effective models.\par
The first problem we analysed was the long-time behaviour and the thermalization of multi-component modified Boltzmann equations. In Section~\ref{MMBE} we motivated this kinetic theory problem as an effect of the surrounding medium to the quasi-particles. The analytical and numerical studies showed, that the modification leads generally to the lack of a detailed balance state. This behaviour can be treated, however, with the dynamical feedback of the average kinetic energy into the parameters characterizing the modification. \par
The second analysis was about the azimuthal asymmetry of the particle yields in HIC, known as the elliptic flow, in Section~\ref{ellflow}. We linked the initial state asymmetry of a non-central event to the elliptic flow through a possible microscopic mechanism, which involves an ensemble of ordered, decelerating source-pairs (dipole-like structures). This radiation is -- at least partly -- induced by the surrounding plasma. The radiation pattern of such an arrangement can be responsible for a major part of the elliptic flow. We also estimated the geometric dimensions of this radiation sources. However, we were not able to clarify the microscopic origin of this effect on the level of QCD degrees of freedom. Also, a detailed analysis on the possible existence of such phenomenological objects is needed.\par
The third problem was the effective field theoretical description of the fluidity of spinless quantum channels. In Section~\ref{etaOs}, we posed the most thoroughly explored question of this thesis, namely which microscopic properties of the quasi-particle spectrum can make the matter more fluent. We firstly built up an EFT framework through the generalization of the concept of QP. Then it was followed by the analysis of model spectral functions to demonstrate that
\begin{enumerate}[\it i)]
\item not only the shortening of the QP-lifetime alone decreases the fluidity measure $\eta/s$.
\item The contribution of the continuum of multi-particle scattering states, and parallel to that, the weakened residue of the QP-pole causes a considerable decrease of the fluidity measure.
\item Even this simple, basically mean-field approximation results a lower bound for $\eta/s$ which is, however, not universal, but constrained by the entropy density.
\end{enumerate}
Many possible directions to continue the investigation of the fluidity measure were also mentioned. The most interesting open question here is, what are the common features of the IR-behaviour of scalar and gauge theories -- if any. Since $\eta/s$ is sensitive to the low-momentum sector of the spectrum, it can signal if a similarity in the hydrodynamical behaviour of such theories exist. The investigation of the infrared physics might be extended to fairly simple yet phenomenologically relevant field theories, as we speculated in the conclusion of Sec.~\ref{etaOs}. \par
The three problems we discussed in this thesis are distinct for several reasons: $i.)$ They are attached to different period of the time evolution of a HIC. $ii.)$ They also use different frameworks to account for the relevant physics, and therefore are effective theories by definition, due to the approximation they apply. However, a common feature can be outlined in all these problems: we delineated the microscopic reasons in all three cases from the presence of medium. The phenomenological picture was, in all three cases, that the QP-like microscopic DoF propagate through some ''medium'', whose dynamical description is beyond the scope of our model. Therefore a different macroscopic behaviour emerges, as the microscopic DoF ''test'' this medium during their motion. This concept of medium-modification often comes up in the theory of HIC. Thus far, neither the complete, first-principle based description of the QGP -- the medium -- nor its interaction with QP-objects -- such as energetic partons -- is known satisfactorily. The theoretical understanding is successful only in the realm of weak jet--medium coupling, where $\eta/s$ is large and therefore the quasi-particle nature of the partonic DoF is not altered significantly. That is why we keep on constructing effective models on the long way we still have to pursue to understand more fundamentally the strongly interacting matter.

\section*{Novel scientific results}
In the following I summarize again the main results of this thesis. All these results are the product of the research I have carried out on my own, or I have significantly contributed to. The corresponding scientific publications are also indicated.
\begin{enumerate}[\bf I.]
\item \label{mmbe1}
\textbf{I analysed the conditions of the existence of a detailed balance solution in a non-extensive modification of \FINAL{the two-component} Boltzmann kinetic equation, motivated by medium effects.} Such a solution balances the gain and loss processes perfectly. \par
With one component, a detailed balance solution always exists compatible with the Jaynes principle. The modification I used alters the kinetic energy constraint of a two-particle collision by adding an ''interaction'' term to the total kinetic energy, i.e.: $E_1+E_2+aE_1E_2 =\text{const.}$ \par
In the case of two components, there are three different types of processes need to be balanced in equilibrium. For example, for particle species $A$ and $B$ there are collisions involving two $A$-type, two $B$-type or an $A$- and a $B$-type particle as well. \textbf{I showed that for non-equal modification parameters $a_{AA}$, $a_{BB}$ and $a_{AB}$ no detailed balance solution exists} -- see Sec.~\ref{detailedBalanceMMBE}. Only a dynamical feedback of the modification parameters can maintain balance between the gain and loss terms, in which these parameters become equal. Thus, an effectively one-component system emerges. \par
I verified these findings also by numerical simulations \cite{MMBEpaper}. Section~\ref{longtimeMMBE} summarizes my findings.

\item \label{mmbe2}
Throughout the numerical analysis of the two-component modified Boltzmann equation, \textbf{I found a scaling family of solutions with stationary shape.} After a short time of isotropisation, each gas component is characterized by an energy distribution depending on time only through the average of the total kinetic energy (which is not conserved in this model). \par
\textbf{I also showed analytically the existence of such solutions. The stationary shape of the distributions, i.e. $f(E,t)=\frac{1}{\avr{E}(t)}\phi(E/\avr{E})$ suggest that the system is in a pre-thermalized state.} Therefore, the gas components can be interpreted as cooling or warming thermodynamical bodies, being in thermal contact with the environment through a medium in which the gas particles are propagating through \cite{MMBEpaper}. See Sec.~\ref{scalingSolMMBE} for the detailed discussion.

\item \label{ellflow1}
\textbf{My colleagues and I constructed a phenomenological model for the description of the azimuthal asymmetry (the so-called elliptic flow) of the non-central heavy-ion collisions.} The key element of this model is the observation, that two charged particles decelerating towards each other can produce photon emission spectra similar to those of observed in proton-proton and proton-nucleon experiments. \par
Using a statistical ensemble of decelerating particle pairs as sources of photon or light particle emission, \textbf{I was able to deduce a fairly simple formula for the elliptic flow, simple enough to fit the experimental data with three independent parameters.} I analysed the elliptic flow of photons and also of charged hadrons using this phenomenological formula. The results can be found in Sec.~\ref{fitsEllFlow}.\par
\textbf{I estimated the geometrical parameters of the emission region, belonging to such bremsstrahlung-induced elliptic flow patterns} \cite{v2paper}, see Sec.~\ref{formFactorEllFlow}.

\item \label{etaos1}
\textbf{I derived the transport coefficients in the linear response approximation using Kubo's formula in an effective field theory framework.} This kind of phenomenological description is motivated by the medium-modified dynamics of a many-body system. It can be also useful to describe the long-wavelength excitations in the vicinity of a critical end point on the phase diagram. \par
The description is parametrized fully by the two-particle correlation functions. \textbf{Using the Keldysh formalism, I gave closed formulae for the thermodynamical quantities and the transport coefficients} \cite{etaOspaper}, for details see Sec.~\ref{linRes}.

\item \label{etaos2}
\textbf{I analysed the ratio of the shear viscosity $\eta$ and the entropy density $s$ using the formulae mentioned in point \ref{etaos1}}. $\eta/s$ characterizes the relaxation of shear deformations to the local equilibrium state of the fluid, therefore it measures the fluidity of the material under consideration. \par
\textbf{I constructed model spectral functions for investigating the liquid--gas crossover parametrically in the previously mentioned effective field theory framework. I found that the continuum part of the spectral function, besides the quasi-particle pole, plays a crucial role in the phenomenology of the fluidity of the system.} Namely, making the spectral function to be less dominated by the quasi-particle pole (by increasing the relative weight of the continuum part) results increased fluidity (decreased value of $\eta/s$) \cite{etaOspaper}, for the detailed discussion see Sec.~\ref{examples}.

\item \label{etaos3}
\textbf{Using the effective field theory framework mentioned in \ref{etaos2}., I examined the lower bound of the shear viscosity $\eta$ with fixed value of the entropy density $s$ \cite{etaOspaper}.} I also analysed the spectral density of states, which minimizes the ratio $\eta/s$. \FINAL{In the spinless case the accessible states seem to accumulate near zero momentum,} as it is discussed in Sec.~\ref{lowerB}.

\end{enumerate}
\clearpage
\appendix
\chapter{Appendices for chapter \ref{MMBE}}

\section{Averaged collision integral}\label{app1MMBE}
In the following we discuss the averaging and the approximations we used calculating $\mathcal{I}^\gamma$ and $\mathcal{I}$ -- see Eq.~(\ref{avrCollIntMMBE}) --, which results the formula (\ref{averaging}) in Sec. \ref{motivationMMBE}. Using compact notations, the 9-dimensional collision integral reads as:
\begin{eqnarray}
\mathcal{I}^\gamma &=& \int_{234}\!\!\int\!\!\mathrm{d}\omega\rho^\gamma(\omega -E_2)\delta(E_1+\omega -E_3-E_4)\times\nonumber \\
& &\times\delta^{(3)}(\ve{p}_3+\ve{p}_4-\ve{p}_1-\ve{p}_2)w_{1234}(f_3f_4-f_1f_2) \nonumber\\
&=& \int_{234}\!\!\rho^\gamma(E_3+E_4-E_1-E_2)\mathcal{K}_{1234},
\end{eqnarray}
where we introduced ${\mathcal{K}_{1234}:=w_{1234}\delta^{(3)}(\ve{p}_3+\ve{p}_4-\ve{p}_1-\ve{p}_2)(f_3f_4-f_1f_2)}$. Now we perform the averaging according to $\gamma$. First we assume the interchangeability of the integrations respect to $\gamma$ and to the phase-space coordinates: \LNNL
\begin{align}
\mathcal{I} &= \llangle\mathcal{I}^\gamma\rrangle= \int\limits_0^\infty\!\!\mathrm{d}\gamma g(\gamma)\mathcal{I}^\gamma =\int_{234}\!\!\mathcal{K}_{1234} \int_0^\infty\!\!\mathrm{d}\gamma g(\gamma)\rho^\gamma(E_3+E_4-E_1-E_2) = \label{1stlineMMBE}\\
&= \int_{234}\!\!\mathcal{K}_{1234} \cdot\llangle\rho^\gamma\rrangle(E_2-\Omega).
\end{align}
Let us suppose the function $\llangle\rho^\gamma\rrangle$ to be a density function with the properties $\int_0^\infty\!\!\mathrm{d}\omega \llangle\rho^\gamma\rrangle(\omega)=1$ and $\int_0^\infty\!\!\mathrm{d}\omega \llangle\rho^\gamma\rrangle(\omega)\omega=\Delta$. Furthermore, let its higher moments be negligible compared to $\Delta\gg \int_0^\infty\!\!\mathrm{d}\omega \llangle\rho^\gamma\rrangle(\omega)\omega^n$ $(n>1)$. These imply that $\llangle\rho^\gamma\rrangle$ is ''peaky'' around $\Delta$. Therefore in Eq.~(\ref{2ndlineMMBE}) we expand the rest of the integral kernel in $E_2$ around $\Omega+\Delta$. After integrating respect to $E_2$, the first term gives the kernel $\mathcal{K}$ evaluated in $E_2=\Omega+\Delta$, the second vanishes.
\begin{align}
\mathcal{I} & \approx \int_{234}\!\!\left\{\left.\mathcal{K}_{1234}\right|_{E_2=\Omega+\Delta} +\partial_{E_2}\left.\mathcal{K}_{1234}\right|_{E_2=\Omega+\Delta}(E_2-\Omega-\Delta)+ \mathcal{O}((E_2-\Omega-\Delta)^2) \right\}\llangle\rho^\gamma\rrangle = \label{2ndlineMMBE}\\
&= \int_{234}\!\!\delta(E_3+E_4-E_1-E_2-\Delta)\mathcal{K}_{1234} + \left\{ \,\,\mathrm{terms\,proportional\,to\,higher\,moments\,of}\,\gamma\,\, \right\}. \label{3rdlineMMBE}
\end{align}
The remaining terms are at least $\mathcal{O}(\Delta^2)$. In the last line (Eq.~(\ref{3rdlineMMBE})) we reimplemented the effect of integrating respect to $E_2$ as resulted by a modified constraint for the kinetic energies.

\newpage
\section{Rejection sampling}\label{rejMethMMBE}
First of all, we briefly discuss how to sample random numbers according to the probability density function (pdf) $f$, ${\intlim{x}{-\infty}{\infty}f(x)=1}$. Suppose, that we are able to generate random numbers in the interval ${[0,1]}$ with uniform distribution. If ${Y\sim \text{Uniform}[0,1]}$, the random variable ${X=F^{-1}(Y)}$ has the cumulative distribution function (cdf) ${F(x)=\intlim{y}{-\infty}{x}f(x)}$:\LNNL
\begin{align}
\mathbb{P}(X\leq x) &= \mathbb{P}(F^{-1}(Y) \leq x) =\mathbb{P}(Y\leq F(x)) =\intlim{y}{0}{F(x)} =F(x).
\end{align}
This is called the inverse sampling method. Now, for problems when the inverse of $F$ is not available in a closed algebraic form, we can use the so called rejection sampling method. Firstly, a majorizing function $g$ to the pdf $f$ is to be found: ${f(x)\leq g(x)}$, in such a way, that the corresponding cdf $F^*$ of the density ${f^*(x):=g(x)/\intlim{y}{-\infty}{\infty}g(y)}$ is to be invertible. Then, the method consists the following steps:
\begin{enumerate}[\it i)]
\item\label{it1rejMMBE} sample a random number $X$ according to $F^*$
\item\label{it2rejMMBE} sample an other random, uniformly: $Y\sim\text{Uniform}[0,1]$
\item\label{it3rejMMBE} if $Y\leq\frac{f(X)}{g(X)}$, then we stop with $Z:=X$, otherwise reject $X$ and go back to \textit{\ref{it1rejMMBE})}.
\end{enumerate}
The cdf of $Z$ is the following:\LNNL
\begin{align}
\mathbb{P}(Z\leq z) &= \sum\limits_{x\leq z}\mathbb{P}(X\leq z\,|\,Y\leq f(x)/g(x))\cdot\mathbb{P}(Y\leq f(x)/g(x)) =\frac{\intlim{x}{-\infty}{z}f^*(x)\intlim{y}{0}{\frac{f(x)}{g(x)}}}{\intlim{x}{-\infty}{\infty}f^*(x)\intlim{y}{0}{\frac{f(x)}{g(x)}}} =\nonumber \\
&= \frac{\intlim{x}{-\infty}{z}g(x)\frac{f(x)}{g(x)}}{\intlim{x}{-\infty}{\infty}g(x)\frac{f(x)}{g(x)}} =\intlim{x}{-\infty}{z}f(x) =F(z),
\end{align}
which is the cdf of the desired random number.

\newpage
\section{Scaling symmetry}\label{app2MMBE}
In this section we show the existence of a special family of solutions of Eq.~(\ref{kinequ3}) of Sec.~\ref{longtimeMMBE} with a scaling property and the joint pattern of the time evolution. For this purpose, we assume that the function $f^\alpha(E,t_0)$ is a solution of Eq. (\ref{kinequ3}). Then, using the properties of the collision integral, we prove that $\avr{E}^{3/2}f^\alpha(\avr{E}E,t_0)$ is also a solution with $\avr{E}=\avr{E}_0e^{\gamma(t-t_0)}$ for arbitrary $t$ with the time-independent constant $\gamma$. We do not investigate here the stability of the scaling solution, only mention that it is indicated by our numerical simulations. \par
First we rewrite the MMBE (\ref{kinequ3}) with energy variables $E'=\frac{(p')^2}{2m}$, $\epsilon=\frac{q^2}{2m}$:\LNNL
\begin{align} \label{kinequ4}
\partial_{t_0} f^\alpha(E,t_0) =& \sum_\beta N^{\alpha\beta}\int_0^1\!\!\mathrm{d}x \int_0^1\!\!\mathrm{d}y \int_0^\infty\!\!\mathrm{d}E' \int_0^\infty\!\!\mathrm{d}\epsilon \sqrt{E'\epsilon}w_{\avr{E}_0}^{\alpha\beta}(E,E',x)\times\nonumber \\
&\times\delta(K^{\alpha\beta}_{\avr{E}_0}(E,E')-\epsilon\oplus^{\alpha\beta}_{\avr{E}_0}E^*(E,E',\epsilon,x,y)) \times \nonumber \\
& \times (f^\alpha(\epsilon,t_0)f^\beta(E^*,t_0)-f^\alpha(E,t_0) f^\beta(E',t_0)) =: \sum_\beta\mathcal{I}_{\avr{E}_0}^{\alpha\beta}[f^{\alpha,\beta}(\,.\,,t_0)](E).
\end{align}
We put an $\avr{E}$ index on every quantity which depends on the modification parameter and therefore on $\avr{E}$. $N^{\alpha\beta}$ is a constant (from the spherical integration and the dispersion relation). The expressions for $E^*$ and $P$ are
\begin{eqnarray}
E^*=E(|\ve{P}-\ve{q}|)=\frac{1}{2m}(P^2+2m\epsilon-2P\sqrt{2m\epsilon}y), & \,\,\,\, & P=\sqrt{2m}\sqrt{E+E'+2EE'x}, \nonumber
\end{eqnarray}
$x$ and $y$ being the cosine of the angle between $\ve{P}$, $\ve{p'}$ and $\ve{P}$, $\ve{q}$, respectively.\par
Using the identity $$E\oplus^{\alpha\beta}_{\avr{E}} E'=\avr{E}\left(\frac{E}{\avr{E}}\oplus^{\alpha\beta}_{\avr{E}}\frac{E'}{\avr{E}}\right)$$ we get the following scaling relation for the rate function: $$w^{\alpha\beta}_{\avr{E}}(E,E',x)= w^{\alpha\beta}_1\left(\frac{E}{\avr{E}},\frac{E'}{\avr{E}},x\right)/\sqrt{\avr{E}}.$$ Changing the variables of the integration, the following holds ($\avr{E}=\avr{E}(t)$, $\avr{E}_0=\avr{E}(t_0)$) for the collision integral ${\mathcal{I}^{\alpha\beta}_{\avr{E}_0}[f]}$ as a functional of the density function $f$:
\begin{equation} \label{scalingI}
\left(\frac{\avr{E}_0}{\avr{E}}\right)^{\frac{3}{2}}\mathcal{I}^{\alpha\beta}_{\avr{E}_0}[f^{\alpha,\beta}(\, .\,,t_0)](E) = \mathcal{I}^{\alpha\beta}_{\avr{E}}\left[ \left(\frac{\avr{E}_0}{\avr{E}}\right)^{\frac{3}{2}}f^{\alpha,\beta}\left(\frac{\avr{E}_0}{\avr{E}}(\, .\,),t_0\right) \right]\left(\frac{\avr{E}}{\avr{E}_0}E \right).
\end{equation}
One takes then Eq. (\ref{kinequ4}) and multiplies it by the factor $(\frac{\avr{E}_0}{\avr{E}})^{\frac{3}{2}}$. Using (\ref{scalingI}) one arrives at\LNNL
\begin{align} \label{scalingSol}
\left(\frac{\avr{E}_0}{\avr{E}}\right)^{\frac{3}{2}}\partial_{t_0}f^{\alpha}(E,t) = & \sum_\beta \mathcal{I}^{\alpha\beta}_{\avr{E}}\left[ \left(\frac{\avr{E}_0}{\avr{E}}\right)^{\frac{3}{2}} f^{\alpha,\beta}\left(\frac{\avr{E}_0}{\avr{E}}(\, .\,),t_0\right) \right]\left(\frac{\avr{E}}{\avr{E}_0}E \right) =: \nonumber \\
=: &\sum_\beta \mathcal{I}^{\alpha\beta}_{\avr{E}}[ f^{\alpha,\beta}((\, .\,),t) ]\left(\frac{\avr{E}}{\avr{E}_0}E \right) = \partial_{t}f^{\alpha}(\avr{E}\avr{E}_0^{-1}E,t).
\end{align}
One realizes, that the first equality in Eq.~(\ref{scalingSol}) is the MMBE (\ref{kinequ4}) with rescaled energy-argument. Eq.~(\ref{scalingSol}) demands the following scaling relation to hold:
\begin{equation} \label{scalingDF0}
f^\alpha(E,t)=\left(\frac{\avr{E}_0}{\avr{E}} \right)^{\frac{3}{2}} f^\alpha\left(\frac{\avr{E}_0}{\avr{E}} E,t\right).
\end{equation}
In other words, the rescaling of the energy variable means a finite step of the time-evolution. We have to check the conditions of the equivalence of Eq.~(\ref{scalingDF0}) and Eq.~(\ref{scalingSol}). Let us differentiate Eq.~(\ref{scalingDF0}) with respect to $t'$:
\begin{eqnarray} \label{checkscaling}
\partial_{t} f^\alpha\left(\frac{\avr{E}_0}{\avr{E}}E,t\right) &=& f^\alpha(E,t)\partial_{t}\left[\left(\frac{\avr{E}_0}{\avr{E}}\right)^{\frac{3}{2}}\right] +\left(\frac{\avr{E}_0}{\avr{E}}\right)^{\frac{3}{2}} \frac{\mathrm{d}t_0}{\mathrm{d}t}\partial_{t_0} f^\alpha(E,t_0) = \nonumber \\
&=& \frac{3}{2}\left(\frac{\avr{E}_0}{\avr{E}}\right)^{\frac{3}{2}}\left(\frac{\partial_{t_0}\avr{E}_0}{\avr{E}_0}\frac{\mathrm{d}t_0}{\mathrm{d}t}-\frac{\partial_{t}\avr{E}}{\avr{E}}\right)f^\alpha(E,t) +\left(\frac{\avr{E}_0}{\avr{E}}\right)^{\frac{3}{2}} \frac{\mathrm{d}t_0}{\mathrm{d}t}\partial_{t_0} f^\alpha(E,t_0) = \nonumber \\
&=& \left(\frac{\avr{E}_0}{\avr{E}}\right)^{\frac{3}{2}} \partial_{t_0} f^\alpha(E,t).
\end{eqnarray}
The last equality in Eq.~(\ref{checkscaling}) holds when $\frac{\mathrm{d}t_0}{\mathrm{d}t} \equiv 1$. and $\partial_t\avr{E}/\avr{E} \equiv \mathrm{const.}$ In summary, the relation in Eq.~(\ref{scalingDF0}) holds if $t=t_0+\Delta t$, with arbitrary constants $t_0$, $\Delta t$ and $\gamma$.\par
In conclusion:
\begin{eqnarray} \label{scalingDF}
f^\alpha(E,t) =\frac{c}{\avr{E}^{3/2}(t)}\psi^\alpha\left(\frac{E}{\avr{E}(t)} \right), & & \avr{E}(t)=\avr{E}_0 e^{\gamma\Delta t},\end{eqnarray}
where $t_0$ is used as reference time, and ${\psi^\alpha(x)=\frac{1}{c}\avr{E}_0^{3/2}f^\alpha(\avr{E}_0 x,t_0)}$. Now computing $\avr{E}$ serves as a consistency relation, which gives us the value of $c$: $$c =\frac{1}{\sum\limits_\alpha 4\pi\sqrt{2}m^{3/2}\intlim{x}{0}{\infty}x^{3/2}\psi^\alpha(x)}.$$

\newpage
\section{Time evolution in laboratory time}\label{app3MMBE}
In this study in chapter~\ref{MMBE} we used the collision frequency to follow the evolution of the density function of the system and to analyse the long-time behaviour of the total kinetic energy per particle $\avr{E}$. Still, the relation of the collision counter $t$ to the laboratory time $t_\mathrm{lab}$ needs clarification. The quantity which gives exactly the probability of a (given kind of) collision in a unit \textit{laboratory} time is nothing but $w^{\alpha\beta}$. So the expected number of collisions in a unit time in a given state of the system is
\begin{eqnarray}
\frac{\mathrm{d}t}{\mathrm{d}t_\mathrm{lab}} &=& \sum_{\alpha\beta}\int\!\!\mathrm{d}^3\ve{p}\int\!\!\mathrm{d}^3\ve{p}'w^{\alpha\beta}(\ve{p},\ve{p}')f^\alpha(E(p))f^\beta(E(p')) \sim \nonumber \\
&\sim & \avr{E}^{-\frac{1}{2}}(t) \int_0^\infty\!\!\mathrm{d}\omega\sqrt{\omega} \int_0^\infty\!\!\mathrm{d}\eta\sqrt{\eta} \int_0^1\!\!\mathrm{d}x w^{\alpha\beta}_1(\omega,\eta,x)\psi(\omega)\psi(\eta),
\end{eqnarray}
for scaling solutions described by Eq.~(\ref{numscalingSOL}). This proportionality leads us to the following separable ODE:
\begin{equation}
\frac{\mathrm{d}t}{\mathrm{d}t_\mathrm{lab}}=\Gamma \avr{E}^{-\frac{1}{2}}(t)=\Gamma\avr{E}_0^{-\frac{1}{2}}e^{-\frac{\gamma}{2}t}
\end{equation}
Here, $\Gamma$ is constant in time, it is however a fairly difficult task to determine it. One has to solve an integro-differential equation for the shape function $\psi(x)$ to get $\Gamma$. But if its value is assumed to be known, the collision time expressed by the laboratory time -- with the initial condition $t_\mathrm{lab}(t=0)=0$ -- reads as:
\begin{equation}\label{colltime_labtime}
t=\frac{2}{\gamma}\ln\left(1+\frac{\Gamma\gamma}{2\avr{E}_0^{1/2}}t_\mathrm{lab}\right).
\end{equation}
Using the expression of Eq.~(\ref{colltime_labtime}), the kinetic energy per particle in laboratory time is
\begin{equation}\label{en_density_labtime}
\avr{E}(t_\mathrm{lab})= \avr{E}_0\left(1+\frac{\Gamma\gamma}{2\avr{E}_0^{1/2}}t_\mathrm{lab}\right)^2.
\end{equation}
$\avr{E}$ grows limitless as $\sim t_\mathrm{lab}^2$ for $\gamma>0$. Let us remark, that for $t_{\mathrm{lab}} \gg \frac{2\avr{E}_0^{1/2}}{\Gamma\gamma}$, $\avr{E}(t_{\mathrm{lab}}) \sim (\Gamma\gamma t_{\mathrm{lab}})^2$ is $\avr{E}_0$-independent. \par
For $\gamma<0$, $\avr{E}$ reaches zero in a finite time in the laboratory system: $t_\mathrm{lab}^*=\frac{2\avr{E}_0^{1/2}}{\Gamma|\gamma|}$. It is interesting, that the number of collisions tends to infinity as $t_\mathrm{lab} \rightarrow t_\mathrm{lab}^*$ in this case. This indicates the presence of the so-called inelastic collapse in the context of non-elastic Boltzmann equation \cite{inelcoll}.
\clearpage
\chapter{Appendices for chapter \ref{etaOs}}
\subsection*{Notations}
Throughout this appendices, the lower index for a space-time or momentum-space dependent quantity means its argument, i.e.\ a four-vector: \LNNL
\begin{align}
\varphi_x\equiv\varphi(x)\equiv\varphi(x^0,\ve{x}).\nonumber
\end{align}
Also an integral sign with lower indexed variable of integration is prescribed on the whole domain of the variable (space or momentum-space):\LNNL
\begin{align}
\int_p(\dots) &= \frac{1}{(2\pi)^4}\intlim{p^0}{-\infty}{\infty}\intlim{^3\ve{p}}{}{}(\dots),\,\, \text{in momentum space,} \nonumber \\
\int_x(\dots) &= \intlim{x^0}{-\infty}{\infty}\intlim{^3\ve{x}}{}{}(\dots),\,\, \text{in space.} \nonumber
\end{align}
For the $\delta$-distributions: ${\int_x\delta_x=1}$, ${\int_p\delta_p=1}$ both in real- and momentum-space.

\section{Propagators in the Keldysh-formalism}\label{propagators}
\begin{figure}
\centering
\includegraphics[width=0.6\linewidth]{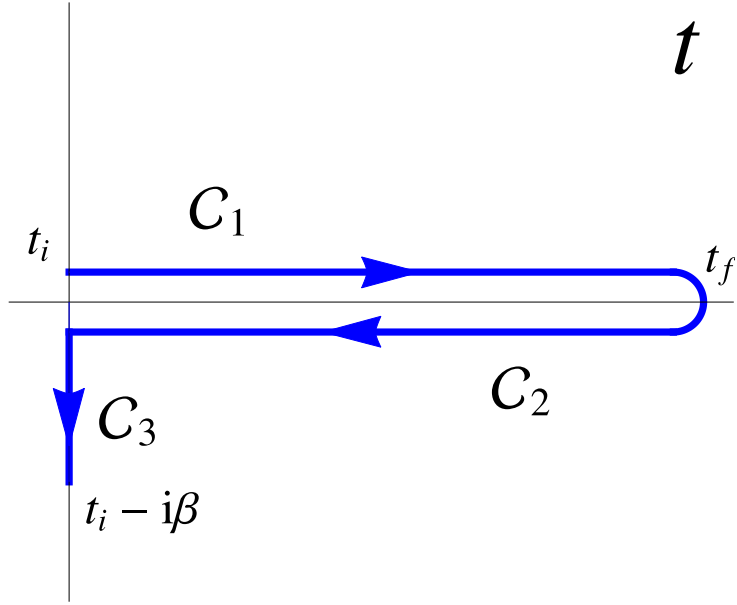}
\LNFIG
\caption{Keldysh-contour $\mathcal{C}$ on the complex time-plane for real-time expectation values of an evolution with initial time $t_i$ an final time $t_f$, connected to a heat bath with inverse temperature $\beta$.}
\label{fig:KeldyshContour}
\end{figure}
The Keldysh-formalism is a useful and elegant way to unify initial value problems, i.e.\! the real-time evolution of the correlation functions, and also thermodynamical averaging (i.e.\! imaginary-time evolution) on the level of the generating functional. It introduces a ''book-keeping'' of field operators by the finite time interval between the argument of those. Using a so-called time-contour on which all the field operators are trivially ordered as the parameter of the curve increases, the time-ordering, anti time-ordering and the imaginary-time ordering of the fields is represented respectively by the segments $\mathcal{C}_1$, $\mathcal{C}_2$ and $\mathcal{C}_3$ of the curve $\mathcal{C}$, see Fig.\ref{fig:KeldyshContour}. With this grouping of the field operators, the path integral representation of the generating functional is the following: \LNNL
\begin{align}
\mathcal{Z}[J] &=\mathcal{Z}_0\avr{e^{\int_pJ_p\varphi_p}} =\text{Tr}\left(e^{-\beta T^{00}}e^{\int_pJ_p\varphi_p}\right) =\\
&= \Int\!\!\mathcal{D}\varphi e^{iS[\varphi^{(1)}]-iS[\varphi^{(2)}]-S_E[\varphi^{(3)}]+\int_pJ_p\varphi_p},
\end{align}
with the compact notation ${J=(J^{(1)},\,J^{(2)},\,J^{(3)})}$ for the source field $J$. $S[\varphi]$ is the classical action of the system, whilst $S_E$ its Euclidean (or Wick-rotated) counterpart. For the derivation and further details, please see Refs.~\cite{JakoBook, keldysh_ctp} and the references therein. On can get the corresponding correlation functions by acting with ${\frac{\delta}{\delta J^{(i)}}}$ on ${\mathcal{Z}[J]}$. Therefore, the Keldysh-propagator is defined as $i\mathcal{G}_{x,y}^{ab}=\avr{\mathcal{T_C}\kelphid{x}{a}\kelphi{y}{b}} \equiv \avr{\kelphid{x}{a}\kelphi{y}{b}}$. From now on we omit $\mathcal{T_C}$ which represents the time-ordering on the Keldysh-contour $\mathcal{C}$. The adjoint sign is also suppressed, since in the case of real scalar fields, it is equivalent with the identification $\varphi^\dagger_p=\varphi_{-p}$. We briefly summarize here the relations between the components of the Keldysh-propagator and expectation values that will be useful later on. \par
In general, the following identities hold:\LNNL
\begin{align}
i\mathcal{G}_{x,y}^{11} &=\theta(x^0-y^0)i\mathcal{G}^{21}_{x,y}+\theta(y^0-x^0)i\mathcal{G}^{12}_{x,y}, \\
i\mathcal{G}_{x,y}^{22} &=\theta(x^0-y^0)i\mathcal{G}^{12}_{x,y}+\theta(y^0-x^0)i\mathcal{G}^{21}_{x,y}, \\
0 &=\mathcal{G}^{12}_{x,y}+\mathcal{G}^{21}_{x,y}-\mathcal{G}^{11}_{x,y}-\mathcal{G}^{22}_{x,y},\\
\rho_{x,y} &=i\mathcal{G}^{21}_{x,y}-i\mathcal{G}^{12}_{x,y}. \label{spectFuncDEF}
\end{align}
In thermal equilibrium, the propagators are translational invariant, thus for the Fourier-transformed ones ${iG_{p,q}^{ab} \equiv \delta_{p-q}iG_p^{ab}}$. We denote the Fourier-transform of a space-time dependent quantity calligraphic $\mathcal{G}$ with an italic $G$. \par
Now let us consider the two-point function of two arbitrary operators $A$ and $B$. The exponential form of the statistical operator allows us the following manipulations: \LNNL
\begin{align}
i\mathcal{G}_{AB}^{12}(t) &= \frac{1}{\mathcal{Z}_0}\text{Tr}\left(e^{-\beta T^{00}} B(0)A(t)\right) =\frac{1}{\mathcal{Z}_0}\text{Tr}\left(e^{-\beta T^{00}} e^{\beta T^{00}}A(t)e^{-\beta T^{00}}B(0)\right) =\nonumber \\
&= \frac{1}{\mathcal{Z}_0}\text{Tr}\left(e^{-\beta T^{00}}A(t-i\beta)B(0)\right) =i\mathcal{G}_{AB}^{21}(t-i\beta),
\end{align}
where we also used that the trace of cyclic permutations is the same. After Fourier transform we get:\LNNL
\begin{align}
iG_{AB}^{12}(\omega) &= \intlim{t}{-\infty}{\infty}e^{i\omega t}i\mathcal{G}_{AB}^{21}(t-i\beta) =e^{-\beta\omega}iG_{AB}^{21}(\omega).
\end{align}
This is the so-called Kubo-Martin-Schwinger (KMS) relation, which connects the otherwise independent propagators $G^{12}_{AB}$ and $G^{21}_{AB}$ in thermal equilibrium. With the definition (\ref{spectFuncDEF}) of the spectral function, it results in:
\[\begin{array}{ccccc}
iG^{12}_p=n_p\rho_p, & & iG^{21}_p=(n_p+1)\rho_p, &\text{with } & n_p=\frac{1}{e^{\omega/T}-1}.
\end{array}\]
The following parity-relations hold:
\[\begin{array}{ccccccc}
iG^{12}_{-p}=iG^{21}_p, & & iG^{11}_{-p}=iG^{11}_p, & & iG^{22}_{-p}=iG^{22}_p, & & \rho_{-p} =-\rho_p.
\end{array}\]
In the special case of a quadratic action, like the one in Eq.~(\ref{EQPactionEtaOs}), the generating functional reads as follows: \LNNL
\begin{align}
\mathcal{Z}[J] &= \mathcal{Z}_0e^{\frac{i}{2}\int_pJ_{-p}^TG_pJ_p},
\end{align}
with the propagator in matrix-form:\LNNL
\begin{align}
iG_p &= \MX{iG^{11}_p}{iG^{12}_p}{iG^{21}_p}{iG^{22}_p} = \,\,(\text{in thermal equilibrium})\,\, =\nonumber \\
&= \intlim{\omega}{-\infty}{\infty}\frac{\rho(\omega,\ve{p})}{p^0-\omega+i0^+}\MX{1}{0}{0}{-1} +\rho_pn_p\MX{1}{1}{1}{1} +\rho_p\MX{0}{0}{1}{1}
\end{align}
Wick's theorem holds with the Keldysh-indices signed properly. We need the four-point function for the viscosity calculation:\LNNL
\begin{align}
\avr{\kelphi{p}{a}\kelphi{q}{b}\kelphi{r}{c}\kelphi{s}{d}} =& \avr{\kelphi{p}{a}\kelphi{q}{b}}\avr{\kelphi{r}{c}\kelphi{s}{d}}
+\avr{\kelphi{p}{a}\kelphi{r}{c}}\avr{\kelphi{q}{b}\kelphi{s}{d}}
+\avr{\kelphi{p}{a}\kelphi{s}{d}}\avr{\kelphi{q}{b}\kelphi{r}{c}}. \label{fourpiontfunc}
\end{align}
It is often convenient working with the Keldysh-propagator in the so-called R/A-basis, which means the linear transformation of the fields into the advanced field ${\varphi^{(a)}:=\varphi^{(1)}-\varphi^{(2)}}$ and the retarded field ${\varphi^{(r)}:=\frac{1}{2}\left(\varphi^{(1)}+\varphi^{(2)}\right)}$. The propagator is then the following in matrix-notation: \LNNL
\begin{align}
G^\text{R/A} &= \left(\begin{array}{cc}
G^{rr} & G^{ra} \\ G^{ar} & 0
\end{array}\right),
\end{align}
where the retarded ($ra$) and advanced ($ar$) propagator components in the momentum space can be expressed as\LNNL
\begin{align}
iG_p^{ra/ar} &= \intlim{\omega}{-\infty}{\infty}\frac{\rho(\omega,\ve{p})}{p^0-\omega\pm i0^+},
\end{align}
and the third non-zero component in thermal equilibrium:\LNNL
\begin{align}
iG^{rr}_p &=\left(n(p^0/T)+\frac{1}{2}\right)\rho_p.
\end{align}

\newpage
\section{Kubo's formula}\label{linresAppEtaOs}
In this section we briefly summarize the concept of the response to an ''external'' perturbation, to linear order in its strength. On the level of the generating functional of a QFT, it means an additional source term, a classical field coupled linearly to a fundamental or a composite operator of the theory. We work out here the response of the average of $B$ to the perturbation ${\mathcal{A}[h]}$, which vanishes for ${h=0}$. We assume local thermal equilibrium, when ${h\equiv 0}$. The term \textit{linear} means that we expand the generating functional around the unperturbed one, ${h\equiv 0}$, to linear order in ${\delta S=\int_y\mathcal{A}_y}$:\LNNL
\begin{align}
\left.\avr{B}\right|_{h}(t,\ve{x}) &= \int\!\!\mathcal{D}\varphi B_x^{(1)}e^{i(S[\varphi^{(1)}]-S[\varphi^{(2)}])+i(\delta S[\varphi^{(1)}]-\delta S[\varphi^{(2)}])} =\\
&=\int\!\!\mathcal{D}\varphi B_x^{(1)}e^{i(S[\varphi^{(1)}]-S[\varphi^{(2)}])+i\int_y(\mathcal{A}_y^{(1)}-\mathcal{A}_y^{(2)})} =\\
&=\sum\limits_{n=0}^\infty\frac{i^n}{n!}\int\!\!\mathcal{D}\varphi B_x^{(1)}e^{i(S[\varphi^{(1)}]-S[\varphi^{(2)}])}\left(\int_y(\mathcal{A}_y^{(1)}-\mathcal{A}_y^{(2)})\right)^n \approx\\
&\approx \left.\avr{B_x}\right|_{h\equiv 0} +i\int_y\underbrace{\left(\left.\avr{B_x^{(1)}\mathcal{A}_y^{(1)}}\right|_{h\equiv 0} -\left.\avr{B_x^{(1)}\mathcal{A}_y^{(2)}}\right|_{h\equiv 0}\right)}_{= \theta(x^0-y^0)\left.\avr{[B_x,\mathcal{A}_y]}\right|_{h\equiv 0}}.
\end{align}
The Keldysh-indices of the operators are explicitly written out. Now we take $\mathcal{A}$ as a linear functional of the classical field $h$ and an operator $A$. The \textit{linear response} $~{\delta\avr{B}}$ is then obtained by the spectral function $\rho_{BA}$ as follows:\LNNL
\begin{align}
\mathcal{A}_x &= \intlim{\tau}{-\infty}{t}A(\tau,\ve{x})h(\tau,\ve{x}), \\
\delta\avr{B_x}[h] &:= \left.\avr{B_x}\right|_{h} -\left.\avr{B_x}\right|_{h\equiv 0} \approx\\
&\approx i\intlim{t'}{-\infty}{t}\intlim{\tau}{-\infty}{t'}\intlim{^3\ve{y}}{}{} \underbrace{\left.\avr{[B,A]}\right|_{h\equiv 0}(t-\tau,\ve{x}-\ve{y})}_{=:\,\rho_{BA}(t-\tau,\ve{x}-\ve{y})} h(\tau,\ve{y}).
\end{align}
After Fourier transform, the response $\delta\avr{B}$ simplifies further:\LNNL
\begin{align}
\left.\delta\avr{B}\right|_{h}(k_0,\ve{k}) &:= \intlim{t}{-\infty}{\infty}\intlim{^3\ve{x}}{}{} e^{itk_0-i\ve{k}\cdot\ve{x}}\left.\delta\avr{B}\right|_{h}(t,\ve{x}) = \nonumber\\
&= \frac{i}{(2\pi)^8}\intlim{\Omega}{-\infty}{\infty}\intlim{\omega}{-\infty}{\infty} \intlim{^3\ve{p}}{}{}\intlim{^3\ve{q}}{}{} h(\Omega,\ve{q})\rho_{BA}(\omega,\ve{p})\times \nonumber \\
&\times \underbrace{\intlim{^3\ve{x}}{}{}\intlim{^3\ve{y}}{}{} e^{i\ve{x}\cdot(\ve{p}-\ve{k})}e^{i\ve{y}\cdot(\ve{q}-\ve{p})}}_{=(2\pi)^6\delta(\ve{p}-\ve{k})\delta(\ve{q}-\ve{p})} \underbrace{\intlim{t}{-\infty}{\infty}e^{it(k^0-\omega)}\intlim{t'}{-\infty}{t}\intlim{\tau}{-\infty}{t'}e^{i\tau(\omega-\Omega)}}_{=-\frac{2\pi\delta(k^0-\Omega)}{(\omega-\Omega)^2}}. \label{KuboDeriv1}
\end{align}\LNNL
The time-integrals in Eq.~(\ref{KuboDeriv1}) can be evaluated, which leads to a spectral representation of the Fourier-transformed linear response:
\begin{align}
\left.\delta\avr{B}\right|_{h}(k_0,\ve{k}) &= -\frac{i}{2\pi}h(k^0,\ve{k})\intlim{\omega}{-\infty}{\infty}\frac{\rho_{BA}(\omega,\ve{k})}{(k^0-\omega+i\epsilon)^2}.
\end{align}
The limit ${k\rightarrow 0}$ specifies the perturbation to be homogeneous in space and slow compared to the microscopic time-scales of the QFT:
\begin{align}
\left.\delta\avr{B}\right|_{h}(k_0,\ve{k}) &\overset{k=0}{=} -\frac{i}{2\pi}\frac{h(k^0=0,\ve{k}=0)}{2}\intlim{\omega}{-\infty}{\infty}\rho_{BA}(\omega,\ve{k}=0)\left(\frac{1}{(\omega-i\epsilon)^2} -\frac{1}{(\omega+i\epsilon)^2}\right) =\label{KuboDeriv3}\\
&= -\frac{i}{2\pi}\frac{h(k=0)}{2}\intlim{\omega}{-\infty}{\infty}\rho_{BA}(\omega,\ve{k}=0)\underbrace{\frac{2i\epsilon}{\omega^2+\epsilon^2}}_{\overset{\epsilon\rightarrow 0}{\longrightarrow}\, 2\pi i\delta(\omega)}\frac{2\omega}{\omega^2+\epsilon^2} \\
&\overset{\epsilon\rightarrow 0}{\longrightarrow} h(k=0)\lim\limits_{\omega\rightarrow 0}\frac{\rho_{BA}(\omega,\ve{k}=0)}{\omega}. \label{KuboDeriv4}
\end{align}
After using in Eq.~(\ref{KuboDeriv3}) that ${\rho_{BA}(\omega)}$ is an odd function, we are finally left with \textit{Kubo's formula} in Eq.~(\ref{KuboDeriv4}).

\subsection*{Hydrodynamical transport coefficients}
As an application, we are going to use the result Eq.~(\ref{KuboDeriv4}) to define the transport coefficients of the relativistic hydrodynamics through the spectral function $\rho_{TT}$, where $T^{\mu\nu}$ is the energy-momentum tensor of the QFT. In Sec.~\ref{linResHydro} we already motivated the linear perturbation $\delta S$, which acts as a source of heat and momentum currents:\LNNL
\FINAL{
\begin{align}
\delta S &= \intlim{^3\ve{x}}{}{} (U_\mu-\overline{u}_\mu) T^{0\mu}.
\end{align}}
The flow velocity in the rest frame of the fluid is ${\overline{u}=(1,0,0,0)}$. The four-vector of the source -- or the strength of the flow-perturbation -- is \FINAL{${U=(1+\delta c)u}$}, and the heat source is linked to the energy density as follows: $\delta c=\frac{\partial\ln \varepsilon}{\partial T}\delta T$, through the inhomogeneous temperature field. Taking the time-derivative of $\delta S$, then using the energy-momentum conservation $~{\partial_\nu T^{\mu\nu}=0}$ one arrives at
\begin{align}
\partial_0\delta S &= \intlim{^3\ve{x}}{}{} [\left(\partial_0 U_\mu\right)T^{0\mu}+(U_\mu-\overline{u}_\mu)\underbrace{\partial_0 T^{0\mu}}_{=-\partial_i T^{i\mu}}], \\
\delta S&=\intlim{\tau}{-\infty}{t}\intlim{^3\ve{x}}{}{} T^{\mu\nu}\partial_{[\mu}U_{\nu]}.
\end{align}
${\partial_{[\mu}U_{\nu]}=(\partial_\mu U_\nu +\partial_\nu U_\mu)/2}$ denotes the symmetric four-derivative. To determine the linear response coefficients, we put $\delta S$ into Kubo's formula:\LNNL
\begin{align}
\delta\avr{T^{\mu\nu}} &= \lim\limits_{\omega\rightarrow 0}\frac{\avr{[T^{\mu\nu},T^{\rho\sigma}]}(\omega,\ve{k}=0)}{\omega}\partial_{[\rho}U_{\sigma]}.
\end{align}
One has to decompose the combination ${[T^{\mu\nu},T^{\rho\sigma}]\partial_{[\rho}U_{\sigma]}}$ in order to identify the transport coefficients. We only impart the result of the analysis here\footnote{The details of this tedious, but otherwise straightforward calculation can be easily reproduced following Ref.~\cite{linres_hydro}.}. Splitting the energy-momentum tensor according to\LNNL
\begin{align}
T^{\mu\nu} &= \varepsilon u^\mu u^\nu -P\Delta^{\mu\nu} +2Q^{[\mu}u^{\nu]} +\pi^{\mu\nu},
\end{align}
where the tensorial components are
\[\begin{array}{ccc}
\varepsilon= u_\mu u_\nu T^{\mu\nu}, & & P=-\frac{1}{3}\Delta_{\mu\nu}T^{\mu\nu}, \\
Q_\mu =\Delta_{\mu\rho}u_\sigma T^{\rho\sigma}, & & \pi_{\mu\nu}= \left(\Delta_{\mu\rho}\Delta_{\nu\sigma}-\frac{1}{3}\Delta_{\mu\nu}\Delta_{\rho\sigma}\right)T^{\rho\sigma},
\end{array}\]
with the energy density $\varepsilon$, pressure $P$, heat current $Q_\mu$ and the traceless viscous stress-tensor $\pi_{\mu\nu}$. The projector ${\Delta^{\mu\nu}=g^{\mu\nu}-u^\mu u^\nu}$ projects to the plane orthogonal to $u$.\par
After the decomposition, we get the following tensor, as the source of the perturbations in linear order in the rest frame of the fluid\footnote{Where we also used the relations of an ideal fluid in thermal equilibrium, to express the heat currents to linear order in the perturbation, see Ref.~\cite{linres_hydro}.}:\LNNL
\begin{align}
T^{\mu\nu}\partial_{[\mu}U_{\nu]} &\overset{\text{rest frame}}{=} P'\partial_iu^i +T^{0i}(\partial_0 u_i +\partial_i\delta c) +\pi^{ij}\left(\partial_{[i}u_{j]}-\frac{1}{3}\delta_{ij}\partial_ku^k\right) \label{enMomDecompRF}
\end{align}
with ${P'=P-\varepsilon\left.\frac{\partial P}{\partial\varepsilon}\right|_\text{lte}}$, where $\left.\frac{\partial P}{\partial \varepsilon}\right|_\text{lte}$ is computed assuming the fluid to be in local thermal equilibrium (i.e.\! ${\pi_{\mu\nu}\equiv 0}$). \par
The decomposition Eq.~(\ref{enMomDecompRF}) together with Kubo's formula result in the following expressions:\LNNL
\begin{align}
\delta\avr{P'} &= \underbrace{\lim\limits_{\omega\rightarrow 0}\frac{\avr{[P',P']}(\omega,\ve{k}=0)}{\omega}}_{=:\,\zeta}\partial_iu^i, \\
\delta\avr{T^{0i}} &= \underbrace{\lim\limits_{\omega\rightarrow 0}\frac{\frac{1}{3}\avr{[T^{0k},T_{0k}]}(\omega,\ve{k}=0)}{\omega}}_{=:\,-\kappa}(\partial_0 u^i-\partial^i\delta c), \\
\delta\avr{\pi^{ij}} &= \underbrace{\lim\limits_{\omega\rightarrow 0}\frac{\frac{1}{5}\avr{[\pi^{lm},\pi_{lm}]}(\omega,\ve{k}=0)}{\omega}}_{=:\,\eta}\left(\partial_{[i}u_{j]}-\frac{1}{3}\delta^{ij}\partial_ku^k\right),
\end{align}
with the bulk viscosity $\zeta$, the heat conductivity $\kappa$ and the shear viscosity $\eta$. The rotational invariance of the local thermal equilibrium state ensures, that the average of other anti-commutators -- with two different tensorial components in it -- vanish.

\newpage
\section{Energy-momentum tensor}\label{enmomTensor}
We discuss the detailed derivation of the energy-momentum tensor in case of a non-local quadratic action. First we translate $\varphi$ by a space-time dependent field $\alpha$. The variation of the action respect to $\alpha$ provides us the gradient of the energy-momentum tensor (in the limit when the translation field is constant). We can argue in the following way. If the system is translational invariant, the Lagrangian is also invariant for space-time-independent translations of its argument. Now let us consider a space-time dependent translation $(\alpha_x)_\mu$. The Lagrangian is changed, but properly ''close'' to the case of constant translation, this variation can be written as ${\delta\mathcal{L}=(\partial_\mu(\alpha_x)_\nu)h^{\mu\nu}(\varphi)}$, since it has to be vanish for constant $\alpha_\mu$. For the variation of the action one gets \LNNL
\begin{align}
\delta S &= \Int{x}\delta\mathcal{L} =-\Int{x}(\alpha_x)_\nu \partial_\mu h^{\mu\nu}.
\end{align}
When the EoM holds, then ${\delta S=0}$ for any variation of the field (including the space-time-dependent translation. Therefore, ${\partial_\mu h^{\mu\nu} =0}$. This relation can be identified with the energy-momentum conservation. \par
We now compute the variation of the action (\ref{EQPactionEtaOs}) for a space-time-dependent translation: \LNNL
\begin{align}
\delta S &= \frac{\mathrm{d}}{\mathrm{d}\varepsilon}\left.\Int{x}S[e^{(\alpha_0+\varepsilon\alpha)\cdot\partial}\varphi]\right|_{\varepsilon\rightarrow 0},
\end{align}
where we used the identity $$\varphi(x+z)=e^{z\cdot\partial_x}\varphi(x).$$
\begin{align}
\delta S &=\frac{\mathrm{d}}{\mathrm{d}\varepsilon}\left[\frac{1}{2}\Int{x}\varphi_{x^\mu+\varepsilon\alpha_x^\mu}\Int{z}\mathcal{K}_ze^{z\cdot\partial_x}\varphi_{x^\mu+\varepsilon\alpha_x^\mu}
\right]_{\varepsilon=0} =\\
&= \frac{1}{2}\Int{x}(\alpha_x)_\mu\partial_x^\mu\varphi_x\Int{z}\mathcal{K}_ze^{z\cdot\partial_x}\varphi_x +\frac{1}{2}\Int{x}\varphi_x\Int{z}\mathcal{K}_ze^{z\cdot\partial_x} (\alpha_x)_\alpha\partial^\mu_x\varphi_x =\\
&= -\Int{x}\alpha_x^\mu(\partial_x\cdot T_x)^\mu.
\end{align}
In order to separate the coefficient $\alpha_\mu$ it is worth to perform a Fourier-transform in $\varphi$, $\varphi^\dagger$, $T^{\mu\nu}$ and $\alpha_\mu$. This results in\LNNL
\begin{align}
\Int{k}(\alpha_k)_\mu(-ik_\nu T^{\mu\nu}_{-k}) =\frac{1}{2}\Int{k}(\alpha_k)_\mu\Int{p}\Int{q}\varphi_p^\dagger\varphi_q \Int{x}\Int{z}&\left[ -e^{ik\cdot z}ip^\mu e^{-ip\cdot x}\mathcal{K}_ze^{z\cdot\partial_x}e^{iq\cdot x} \right. \nonumber \\
& \left.+e^{-ip\cdot x}\mathcal{K}_ze^{z\cdot\partial_x}e^{ik\cdot x}iq^\mu e^{i\cdot x}\right],
\end{align}
where the difference between the field variable and its conjugate is indicated. Using the identity $\delta_{k+p-q}=\delta_{k+p-q}\frac{k\cdot(p+q)}{q^2-p^2}$ and collecting the terms result in\LNNL
\begin{align}
ik_\nu T^{\mu\nu}_k =& \frac{1}{2}\Int{p}\Int{q}\varphi_p^\dagger\varphi_q \delta_{k+p-q}\left(ip^\mu K_q-iq^\mu K_p \right)=\\
=&\frac{1}{2}\Int{p}\Int{q}\varphi_p^\dagger\varphi_q \delta_{k+p-q}\left( ip^\mu\frac{k\cdot(p+q)}{q^2-p^2}( K_q- K_p) -ik^\mu K_p\right) = \label{preDkernel} \\
=& \frac{1}{2}\Int{p}\Int{q}\varphi_p^\dagger\varphi_q \delta_{k+p-q} ip^\mu\frac{k\cdot(p+q)}{q^2-p^2}( K_q- K_p)=\\
=:& ik_\nu \frac{1}{2}\Int{p}\Int{q} \varphi_p^\dagger\varphi_q \delta_{k+p-q}D^{\mu\nu}_{p,q}. \label{Dkernel}
\end{align}
Here in Eq.~(\ref{Dkernel}) we left the last term in the parenthesis of Eq.~(\ref{preDkernel}). This can be done because of the EoM ${~K_p\varphi_p=0}$. We then take the non-$k$-orthogonal part of $T^{\mu\nu}$, such that ${ik_\mu T^{\mu\nu}_k=ik\cdot(\dots)}$ $\Rightarrow$ ${T^{\mu\nu}_k=(\dots)}$. Averaging over the equilibrium ensemble, we arrive at the energy-momentum density $\varepsilon^{\mu\nu}$:\LNNL
\begin{align}
\varepsilon^{\mu\nu} &= \Int{k}\avr{T_k^{\mu\nu}}=\frac{1}{2}\Int{k}\Int{p}\Int{q}\avr{(\varphi_p^\dagger)^{(1)}\varphi^{(2)}_q} \delta_{k+p-q}D^{\mu\nu}_{p,q}= \\
&=\frac{1}{2}\Int{k}\Int{p}\Int{q}iG^{12}_p\delta_{p-q} \delta_{k+p-q}D^{\mu\nu}_{p,q}=\frac{1}{2}\Int{p}D^{\mu\nu}_{p,p}\rho_p n_p,\,\,\text{where} \\
D^{\mu\nu}_{p,p} &= \lim\limits_{q\rightarrow p} \frac{p^\mu(p+q)^\nu}{q^2-p^2}(K_q- K_p) \overset{q=p+\zeta n}{=} \frac{p^\mu p^\nu}{n\cdot p}\lim\limits_{\zeta\rightarrow 0}\frac{ K_{p+\zeta n}- K_p}{\zeta} =\\
&\overset{K_p\equiv K_{|p|}}{=} \frac{p^\mu p^\nu}{|p|}\frac{\partial K_{|p|}}{\partial |p|} =\frac{p^\mu p^\nu}{\omega}\frac{\partial K_{|p|}}{\partial\omega}.
\end{align}
\remB{
Now we can use the parity properties ${\rho_{-p}=-\rho_p}$ and ${n_{-p}=-1-n_p}$ to rewrite the above formula into the following form:
\begin{align}
\varepsilon^{\mu\nu} &= \intpos{p}D^{\mu\nu}_{p,p}\rho_pn_p +\underbrace{\frac{1}{2}\intpos{p}D^{\mu\nu}_{p,p}\rho_p}_{=:\varepsilon^{\mu\nu}_0}.
\end{align}
The first term vanishes for zero temperature -- the integrand is suppressed by the Bose--Einstein-distribution --, whilst the second term $\varepsilon^{\mu\nu}_0$ remains also for $T=0$. In fact, this \textit{vacuum contribution} turns out to be divergent. For example, in case of a free particle (${\rho(p)\sim\delta(p^2-m^2)}$), the divergent integral is: ${\intlim{^3\ve{p}}{}{}\int_{-\infty}^\infty\mathrm{d}\omega\delta(\omega^2-\ve{p}^2-m^2) \sim \intlim{^3\ve{p}}{}{}\frac{1}{\sqrt{\ve{p}^2+m^2}}}$. \par
We avoid this contribution by simply prescribing the values of the thermodynamic observables to be zero at $T=0$:
\begin{align}
\varepsilon^{\mu\nu}_\text{ren.} &:=\varepsilon^{\mu\nu}-\varepsilon^{\mu\nu}_0.
\end{align}
This ''renormalization'' makes also the transport coefficients free from divergent vacuum contributions. In case of temperature-dependent spectral functions, however, it is not clear that this subtraction leads always to a finite result. That has to be checked separately in any examples. A possible finite contribution from $\varepsilon^{\mu\nu}_0$ -- after the subtraction -- can always be incorporated into a temperature dependent background-contribution.
}

\newpage
\section{Shear viscosity}\label{visco}
Using the definition of the spectral function of an operator, we derive $\rho_{T^\dagger T}$ of the energy-momentum tensor $T^{ij}$. With the renormalized energy-momentum tensor in Eq.~(\ref{Dkernel}) and using the relation in Eq.~(\ref{fourpiontfunc}) the computation is straightforward:\LNNL
\begin{align}
\rho_{(T^\dagger)^{ij}T^{ij},k} =& iG^{21}_{(T^\dagger)^{ij}T^{ij},k}-iG^{12}_{(T^\dagger)^{ij}T^{ij},k} = \\
=& \frac{1}{4}\Int{k'}\Int{p}\Int{q}\Int{r}\Int{s} \delta_{k+p-q}\delta_{k'+r-s}D^{ij}_{p,q}D^{ij}_{r,s}\times \nonumber \\
&\times \left(\avr{\kelphi{p}{2}\kelphi{-q}{2}\kelphi{-r}{1}\kelphi{s}{1}}-\avr{\kelphi{p}{1}\kelphi{-q}{1}\kelphi{-r}{2}\kelphi{s}{2}}\right)=\\
=& \frac{1}{4}\int_{k',p,q,r,s}\!\!\!\!\!\!\!\! \delta_{k+p-q}\delta_{k'+r-s}D^{ij}_{p,q}D^{ij}_{r,s} \times\nonumber \\
&\times \left[\delta_{p-q}\delta_{r-s}\left(iG^{22}_piG^{11}_r-iG^{11}_piG^{22}_r\right)+ \right.\nonumber \\
&\left. +\left(\delta_{p-r}\delta_{q-s}+\delta_{p+s}\delta_{q+r}\right)\left(iG^{12}_piG^{21}_q-iG^{21}_piG^{12}_q\right)\right]= \\
=& \frac{1}{4}\Int{p}\left((D^{ij}_{p,p+k})^2+D^{ij}_{p,p+k}D^{ij}_{p+k,p}\right)\rho_p\rho_{p+k}(n_p-n_{p+k}). \label{rhoTT}
\end{align}
Now we take $\ve{k}=0$ and expand the first factor of the integral kernel in Eq.~(\ref{rhoTT}):\LNNL
\begin{align}
\left. (D^{ij}_{p,p+k})^2\right|_{\ve{k}=0} =& \left. D^{ij}_{p,p+k}D^{ij}_{p+k,p}\right|_{\ve{k}=0} =\nonumber \\
=&\left[\frac{2p^ip^j}{\omega^2-2\omega\tilde{\omega}}( K_{\tilde{\omega}+\omega,\ve{p}}- K_{\tilde{\omega},\ve{p}})\right]^2 \overset{\omega\rightarrow 0}{\approx} \left(\frac{p^ip^j}{\tilde{\omega}}\frac{\partial K_{\tilde{\omega},\ve{p}}}{\partial\tilde{\omega}}\right)^2 +\mathcal{O}(\omega). \label{expandDsqu}
\end{align}
The linear term of $\rho_{T^\dagger T}$ in $\omega$ in the long-wavelength limit is the shear viscosity $\eta$. Using Eq.~(\ref{expandDsqu}) and also expanding the spectral function $\rho$ and the thermal factor ${n_p-n_{p+k}}$ to first-order in $\omega$ we get: \LNNL
\begin{align}
\eta =& \lim\limits_{\omega\rightarrow 0}\frac{\rho_{(T^\dagger)^{12}T^{12}}(\omega,\ve{k}=0)}{\omega} =\nonumber \\
=& \lim\limits_{\omega\rightarrow 0}\frac{1}{2\omega}\Int{p}\left[\left(\frac{p^1p^2}{\tilde{\omega}}\frac{\partial K_{\tilde{\omega},\ve{p}}}{\partial\tilde{\omega}}\right)^2 +\mathcal{O}(\omega)\right]\left[\rho_{\tilde{\omega},\ve{p}}^2+\omega\rho_{\tilde{\omega},\ve{p}}\frac{\partial\rho_{\tilde{\omega},\ve{p}}}{\partial\tilde{\omega}}+\mathcal{O}(\omega^2)\right]\left(-\omega \frac{\partial n_{\tilde{\omega}}}{\partial\tilde{\omega}}+\mathcal{O}(\omega^2)\right) = \\
=& \frac{1}{2}\Int{p}\left(\frac{p^1p^2}{\tilde{\omega}}\frac{\partial K_p}{\partial\tilde{\omega}}\rho_p\right)^2(-n'_{\tilde{\omega}}).
\end{align}

\newpage
\section{Scalar source term}\label{sourceTerm}
To explore the effect of non-zero vacuum-expectation value of $\varphi$, we make the identification $\varphi=\xi+\phi$ in the formulae of appendices \ref{propagators}, \ref{enmomTensor}, \ref{visco} and handle $\phi$ as a classical field, i.e.\ without Keldysh-indices. We wish to prescribe the condition $\avr{\varphi}=\phi$. Substituting ${\varphi=\xi+\phi}$ into the action with the source field $J$ we get\LNNL
\begin{align}
S[\varphi] &= \frac{1}{2}\Int{x}\Int{y}\varphi_x^\dagger\mathcal{K}_{x-y}\varphi_y +\underbrace{\frac{1}{2}\Int{x}\left(\varphi_x^\dagger J_x+\varphi_xJ_x^\dagger\right)}_{=:S_J[\varphi]}= \frac{1}{2}\Int{p}\varphi_{-p}\varphi_pK_p +\frac{1}{2}\Int{p}\left(\varphi_{-p}J_p+\varphi_pJ_{-p}\right) \\
&\overset{\varphi=\xi+\phi}{=} \frac{1}{2}\Int{p}\xi_{-p}\xi_pK_p +\frac{1}{2}\Int{p}\left(\phi_{-p}J_p+\phi_pJ_{-p}\right)+ \frac{1}{2}\Int{p}\underbrace{\left(\xi_{-p}\phi_pK_p+\phi_{-p}\xi_pK_p+\xi_{-p}J_p+\xi_pJ_{-p}\right)}_{\overset{!}{=}0}.
\end{align}
The elimination of the $\xi$-linear terms imposes the constraint $K_p\phi_p=-J_p$. The energy-momentum tensor has an additional term coming from ${~S_J[\varphi]}$. Collecting the terms with the field-combinations $\xi\xi$, $\phi\phi$ and $\xi\phi$, we arrive at\LNNL
\begin{align}
T^{\mu\nu}_k &=\frac{1}{2}\Int{p}\Int{q}\delta_{k+p-q}D^{\mu\nu}_{p,q}(\xi_{-p}+\phi_{-p})(\xi_q+\phi_q) -\Int{p}\Int{q}\delta_{k+p-q}\frac{p^\mu(p+q)^\nu}{q^2-p^2}J_q(\xi_{-p}+\phi_{-p}) \\
&=\Int{p}\Int{q}\delta_{k+p-q}\left(\mathcal{D}^{\mu\nu}_{p,q}\xi_{-p}\xi_q +\mathcal{E}^{\mu\nu}_{p,q}\phi_{-p}\phi_q +\mathcal{F}^{\mu\nu}_{p,q}\xi_{-p}\phi_q \right), \label{enMomWsource}
\end{align}
where the corresponding kernel functions read as\LNNL
\begin{align}
\mathcal{D}^{\mu\nu}_{p,q} &= \frac{1}{2}D^{\mu\nu}_{p,q}, &\\
\mathcal{E}^{\mu\nu}_{p,q} &= \frac{1}{2}D^{\mu\nu}_{p,q}+\frac{p^\mu(p+q)^\nu}{q^2-p^2}K_q, &\\
\mathcal{F}^{\mu\nu}_{p,q} &= \frac{D^{\mu\nu}_{p,q}+D^{\mu\nu}_{-q,-p}}{2}+\frac{p^\mu(p+q)^\nu}{q^2-p^2}K_q.
\end{align}
Only the terms proportional to $\xi\xi$ and $\phi\phi$ contribute to the average $\avr{T^{\mu\nu}}$, resulting an extra term compared to the case of $\phi\equiv 0$. Now, we choose a spatially homogeneous and temperature dependent background as follows:\LNNL
\begin{align}
\phi_p=\sqrt{\frac{B(T)}{K_{p=0}}}\delta_p. \label{homBGfield}
\end{align}
With this choice, we are left with the expression\LNNL
\begin{align}
\Int{k}\avr{T^{\mu\nu}_k} &= \avr{T^{\mu\nu}_{0,k}} +B,
\end{align}
which leads exactly the results we mentioned in Sec.~\ref{thermoConsist}:\LNNL
\begin{align}
\varepsilon=\varepsilon_{\phi\equiv 0} +B,&\,\, &P=P_{\phi\equiv 0}-B.
\end{align}
For calculating the spectral function $\rho_{T^\dagger T}$, first we re-observe Eq.~(\ref{enMomWsource}). The expectation value of those terms containing odd number of $\xi$ or $\phi$ fields vanishes. Terms with only $\phi$ fields cancel each other in the anti-commutator, since those do not carry Keldysh-indices. Writing out the remaining ones explicitly: \LNNL
\begin{align}
iG^{21}_{T^\dagger T} = \Int{p}\Int{q}\Int{r}\Int{s}\delta_{k+p-q}\delta_{k'+r-s} &\left( \mathcal{D}^{\mu\nu}_{p,q}\mathcal{D}^{\mu\nu}_{r,s} \avr{\xi_p^{(2)}\xi_{-q}^{(2)}\xi_{-r}^{(1)}\xi_s^{(1)}}+ \mathcal{F}^{\mu\nu}_{p,q}\mathcal{F}^{\mu\nu}_{r,s}\avr{\xi_p^{(2)}\xi_{-r}^{(2)}}\phi_{-q}\phi_s +\right.\\
&\left. +\mathcal{D}^{\mu\nu}_{p,q}\mathcal{E}^{\mu\nu}_{r,s}\avr{\xi_p^{(2)}\xi_{-q}^{(2)}}\phi_{-r}\phi_s +\mathcal{D}^{\mu\nu}_{r,s}\mathcal{E}^{\mu\nu}_{p,q}\phi_p\phi_{-q}\avr{\xi_{-r}^{(1)}\xi_s^{(1)}}\right), \\
iG^{12}_{T^\dagger T} = \Int{p}\Int{q}\Int{r}\Int{s}\delta_{k+p-q}\delta_{k'+r-s} &\left( \mathcal{D}^{\mu\nu}_{p,q}\mathcal{D}^{\mu\nu}_{r,s} \avr{\xi_p^{(1)}\xi_{-q}^{(1)}\xi_{-r}^{(2)}\xi_s^{(2)}}+ \mathcal{F}^{\mu\nu}_{p,q}\mathcal{F}^{\mu\nu}_{r,s}\avr{\xi_p^{(1)}\xi_{-r}^{(2)}}\phi_{-q}\phi_s +\right.\\
&\left. +\mathcal{D}^{\mu\nu}_{p,q}\mathcal{E}^{\mu\nu}_{r,s}\avr{\xi_p^{(1)}\xi_{-q}^{(1)}}\phi_{-r}\phi_s +\mathcal{D}^{\mu\nu}_{r,s}\mathcal{E}^{\mu\nu}_{p,q}\phi_p\phi_{-q}\avr{\xi_{-r}^{(2)}\xi_s^{(2)}}\right).
\end{align}
The spectral function of the composite operator ${~(T^\dagger)^{\mu\nu}T^{\mu\nu}}$ is the difference of the two above-written formulae:\LNNL
\begin{align}
\rho_{T^\dagger T} &=iG^{21}_{T^\dagger T}-iG^{12}_{T^\dagger T} = \rho_{T^\dagger T,\,0}+ \\
& +\Int{p}\Int{q}\Int{r}\Int{s}\delta_{k+p-q}\delta_{k'+r-s}\mathcal{F}^{\mu\nu}_{p,q}\mathcal{F}^{\mu\nu}_{r,s}\phi_{-q}\phi_s \delta_{p-r}\rho_p + \label{rhoTTSourcePlusTerm1}\\
&+\Int{p}\Int{q}\Int{r}\Int{s}\delta_{k+p-q}\delta_{k'+r-s}\left(\mathcal{D}^{\mu\nu}_{r,s}\mathcal{E}^{\mu\nu}_{p,q}\phi_p\phi_{-q}\delta_{r-s}(iG^{11}_r-iG^{22}_r)+ \right.\nonumber \\
& \left. +\mathcal{D}^{\mu\nu}_{p,q}\mathcal{E}^{\mu\nu}_{r,s}\phi_{-r}\phi_s\delta_{p-q}(iG^{22}_p-iG^{11}_p)\right). \label{rhoTTSourcePlusTerm2}
\end{align}
The first additional term compared to the case of ${\phi\equiv 0}$ is Eq.~(\ref{rhoTTSourcePlusTerm1}). In case of the homogeneous background in Eq.~(\ref{homBGfield}), it simplifies to\LNNL
\begin{align}
\sim (\mathcal{F}^{\mu\nu}_{k,0})^2\rho_k\frac{B}{K_0} \overset{\ve{k}=0}{\rightarrow} 0,
\end{align}
since for space-space indices all the three kernel functions vanish in the long-wavelength limit ${p=0}$, if either of their arguments vanishes:
\[\begin{array}{ccccc}
\left.\mathcal{D}^{ij}_{p,q=0}\right|_{\ve{p}=0}=0,& \,\, & \left.\mathcal{E}^{ij}_{p,q=0}\right|_{\ve{p}=0}=0, &\,\, &\left.\mathcal{F}^{ij}_{p,q=0}\right|_{\ve{p}=0}=0.
\end{array}\]
Eq.~(\ref{rhoTTSourcePlusTerm2}) is the second additional term to $\rho_{T^\dagger T,\,0}$. For the background Eq.~(\ref{homBGfield}) it reads as\LNNL
\begin{align}
\sim & \delta_k\mathcal{E}_{0,0}^{\mu\nu}\underbrace{\left(\Int{r}\mathcal{D}_{r,r}^{\mu\nu}(iG^{11}_r-iG^{22}_r) +\Int{p}\mathcal{D}_{p,p}^{\mu\nu}(iG^{22}_p-iG^{11}_p)\right)}_{\equiv\, 0,\,\,\text{for any }\mu\text{ and }\nu}\frac{B}{K_0}=0,
\end{align}
which vanishes for any $\mu$ and $\nu$ pairs. Whereas neither Eq.~(\ref{rhoTTSourcePlusTerm1}) nor Eq.~(\ref{rhoTTSourcePlusTerm2}) contribute to the spectral function ${~\rho_{T^\dagger T}}$, the expression of the shear viscosity does not modify in the case of a homogeneous, temperature dependent background.

\newpage
\section{QP-approximation with square-root cut self-energy}\label{QPsumruleAnal}
We begin with the analysis of the sum rule Eq.~(\ref{QPsumrule}), together with the self-energy in Eqs.~(\ref{cutImSigma}),~(\ref{cutReSigma}). In the case of ${\gamma=0}$, the following two equations hold:\LNNL
\begin{align}
M_*^2&=m^2+2\zeta\,\sqrt{1-\frac{M^2}{M_*^2}}\text{arcsin}\frac{M_*}{M}, \label{poleMassEq1}\\
\frac{\partial\re\Sigma_s(M_*)}{\partial p} &= \frac{2\zeta}{M_*}-M_*\frac{1-\frac{m^2}{M_*^2}}{1-\frac{M_*^2}{M^2}}, \label{poleMassEq2}
\end{align}
where we already used Eq.~(\ref{poleMassEq1}) to get the right-hand side of Eq.~(\ref{poleMassEq2}). The sum rule takes the form:\LNNL
\begin{align}
1 =& \underbrace{\frac{1}{1-\frac{\zeta}{M_*^2}+\frac{1}{2}\frac{1-m^2/M_*^2}{1-M_*^2/M^2}}}_{=:Z_p} +\nonumber \\
&+\underbrace{\frac{\zeta}{M^2}\intlim{y}{1}{\infty}\frac{\sqrt{y^2-1}}{\left(y^2-\frac{m^2}{M^2}+2\frac{\zeta}{M^2}\text{ln}\left(y-\sqrt{y^2-1}\right)\sqrt{1-\frac{1}{y^2}}\right)^2+\pi^2\left(\frac{\zeta}{M^2}\right)^2\left(1-\frac{1}{y^2}\right)}}_{=:Z_c}. \label{QPsumruleApp}
\end{align}
\begin{figure}
\centering
\subfloat[]{
\includegraphics[width=0.5\linewidth]{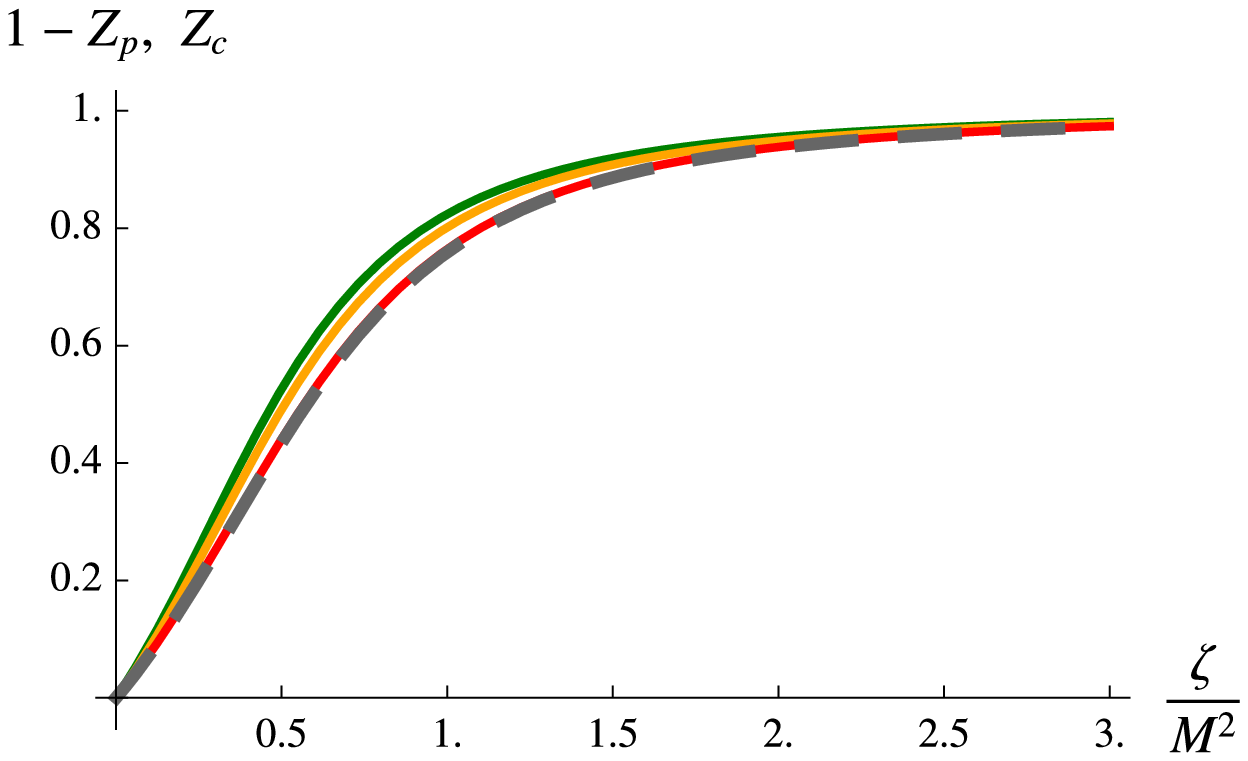}
}
\subfloat[]{
\includegraphics[width=0.5\linewidth]{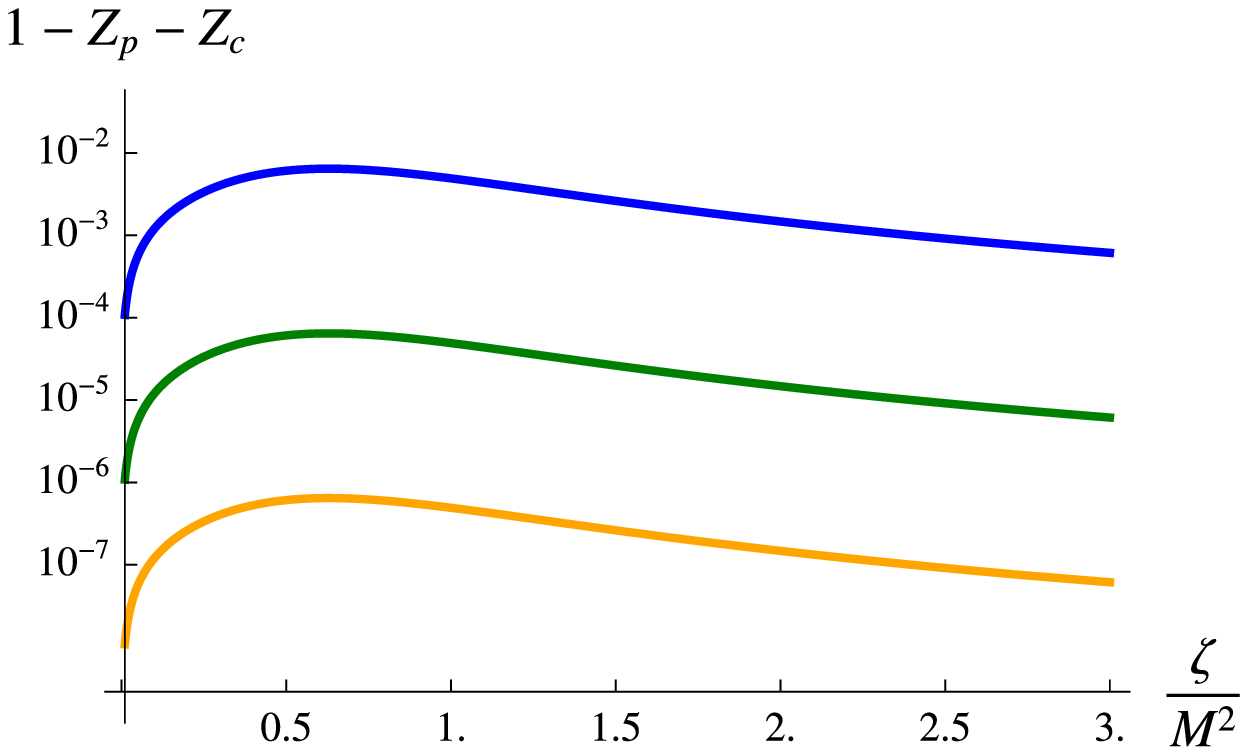}
}
\LNFIG
\caption{(a): The relationship between $1-Z_p$ and $Z_c$ versus $\zeta/M^2$ for various values of $m^2/M^2$. The dashed curve corresponds to $Z_c$ with $m^2/M^2=0.1$. The continuous curves correspond to $1-Z_p$: $m/M=0.5$ (green), $m/M=0.4$ (orange), $m/M=0.1$ (red). For $m/M=0.1$ the two curves are cover each other, the difference is invisible for the naked eye. (b): The fulfilment of the sum rule $Z_p+Z_c=1$ regardless of the value of $\zeta$, in various values of $m/M$: 0.1 (blue), 0.01 (green), 0.001 (orange).}
\label{fig:QPapprSumrule}
\end{figure}
In zeroth order in ${\mathcal{O}(\zeta/M^2,\,m^2/M^2)}$, i.e.\! ${Z_p\approx 1}$ we get\LNNL
\begin{align}
M_*^2 &\approx m^2+2\zeta +\mathcal{O}\left(\frac{\zeta}{M^2},\,\frac{m^2}{M^2}\right). \label{poleMassAppr}
\end{align}
The error made by the previous approximation is suppressed at least by a factor of $M^{-2}$:\LNNL
\begin{align}
Z_c &\approx \frac{\zeta}{M^2}\intlim{y}{1}{\infty}\frac{\sqrt{y^2-1}}{y^4} +\frac{\zeta}{M^4}\intlim{y}{1}{\infty}\frac{-2\sqrt{y^2-1}\left(2\zeta\sqrt{1-\frac{1}{y^2}}\ln(y-\sqrt{y^2-1})-m^2\right)}{y^6} +\mathcal{O}(M^{-6}) = \nonumber \\
&=\frac{1}{3}\frac{\zeta}{M^2} +\frac{14}{45}\frac{\zeta^2}{M^4}+\frac{12}{45}\frac{m^2\zeta}{M^4} +\mathcal{O}\left(\frac{\zeta^3}{M^6},\,\frac{\zeta m^4}{M^6},\,\frac{\zeta^2m^2}{M^6}\right)
\end{align}
This effect is demonstrated on Fig.~\ref{fig:QPapprSumrule}. Let us write down the viscosity kernel $g^2(p)$ for the QP-approximation:\LNNL
\begin{align}
g^2(p) =& \left[\frac{\im\Sigma_s}{(p^2-m^2-\re\Sigma_s)^2+(\im\Sigma_s)^2}\left(2p-\frac{\partial\re\Sigma_s}{\partial p}\right)\right]^2 =\nonumber \\
&\overset{\im\Sigma_s\ll m^2}{\approx} \frac{1}{2}\delta(p-M_*) \frac{2M_*-\frac{\partial\re\Sigma_s(M_*)}{\partial p}}{ \im\Sigma_s(M_*)},
\end{align}
where we performed the QP-approximation in the last step, by taking the leading order of the expansion by the vanishing imaginary part of the self-energy. With the entropy density ${s_{QP}(T,M_*)}$, the ratio $\eta$ over $s$ takes the following form:\LNNL
\begin{align}
\frac{\eta}{s} &\overset{M\gg m,\,M_*}{\approx} \frac{1}{2}\frac{2M_*-\frac{\partial\re\Sigma_s(M_*)}{\partial p}}{ \im\Sigma_s(M_*)}\frac{T^4\lambda_\eta(M_*/T)}{T^3\chi_s(M_*/T)} =\nonumber \\
&=\frac{T^2}{\im\Sigma_s(M_*)}\left(1- \frac{\zeta}{M_*^2}+\frac{1-\frac{m^2}{M_*^2}}{2}\right) \overset{\text{Eq.}}{\underset{(\ref{poleMassAppr})}{=}} \frac{T^2}{\im\Sigma_s(M_*)}. \label{QPetaOslargeM1}
\end{align}
As a next step, we give an estimation for $\im\Sigma_s(M_*)$ for large $M$. Since it vanishes for ${\gamma=0}$, we have to choose a small, non-zero $\gamma$. We link its value to the ''non-interacting'' QP-limit, ${\zeta=0}$, in a physically motivated way: demanding ${\eta/s}$ to be finite\footnote{Note here, that the $p$-dependence of the self-energy does not make $\eta/s$ finite. In fact, $\eta$ will always diverge if there is a lone Dirac-delta in $\overline{\rho}$.}. Identifying the ${\zeta=0}$ case with the propagator ${\frac{1}{p^2-m^2-i\gamma_0p}}$, in the large-$M$/small-$p$ limit we get:\LNNL
\begin{align}
\im\Sigma_s(M\rightarrow\infty) &\approx \zeta\frac{\gamma^2p}{M^3}+\mathcal{O}(p^3M^{-5}) \overset{!}{=} \gamma_0 p. \label{QPetaOslargeMgamma}
\end{align}
The large-$M$ estimation in Eq.~(\ref{QPetaOslargeM1}) then can be written further, and finally resembles the result of our QP-analysis done in Sec.~\ref{exQPapprox}, cf.\! Eq.~(\ref{etaOsQP}):\LNNL
\begin{align}
\frac{\eta}{s} &\overset{M\gg m,\,M_*}{\approx} \frac{T^2M^3}{\zeta M_*\gamma^2} \overset{\text{Eq.}}{\underset{(\ref{QPetaOslargeMgamma})}{=}} \frac{T^2}{\gamma_0 M_*}= \frac{T^2}{\gamma_0\sqrt{m^2+2\zeta}}, \label{QPetaOslargeM2}
\end{align}
\clearpage
\chapter*{Acknowledgements}

First and foremost I would like to express my sincere gratitude to my supervisor Tam\'as S\'andor Bir\'o, who was always open to discuss my ideas and gave me guidance whenever my work got off-track or I lost my motivation. I learned a lot from him about how to become a better scientist and -- I think -- also about myself. I am also truly grateful to my consultant Antal Jakov\'ac, who is always ready to discuss interesting things not just about physics. But I cannot deny how much my thinking changed about physics, mostly due to our acquaintance. \par
I am really thankful for all the help, encouragement and advices I got from my closest colleagues in the Heavy-Ion Research Group at the Wigner RCP: Gergely G\'abor Barnaf\"oldi, P\'eter V\'an and K\'aroly \"Urm\"ossy.\par
This dissertation would be much less elaborated without the careful reading and thoughtful criticism by P\'eter Kov\'acs and Gergely Mark\'o. I can only hope that their invested time has paid off as the improvement of this work. \par
I am very lucky to have such friends like P\'eter Mati, M\'arton Vargyas, L\'aszl\'o Holl\'o, J\'ozsef Konczer and M\'ate Lencs\'es (and also those I have indelicately forgot to mention here). They cheered me up and encouraged me when I was down, so many times that I am sure I would not be here, where I am, without them. \par
\FINAL{Csak rem\'elhetem, hogy a v\'eg n\'elk\"uli seg\'its\'eget, szeretetet \'es \'erdekl\H{o}d\'est, amit a csal\'adomt\'ol kaptam mindeddig, alkalmam ny\'ilik viszonozni. K\"osz\"on\"om a t\'amogat\'asukat, ami annyi k\"ul\"onb\"oz\H{o} m\'odon seg\'itett, hogy p\'ar sz\'oban k\"osz\"onetet mondani \'erte lehetetlen.} \par
Last, but not least I thank Defu Hou and Xin-Nian Wang for giving me a job at CCNU, China!\\[10\baselineskip]
{\small The following sources of financial support were a great help during my PhD years: Hungarian Research Fund (OTKA) under contract Numbers K104260, K104292 and K116197; the Campus Hungary Mobility Scholarship (with which I spent 2 months in Frankfurt in the end of 2013) and the ''Young Talents of the Nation Scholarship'' NTP-EF\"O-P-15 of the Human Capacities Grant Management Office, Hungary (that I was finally able to get a new laptop).}
\clearpage
\nocite{apsrev41Control}
\bibliographystyle{apsrev4-1} 
\raggedright
\end{document}